\documentclass[preprint2]{aastex}
\usepackage{epsfig}

\begin{document}


\newcommand{\bq}{\begin{equation}}
\newcommand{\eq}{\end{equation}}
\newcommand{\kcorr}{$K$-correction}
\newcommand{\kcorrs}{$K$-corrections}
\newcommand{\dmm}{\mbox{$\Delta$m$_{15}(B)$}}
\def\Lbsun{\hbox{$\rm\thinspace L_{B\odot}$}}
\def\Mbsun{\hbox{$\rm\thinspace M_{B\odot}$}}
\newcommand{\gapprox}{{_> \atop{^\sim}}} 
\newcommand{\lapprox}{{_< \atop{^\sim}}} 

\bibliographystyle{aj}

\title{Spectra of High-Redshift Type Ia Supernovae and a
Comparison with their Low-Redshift Counterparts\footnote{Much of the
data presented herein were obtained at the W.M. Keck Observatory,
which is operated as a scientific partnership among the California
Institute of Technology, the University of California and the National
Aeronautics and Space Administration. The Observatory was made
possible by the generous financial support of the W.M. Keck
Foundation.  Also based in part on observations made with the European
Southern Observatory telescopes (ESO programs 58.A-0745 and
59.A-0745).}}

\author{I.~M.~Hook\altaffilmark{1},
D.~A.~Howell\altaffilmark{2},
G.~Aldering\altaffilmark{3},
R.~Amanullah\altaffilmark{4},
M.~S.~Burns\altaffilmark{5},
A.~Conley\altaffilmark{3,6},
S.~E.~Deustua\altaffilmark{7},
R.~Ellis\altaffilmark{8},
S.~Fabbro\altaffilmark{9},
V.~Fadeyev\altaffilmark{3},
G.~Folatelli\altaffilmark{4},
G.~Garavini\altaffilmark{4,10},
R.~Gibbons\altaffilmark{11},
G.~Goldhaber\altaffilmark{3,6},
A.~Goobar\altaffilmark{4},
D.~E.~Groom\altaffilmark{3},
A.~G.~Kim\altaffilmark{3},
R.~A.~Knop\altaffilmark{11},
M.~Kowalski\altaffilmark{3},
C.~Lidman\altaffilmark{12},
S.~Nobili\altaffilmark{4,10},
P.~E.~Nugent\altaffilmark{3},
R.~Pain\altaffilmark{10},
C.~R.~Pennypacker\altaffilmark{3},
S.~Perlmutter\altaffilmark{3,6},
P.~Ruiz-Lapuente\altaffilmark{13},
G.~Sainton\altaffilmark{10},
B.~E.~Schaefer\altaffilmark{14},
E.~Smith\altaffilmark{11},
A.~L.~Spadafora\altaffilmark{3},
V.~Stanishev\altaffilmark{4},
R.~C.~Thomas\altaffilmark{3},
N.~A.~Walton\altaffilmark{15},
L.~Wang\altaffilmark{3}, and
W.~M.~Wood-Vasey\altaffilmark{3,6}\\
(THE SUPERNOVA COSMOLOGY PROJECT)}
\altaffiltext{1}{Department of Physics, University of Oxford, Nuclear 
\& Astrophysics Laboratory,  Keble Road, Oxford, OX1 3RH, UK}
\altaffiltext{2}{Department of Astronomy and Astrophysics, University 
of Toronto, 60 St. George St., Toronto, Ontario M5S 3H8, Canada}
\altaffiltext{3}{E. O. Lawrence Berkeley National Laboratory, 1 
Cyclotron Rd., Berkeley, CA 94720, USA }
\altaffiltext{4}{Department of Physics, Stockholm University,  Albanova 
University Center, S-106 91 Stockholm, Sweden}
\altaffiltext{5}{Colorado College, 14 East Cache La Poudre St., 
Colorado Springs, CO 80903}
\altaffiltext{6}{Department of Physics, University of California 
Berkeley, Berkeley, 94720-7300 CA, USA}
\altaffiltext{7}{American Astronomical Society,  2000 Florida Ave, NW, 
Suite 400, Washington, DC, 20009 USA.}
\altaffiltext{8}{California Institute of Technology, E. California 
Blvd, Pasadena,  CA 91125, USA}
\altaffiltext{9}{CENTRA-Centro M. de Astrof\'{\i}sica and Department of 
Physics,    IST, Lisbon, Portugal }
\altaffiltext{10}{LPNHE, CNRS-IN2P3, University of Paris VI \& VII, 
Paris, France }
\altaffiltext{11}{Department of Physics and Astronomy, Vanderbilt 
University, Nashville, TN 37240, USA}
\altaffiltext{12}{European Southern Observatory, Alonso de Cordova 
3107, Vitacura, Casilla 19001, Santiago 19, Chile }
\altaffiltext{13}{Department of Astronomy, University of Barcelona, 
Barcelona, Spain }
\altaffiltext{14}{Louisiana State University, Department of Physics and 
Astronomy,
Baton Rouge, LA, 70803, USA}
\altaffiltext{15}{Institute of Astronomy, Madingley Road, Cambridge CB3 
0HA, UK }

\begin{abstract}

We present spectra for 14 high-redshift ($0.17 < z < 0.83$)
supernovae, which were discovered by the Supernova Cosmology Project
as part of a campaign to measure cosmological parameters. The spectra
are used to determine the redshift and classify the supernova type,
essential information if the supernovae are to be used for
cosmological studies. Redshifts were derived either from the spectrum
of the host galaxy or from the spectrum of the supernova itself. We
present evidence that these supernovae are of Type~Ia by matching to
spectra of nearby supernovae. We find that the dates of the spectra
relative to maximum light determined from this fitting process are
consistent with the dates determined from the photometric light
curves, and moreover the spectral time-sequence for SNe Type~Ia at low
and high redshift is indistinguishable. We also show that the
expansion velocities measured from blueshifted Ca~H\&K are consistent
with those measured for low-redshift Type Ia supernovae. From these
first-level quantitative comparisons we find no evidence for evolution
in SNIa properties between these low- and high-redshift samples. Thus
even though our samples may not be complete, we conclude that there is
a population of SNe Ia at high redshift whose spectral properties
match those at low redshift.

\end{abstract}


\section{Introduction}

The peak magnitudes of Type Ia supernovae (SNe~Ia) are one of the best
distance indicators at high redshifts, where few reliable distance
indicators are available to study the cosmological parameters.
Beginning with the discovery of SN 1992bi \citep{sn92bi}, the
Supernova Cosmology Project (SCP) has developed search techniques and
rapid analysis methods that allow systematic discovery and follow up
of ``batches'' of high-redshift supernovae. These searches and those
of the High-z SN team have resulted in $\sim100$ published SNe ($0.15
< z < 1.2$), which have been used for measurements of the cosmological
parameters $\rm \Omega_M$ and $\rm \Omega_{\Lambda}$ and to provide
initial constraints on the equation of state of the Universe, $w$
\citep{nature98,schmidt_98,42SNe_98,garn_w_98,riess_acc_98,knop03,barris04,
tonry03,reiss04}.

Spectra have been obtained for as many of the SCP supernovae as
possible, and as close as possible to maximum light. The spectra are
used to determine the redshift of the event and to confirm its
spectral sub-type, both crucial if the supernova is to be used in a
determination of cosmological parameters. In addition the spectra
provide a check that SNe~Ia at high redshift are similar to those at
low redshift, an assumption central to the use of SNe~Ia as
`standardized candles'.  This paper represents a significant
contribution to the amount of published high-redshift SN Ia
spectroscopy \citep{lidman05,barris04,tonry03,coil00,Matheson}.

In this paper we present spectra for a subset of our distant
supernovae, discovered during search campaigns in January and March
1997. Since the goal of this paper is a first quantitative comparison
of high- and low-redshift SN spectra, spectra were chosen where the SN
features were relatively clear. Spectra of 33 objects were taken
during the two campaigns, of which four were found to be clear
broad-lined QSOs and two were ``featureless'' blue objects (possibly BL
Lacs) for which a redshift could not be measured. The other objects
all have measured redshifts and of these about half (13) were rejected
for the purposes of this paper based on one or both of the following
criteria: low signal-to-noise ratio (i.e. less than about 10 per
20\AA\ bin over the observed wavelength range 6000-8000\AA) or large
contamination by the host galaxy, roughly corresponding to a
percentage increase of less than 50\% in $R$-band photometry between
reference and ``new'' images (although the latter is only a rough guide
since the phase of the SN, its location relative to the core of the
galaxy and the seeing all strongly affect how clearly the SN appears
in the spectrum).

The spectra presented here provide the basis for the identification of
these objects as Type Ia, as quoted in \cite{42SNe_98} and
\cite{knop03}.  Future papers by G. Garavini et al. and E. Smith et
al. will present spectroscopic results and analysis on other SCP
datasets.

For each object in our sample we give the redshift and present the
results of matching the spectrum to various supernova templates
(including SNe of Types other than Ia), and hence give spectroscopic
evidence that these distant supernovae are of Type Ia.  The spectral
dates derived from this matching process are compared with the dates
derived from the light curves.  Finally we present the set of
high-redshift SN spectra as a time sequence showing that the spectrum
changes with light curve phase in a similar way at low and high
redshift.

\section{Discovery of the Supernovae}

The observing strategy developed for these high-redshift SN search
runs involved comparison of large numbers of galaxies in each of
$\sim$50 $14.7'\times14.7'$ fields observed twice with a separation of
3--4 weeks \citep{perl96}. This strategy ensures that almost all the
supernovae are discovered before maximum light, and, since it could be
guaranteed that at least a dozen supernovae would be found on a given
search run, the follow-up observing time could be scheduled in
advance. This strategy made it possible to schedule spectroscopy time
on large telescopes while the supernovae were still close to maximum
light, and thus the supernova features can be observed even at
redshifts $>$1. Follow-up photometric measurements were also made of
the supernova light-curves, from which the maximum brightness and date
of maximum is derived.

The supernovae described in this paper were identified on $R$-band
images taken in two searches on the CTIO 4-m Blanco telescope in
January and March 1997. Candidate supernovae were identified on the
difference images (see \cite{perl96} for details on the identification
of candidates). Those supernovae identified as Type Ia were followed
photometrically in the $R$- and $I$-bands at the WIYN telescope at
KPNO, the ESO 3.6m and one (SN 1997ap) was followed with the {\it
Hubble Space Telescope} \citep{nature98}.

Parameterised light-curve fits for all but one of the objects
presented in this paper (SN 1997ag, which does not have sufficient
photometric light curve measurements for cosmological use) are
published in \citet{42SNe_98}.  These SNe form part of the set that
provided the first evidence for the accelerating expansion of the
universe and the presence of some form of dark energy driving this
expansion. Many of these objects (but excluding SN~1997ag, SN~1997G,
SN~1997J and SN~1997S because of poor colour measurements,
$\sigma_{R-I}>0.25$) were also used in the more recent measurements of
cosmological parameters by \citet{knop03}.

\section{Spectroscopic Observations}

The spectra were obtained during two observing runs at the Keck-II
10-m telescope (two nights in January and three in March 1997) and one
run at the ESO 3.6m (1 night in January 1997). The spectroscopic runs
were scheduled to occur within one week of the corresponding search
run at CTIO. At the time of observations, close to maximum light, the
typical magnitudes of the supernova candidates were 22-24 magnitude in
$R$-band and in many cases their host galaxies had similar apparent
magnitudes. In all cases where the galaxy was bright enough, the slit
was aligned at a position angle on the sky such that the
supernova and the center of the host galaxy were both in the
slit. This was done to allow redshift determination from features in
the host galaxy spectrum.

Observations of the spectrophotometric standards HD84937, HD19445 and
BD262606 \citep{oke_abmag_83} were taken at Keck-II, and LTT 1788
\citep{SB83,BS84,Hamuy94} was observed at ESO, in order to obtain
approximate relative flux calibration of the spectra. The standards
were observed at the parallactic angle, although note that the SN
spectra were not necessarily taken at the parallactic angle as
described above. Although the effects of atmospheric dispersion are
small when observing in the red (as is the case for these spectra),
some small wavelength-dependent slit losses may occur.  Therefore the
overall slope of the spectra is not considered to be reliable and the
slope is left as a free parameter in the analysis that follows.

\paragraph{Keck data} 
The LRIS spectrograph \citep{B095} with the 300~l/mm grating was used
at the Cassegrain focus of the Keck-II 10m telescope. The spectra
cover the wavelength range 5000 to 10000\AA\ with dispersion of $\sim$
2.5\AA\ per pixel.  Typical exposure times were 0.5 to 1.5 hours.

During the January run, dome flats were taken at the position of
targets when possible. This was necessary because flexure causes the
fringe pattern to shift depending on the zenith distance and rotator
angle. Previous tests showed that dome flats gave marginally better
results compared to internal flats when trying to subtract sky lines
at the red end of the spectrum.  During the March run, bad weather and
technical problems severely limited the amount of usable
time. Therefore, to allow us to observe all the candidates, the
exposure times for each object were reduced and only internal flats
were taken. Hg-Ne-Ar arc spectra for wavelength calibration were
obtained at the same position as the flats.

\paragraph{ESO 3.6m data} The EFOSC1 spectrograph \citep{Buzzoni84} 
was used at the cassegrain focus of the ESO 3.6m. The R300 grism was
used, giving a dispersion of 3.5\AA/pixel over the range 5900 to
10000\AA.  Dome flats were used to flat-field the data, and wavelength
calibration was carried out using observations of He-Ar (ESO) arc
lamps.

Table~\ref{snobstab} gives the dates and exposure times for the
observations of the various supernova candidates.

\begin{deluxetable}{lcllcr}
\tablecaption{Summary of spectral observations.\label{snobstab}}
\tablewidth{0pt}
\tablehead{
\colhead{IAU name}&\colhead{R mag\tablenotemark{a}}&\colhead{Telescope}&\colhead{Exp}    &\colhead{Date} \\
         &               &                   &\colhead{(hours)}&\colhead{(UT)} }
\startdata 
1997F & 23.9 & Keck-II     & 1.0  & 1997 Jan 12   \\
1997G & 23.7 & Keck-II     & 1.3  & 1997 Jan 13   \\
1997I & 20.9 & ESO 3.6m    & 1.0  & 1997 Jan 13   \\
1997J & 23.4 & Keck-II     & 1.3  & 1997 Jan 13   \\
1997N & 21.0 & ESO 3.6m    & 0.5  & 1997 Jan 13   \\
1997R & 24.4 & Keck-II     & 1.4  & 1997 Jan 13   \\
1997S & 23.6 & Keck-II     & 1.8  & 1997 Jan 13   \\
1997ac & 23.1 & Keck-II    & 0.17 & 1997 Mar 14   \\ 
1997af & 23.8 & Keck-II    & 0.58 & 1997 Mar 14   \\
1997ag & 23.2 & Keck-II    & 0.25 & 1997 Mar 14   \\
1997ai & 22.3 & Keck-II    & 0.35 & 1997 Mar 13   \\
1997aj & 22.3 & Keck-II    & 0.67 & 1997 Mar 13   \\
1997am & 22.9 & Keck-II    & 0.31 & 1997 Mar 13   \\
1997ap & 24.2 & Keck-II    & 1.5  & 1997 Mar 14   \\
\enddata
\tablenotetext{a}{The magnitudes given are for the supernovae at the
time of the spectroscopic observations, and were estimated from nearby
light-curve photometry points.  These are accurate to about
0.2 mag.
}
\end{deluxetable}

\section{Data Reduction}

The spectra were reduced using the IRAF\footnote{IRAF is distributed
by the National Optical Astronomy Observatories, which are operated by
the Association of Universities for Research in Astronomy, Inc., under
cooperative agreement with the National Science Foundation.} spectral
reduction package. The data were first overscan-subtracted, bias
corrected and flat-fielded using the flats described above.  The
spectra were then extracted to provide 1-D spectra of the targets. In
some cases the supernova was sufficiently offset from the core of its
host galaxy that it was possible to extract the supernova spectrum and
that of the host galaxy separately. In other cases it was only
possible to extract a single spectrum containing a mix of supernova
and host galaxy light.  These spectra were then wavelength calibrated
and flux calibrated.  Except where otherwise stated explicitly, a
correction for atmospheric absorption features was made.

\section{Identification}

SNe~Ia are classified spectroscopically by a lack of hydrogen and the
presence of a strong \ion{Si}{2} feature at $\sim$6150\AA\ in the
rest frame (see \cite{1997ARAA..35..309F} for a review). However, at
redshifts above 0.5 this line moves into the infrared and can no
longer be seen with typical CCDs. Therefore we use the following
criteria to classify a spectrum as a SN~Ia when this line is not seen:

\begin{itemize}

\item Presence of the SII ``W'' feature at $\sim$5500\AA. This is
only seen in Type Ia supernovae (see Figure~\ref{Imax}).

\begin{figure*} 
\includegraphics[scale=0.6,angle=270]{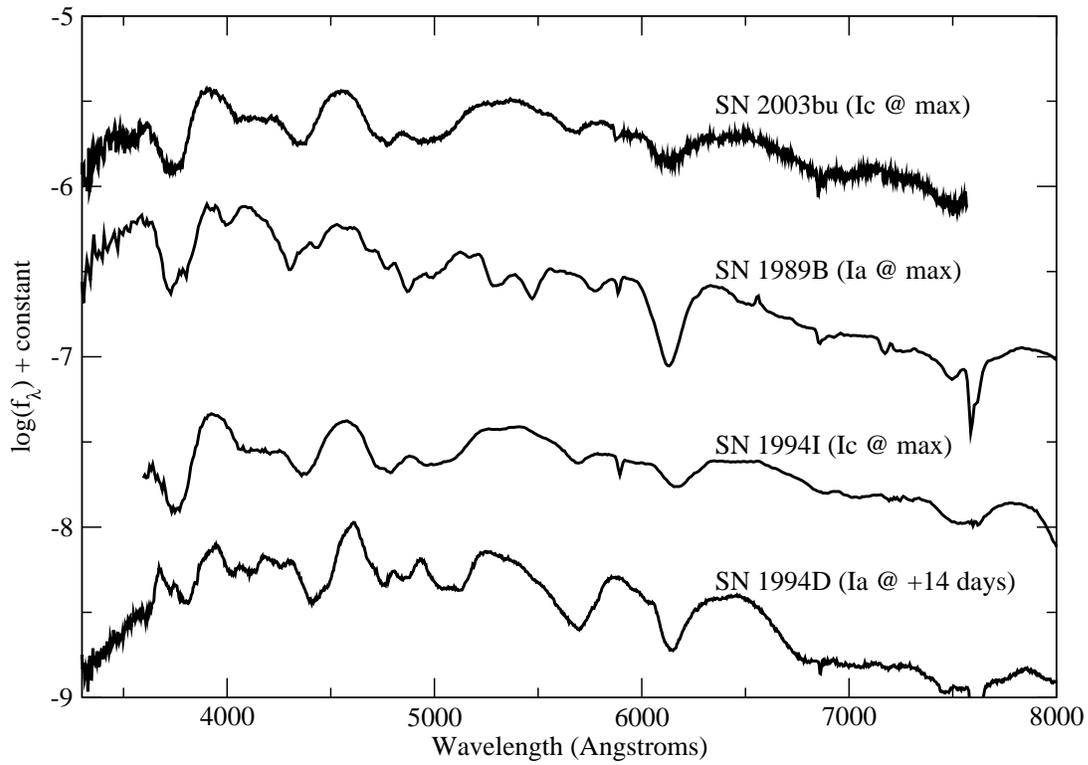}
\caption{Comparison of nearby SNe Ia and Ic at maximum light.  SNe Ic can
closely resemble SNe Ia blueward of 4500\AA\ with the exception of the
4000\AA\ \ion{Si}{2} feature.  Also, SNe Ic have a group of Fe
features spanning 4500--5200\AA\ that resemble the same features in a
Type Ia spectrum $\sim$2 weeks after maximum light. The data shown are
SN 1989B \citep{1994AJ....108.2233W}, SN 2003bu (P. Nugent, private
communication), SN 1994I \citep{1996ApJ...462..462C} and SN 1994D
\citep{1996MNRAS.281..263M}.}
\label{Imax}
\end{figure*}

\item Presence of \ion{Si}{2} feature at $\sim$4000\AA. If seen,
this feature indicates that the supernova is definitely of Type Ia,
although for spectra prior to maximum light or for SN 1991T-like
SNe~Ia, this feature can be weak. Note that it is easy to mistake Ca
H\&K from the host galaxy for this feature.

\item The combination of both the light curve age with the temporal
evolution of the \ion{Fe}{2} features at $\sim$4500 and 5100\AA.  In a SN
Ib/c at maximum light, these \ion{Fe}{2} features resemble those in a
SN Ia about two weeks after maximum light (Figure~\ref{Imax}), but the
appearance of a Type Ia spectrum at maximum light is unique.

\end{itemize}

\subsection{Matching Technique}

The spectra in this paper were matched against template SN spectra
using a spectrum matching code developed by \citet{howell02}.  All
fits were performed after rebinning the data to 20\AA\ to allow more
rapid runs of the fitting program.  At a given redshift, the code
computes:
$$\chi^2=\frac{w(\lambda)(O(\lambda)-aT(\lambda)10^{cA_{\lambda}}-bG(\lambda))^2}{\sigma(\lambda)^2},$$
where $w(\lambda)$ is a weighting function, $O$ is the observed
spectrum, $T$ is the SN template spectrum, $G$ is the host galaxy
template spectrum, $A_{\lambda}$ is the redding law, $\sigma$ is the
error on the spectrum, and $a$, $b$, and $c$ are constants that are
varied to find the best fit.  The weighting function was set to be
unity across the spectrum except at the telluric features, which were
given zero weight.  This equation is iterated over a range of
redshifts to find the minimum $\chi^2$ in redshift, host galaxy,
template supernova, and reddening space.  In this analysis, the
reddening law (\citet{card89,ODonn94}) was fixed at $R_V=3.1$,
although the spectrum matching code can handle other values of $R_V$.

When the redshift was well determined from narrow galaxy lines, it was
constrained in the fits.  Likewise, when the host galaxy type was
known from \citet{sullivan03}, it was fixed in the following manner.
For Sullivan type 0, E and S0 galaxies were tested.  For type 1, Sa and
Sb galaxies were allowed, and for type 2, Sb, Sc, and starburst galaxy
types were used.  The galaxy templates subtracted here were those of
\citet{kinney93} which have been smoothed and had their narrow lines
removed.

In three cases, SN 1997ac, SN 1997ai and SN 1997ap, it was not
possible to obtain a host galaxy redshift.  In these cases the
redshift was determined from fits to SN templates.  For each template,
the best-fit redshift was determined by stepping through redshift
space in 0.001 increments. In section 5.3 we describe the estimation
of redshift uncertainties for these cases.

The fitting program is intended to aid human classifiers of SN spectra
--- it is not a replacement for them.  The program returns a list of
fits in decreasing order of goodness-of-fit.  The output is inspected
and a ``best fit'' is chosen.  A good indication of the level of
certainty of a match is the amount of agreement between the best fits.
For data with good signal-to-noise ratio (S/N) and little host
contamination, there is almost always excellent agreement between the
top five fits.

The program is also limited by the available template data.  For
example, less than a dozen SNe have UV spectroscopy near or before
maximum light \citep{IUE}, and core collapse SNe are particularly
underrepresented in terms of UV observations.  Thus at high redshift,
where the observed optical band corresponds to rest-frame UV, it can
be hard to obtain conclusive results.  One other problem area is at
very early times.  From day $-18$ to $-8$ there are few spectra of SNe
Ia, and these are the epochs where SNe Ia show the greatest diversity.
Furthermore, \ion{Si}{2} 4000\AA\ can be weak or absent at very early
times, making a secure classification difficult.  Finally, a week
after maximum light SNe Ib/c and SNe Ia show the greatest similarity
in their spectra (see Figure~\ref{Imax}).  We note that more UV
spectra of SNe Ib/c would be beneficial to the classification
process. With such spectra it would be possible to confirm some cases
which are probable Type Ia but for which Type Ib/c cannot be
eliminated.  If \ion{Si}{2} 6150\AA\ is not in the observed spectral
range, or there is host contamination, classification at this phase
can be harder than at other phases. The list of nearby supernovae used
in the fitting process is given in Table~\ref{library}. A full list of
the dates of these spectra and their references will be presented by
Howell (in preparation).

\begin{deluxetable}{ll}
\tablecaption{A list of the nearby supernovae used as templates in the fitting
procedure. Details of the dates of these spectra and their references
will be presented by Howell (in preparation). The SUSPECT SN archive
\citep{Richardson02} was used in the creation of this
library.\label{library}}
\tablehead{
\colhead{SN Type}& \colhead{Supernova Name}}
\startdata
Ia & SN 1981B, SN 1986G, SN 1989B, SN 1990N, SN 1991T, SN 1991bg, SN 1992A, \\
   & SN 1994D, SN 1996X, SN 1998bu, SN 1999aa, SN 1999ac, SN 1999ao, SN 1999av,\\
   & SN 1999aw, SN 1999be, SN 1999bk, SN 1999bm, SN 1999bn, SN 1999bp, SN 1999ee \\

II & SN 1979C, SN 1980K, SN 1984E, SN 1986I, SN 1987A, SN 1987B, SN 1987K,\\
   & SN 1988A, SN 1988Z, SN 1993J, SN 1993W, SN 1997cy, SN 1999em, SN 1999em\tablenotemark{a} \\

Ibc & SN 1983N, SN 1983V, SN 1984L, SN 1987M, SN 1988L, SN 1990B, SN 1990U,\\ 
    & SN 1990aa, SN 1991A, SN 1991K, SN 1991L, SN 1991N, SN 1991ar, SN 1994I,\\
    & SN 1995F, SN 1995bb, SN 1996cb, SN 1997C, SN 1997dd, SN 1997dq, SN 1997ei,\\
    & SN 1997ef, SN 1998bw, SN 1998dt, SN 1999P, SN 1999bv, SN 1999di, SN 1999dn,\\
    & SN 1999ex, SN 2000H \\
\enddata
\tablenotetext{a}{Theoretical SN 1999em spectra provided by
Peter Nugent}
\end{deluxetable}

Near maximum light, normal SNe Ia show a characteristic pattern of
absorption features at rest-frame 3800\AA\ (\ion{Ca}{2}), 4000\AA\
(\ion{Si}{2}), 4300\AA\ (\ion{Mg}{2} and Iron-peak lines), 4900\AA\
(various Iron-peak lines), 5400\AA\ (SII 'W'), 5800\AA\ (\ion{Si}{2}
and \ion{Ti}{2}), and 6150\AA\ (\ion{Si}{2}).  If the S/N of the
spectrum is low, typical Ia-indicator lines such as \ion{Si}{2} and
SII may be hard to identify definitively.  However, the overall
pattern of features in a SN Ia at a given epoch is unique, so fits to
the overall spectrum (with sufficient S/N) can allow a secure
classification of the SN type, especially when there is an independent
confirmation of the SN epoch from the light curve.

In summary, this procedure provides a more robust measure of the SN
type than unaided human typing by eye. Furthermore, it uses the shape
of the entire spectrum to classify SNe, not just one or two key
features.  The technique also provides a spectroscopic determination
of the rest-frame date the spectrum was taken (relative to maximum
light), which can be checked against the light curve, yielding an
independent consistency check as shown in the next section.

\subsection{Matching Results}
\begin{figure*} 
\epsscale{1.4}
\plotone{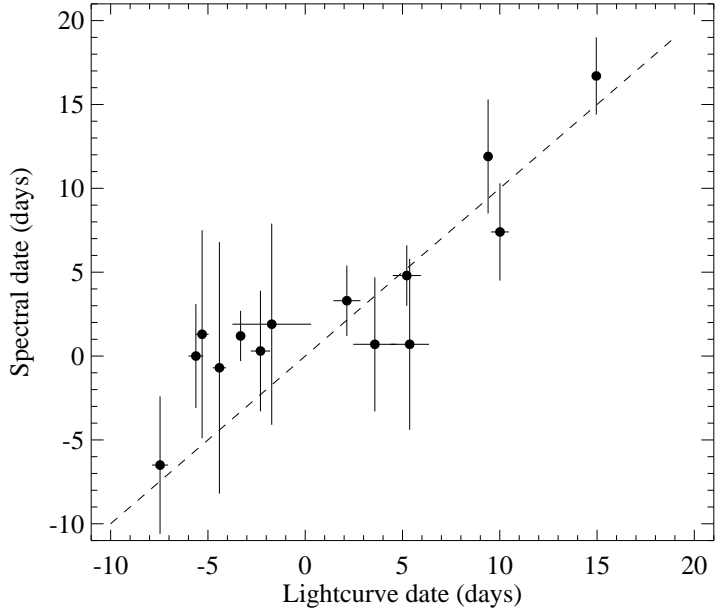}
\caption{Comparison of the epoch of the spectrum as determined from
the light curve ($\tau_{lc}$) with that determined from the weighted
average of the 5 best template fits to the spectrum ($\tau_{s}$). In
both cases the epochs are expressed as rest-frame days relative to
maximum light. The dashed line shows the locus $\tau_{lc}$ =
$\tau_{s}$.}
\label{specday}
\end{figure*}

The results of the fitting procedure are given in Table
\ref{fittable}.  We show the best fit SN template, the type of host
galaxy that was subtracted (if any), the redshift, and the epoch of
the template spectrum.  In addition, we give a ``spectroscopic
epoch,'' which is the weighted average of the epochs from the five
best-fit spectra.  This spectroscopic determination of the epoch of
the SN can be compared with the epoch determined from the light curve,
also presented in Table \ref{fittable}.  Figure \ref{specday} shows
that the two numbers are in good agreement, with an RMS difference of
3.3 days.

\begin{deluxetable}{llrrrlcl}
\tablecaption{Summary of fitting results.\label{fittable}
}
\tablehead{
\colhead{IAU name} &\colhead{$z$} &\colhead{$\tau_{\rm lc}\tablenotemark{a}$} &\colhead{$\tau_{\rm s}\tablenotemark{b}$} & \colhead{$\tau_{\rm t}\tablenotemark{c}$} & \colhead{Match} &\colhead{G\tablenotemark{d}} &\colhead{G\tablenotemark{e}} }
\startdata
1997F  &$0.580 \pm 0.001$ &$ -7.5\pm 0.4 $ & $ -6.5 \pm 4.1$&$ -8$&1999ee &1&Sa \\
1997G  &$0.763 \pm 0.001$ &$ +5.4\pm 1.0 $ & $ +0.7 \pm 5.1$&$ +4$&1999bk &2&-  \\
1997I  &$0.172 \pm 0.001$ &$ -3.3\pm 0.1 $ & $ +1.2 \pm 1.5$&$ +1$&1999bp &1&Sb \\
1997J  &$0.619 \pm 0.001$ &$ +3.6\pm 1.1 $ & $ +0.7 \pm 4.0$&$ +2$&1999bn &0&S0 \\
1997N  &$0.180 \pm 0.001$ &$+15.0\pm 0.1 $ & $+16.7 \pm 2.3$&$+15$&1991T  &2&SB2\\
1997R  &$0.657 \pm 0.001$ &$ -5.6\pm 0.4 $ & $ +0.0 \pm 3.1$&$ -5$&1989B  &1&Sa \\
1997S  &$0.612 \pm 0.001$ &$ +2.1\pm 0.7 $ & $ +3.3 \pm 2.1$&$ +5$&1992A  &-&SB1\\
1997ac &$0.323 \pm 0.005$\tablenotemark{f} &$ +9.4\pm 0.1 $ & $+11.9 \pm 3.4$&$+11$&1998bu &-&-  \\
1997af &$0.579 \pm 0.001$ &$ -5.3\pm 0.3 $ & $ +1.3 \pm 6.2$&$ -2$&1999bp &2&SB6\\
1997ag &$0.592 \pm 0.001$ &$ -1.7\pm 2.0 \tablenotemark{g} $ & $ +1.9 \pm 6.0$&$ -7$&1990N  &-&Sa \\
1997ai &$0.454 \pm 0.006$\tablenotemark{f} &$ +5.2\pm 0.7 $ & $ +4.8 \pm 1.8$&$ +5$&1992A  &-&-  \\
1997aj &$0.581 \pm 0.001$ &$ -4.4\pm 0.3 $ & $ -0.7 \pm 7.5$&$ -3$&1999aa &2&Sb \\
1997am &$0.416 \pm 0.001$ &$+10.0\pm 0.5 $ & $ +7.4 \pm 2.9$&$ +9$&1992A  &2&SB1\\
1997ap &$0.831 \pm 0.007$\tablenotemark{f} &$ -2.3\pm 0.5 $ & $ +0.3 \pm 3.6$&$ -5$&1989B  &2&-  \\
\enddata
\tablenotetext{a}{Spectral epoch relative to $B$ light curve maximum (in the rest frame) as determined from light curve
fitting \citep{knop03}}
\tablenotetext{b}{Spectral epoch from weighted average of 5 best-fit
spectra. The uncertainty quoted is the weighted standard deviation of
the epochs of the 5 best SN Ia fits.}
\tablenotetext{c}{Spectral epoch of best matching comparison template spectrum.}
\tablenotetext{d}{Host galaxy type from \citet{sullivan03}; 0: E-S0, 1: Sa-Sb, 2: Sc and later}
\tablenotetext{e}{Template host galaxy spectrum subtracted from the data for the fits.}

\tablenotetext{f} {Redshift determined from the supernova alone. These
values are derived based on the matching procedure described in
section 5.1 and supercede those in \cite{42SNe_98,knop03}.} The
uncertainties in these redshifts were estimated as described in
section 5.3.

\tablenotetext{g} {This SN has a poorly-constrained light-curve
because only a small number of photometric measurements were made. In
order to estimate the time of maximum it was necessary to constrain
the stretch parameter to s=1 when fitting the light-curve.}

\end{deluxetable}

The program also formally returns an estimate of the amount of
reddening (or de-reddening) required to match the observed spectrum,
but this number may not reflect the actual reddening toward the SN.
This parameter accounts for all differences in color between the
observation and the template, so differential slit losses due to
observations not taken at the parallactic angle, wavelength-dependent
seeing, errors in flux calibration, and uncorrected reddening in the
templates, all make an accurate determination of the reddening to the
SN from this spectroscopy alone unfeasible.

\subsection{Measurement of redshifts}

In most cases the slit angle for the spectroscopic observations was
chosen so that light from both the SN and the host galaxy (when
visible on the CTIO images) fell down the slit. Since features in the
galaxy spectra are typically much narrower than those in supernova
spectra it is often possible to obtain a redshift from the galaxy
features, even in cases where the supernova and host galaxy light are
blended. Table~\ref{fittable} summarises the measured redshifts.

In the cases where galaxy features were visible, the centroids of the
lines were measured, and the redshift calculated by taking the mean of
the redshifts derived from the individual lines. The uncertainty in
the mean redshift $z$ is estimated as $\pm 0.001$.

In the cases where there were no identifiable host galaxy features to
determine the redshift (SN 1997ac, SN 1997ai and SN 1997ap), the
redshift was determined from fits to SN templates as described in
section 5.1. In order to estimate the uncertainty in these redshifts
we used the other 11 cases, whose redshifts were known from the host
galaxies, as follows. Each high-redshift spectrum was cross-correlated
with the corresponding best-fit template (using the IRAF
cross-correlation routine ``fxcor'') allowing redshift to vary. In
this sample of 11 cases, we found a mean redshift difference between
SN and host redshift of -0.0012 and an RMS difference of 0.005. Thus
there does not appear to be any significant systematic redshift error
and the uncertainty in redshift is about 0.005. This test measures the
uncertainty in redshift determination when the ``correct'' template is
used. For cases where the redshift is not known from the host galaxy
there is an additional uncertainty from the diversity of velocities
seen in nearby template spectra. To estimate the size of this effect
we took the standard deviation of the velocity of the \ion{Si}{2}
6150\AA\ feature for the SNe Ia presented in Benetti et
al. (2005). Our calculated value of 1300 km s$^{-1}$ corresponds to a
redshift uncertainty of 0.004. Combining these two effects in quadrature
gives an estimated uncertainty in redshift of about 0.006 on
average. However we also note the uncertainties should increase with
redshift (and indeed this general trend is seen in our sample of 11
objects) because the lines become broader, the S/N becomes poorer, and
the spectrum corresponds more to the rest-frame UV where there are
fewer templates available. Thus we use a simple relation of
$0.004\times(1+z)$ to estimate the errors (which gives an error of
0.006 at our average redshift of $z\sim 0.5$).

\section{Results}

\begin{figure} 
\epsscale{1.0}
\plotone{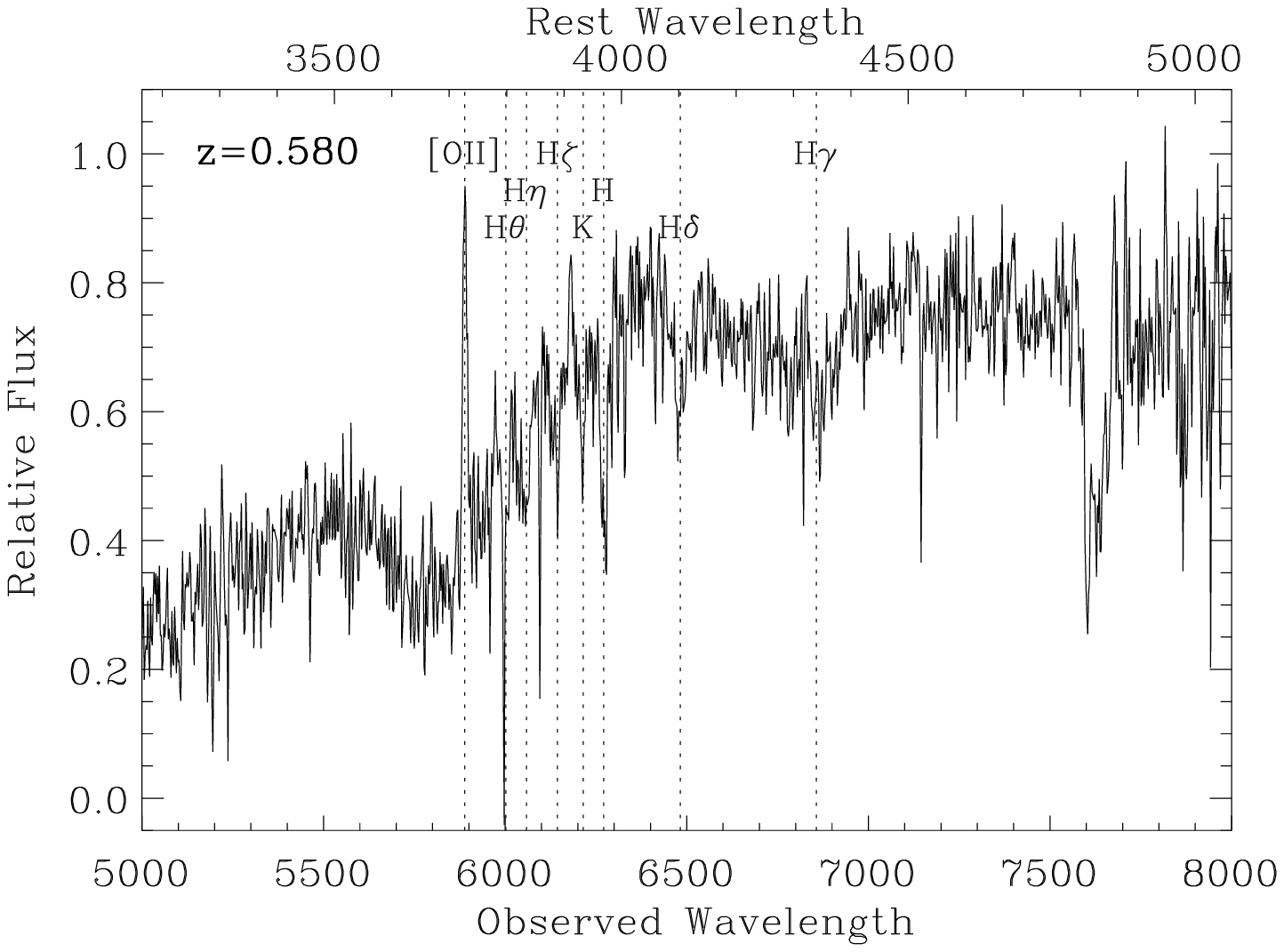}
\plotone{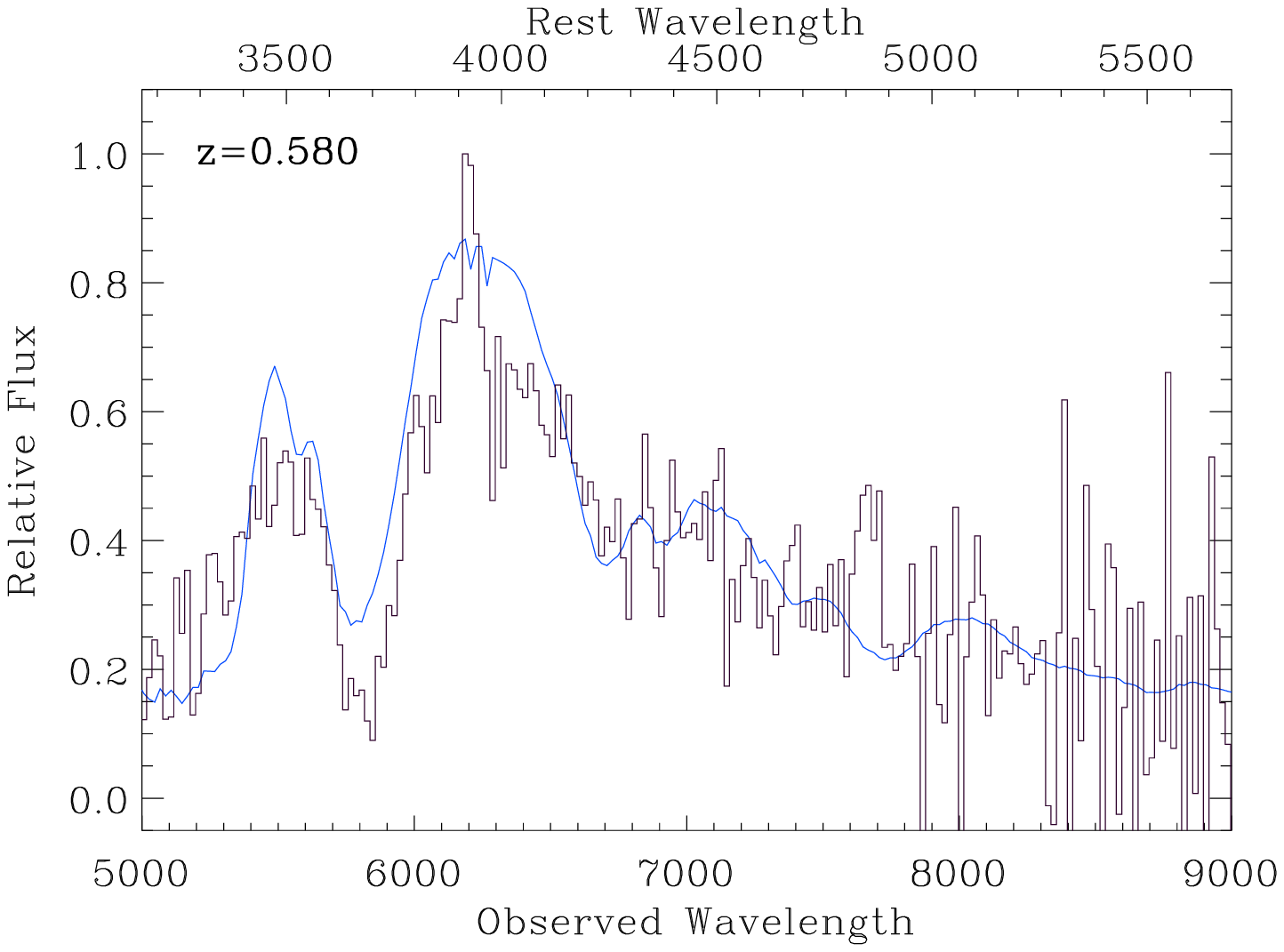}
\caption{(Upper) Unsmoothed spectrum of SN 1997F showing narrow features
used to determine the redshift (note that atmospheric absorption has
not been corrected in this spectrum). (Lower) Rebinned spectrum of SN 1997F
(histogram) after interpolating over narrow galaxy lines and sky
absorption feature at 7600\AA, and subtracting template Sa galaxy,
compared with SN 1999ee at $-8$ days (smooth curve).}
\label{974spec}
\end{figure}

\begin{figure}
\plotone{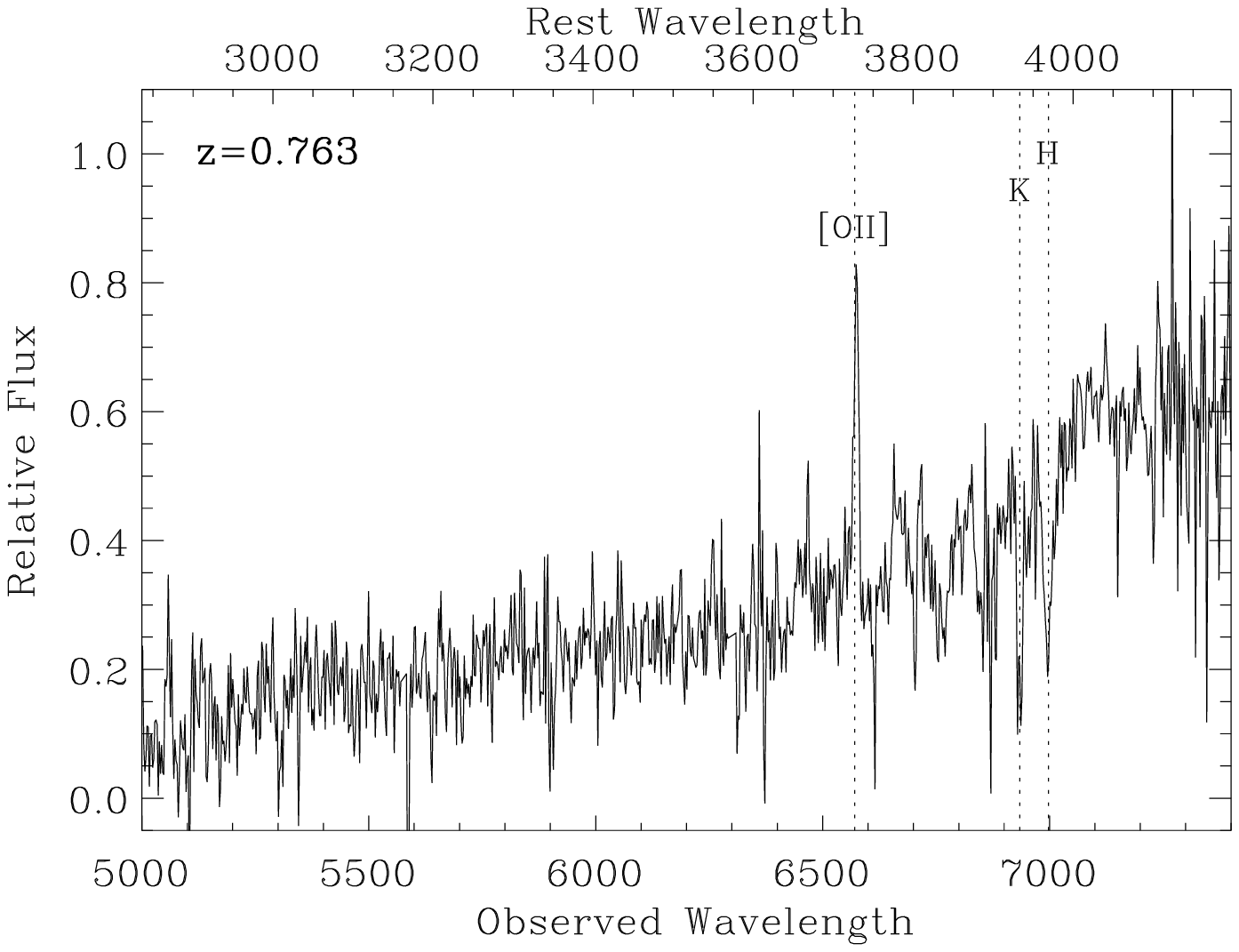}
\plotone{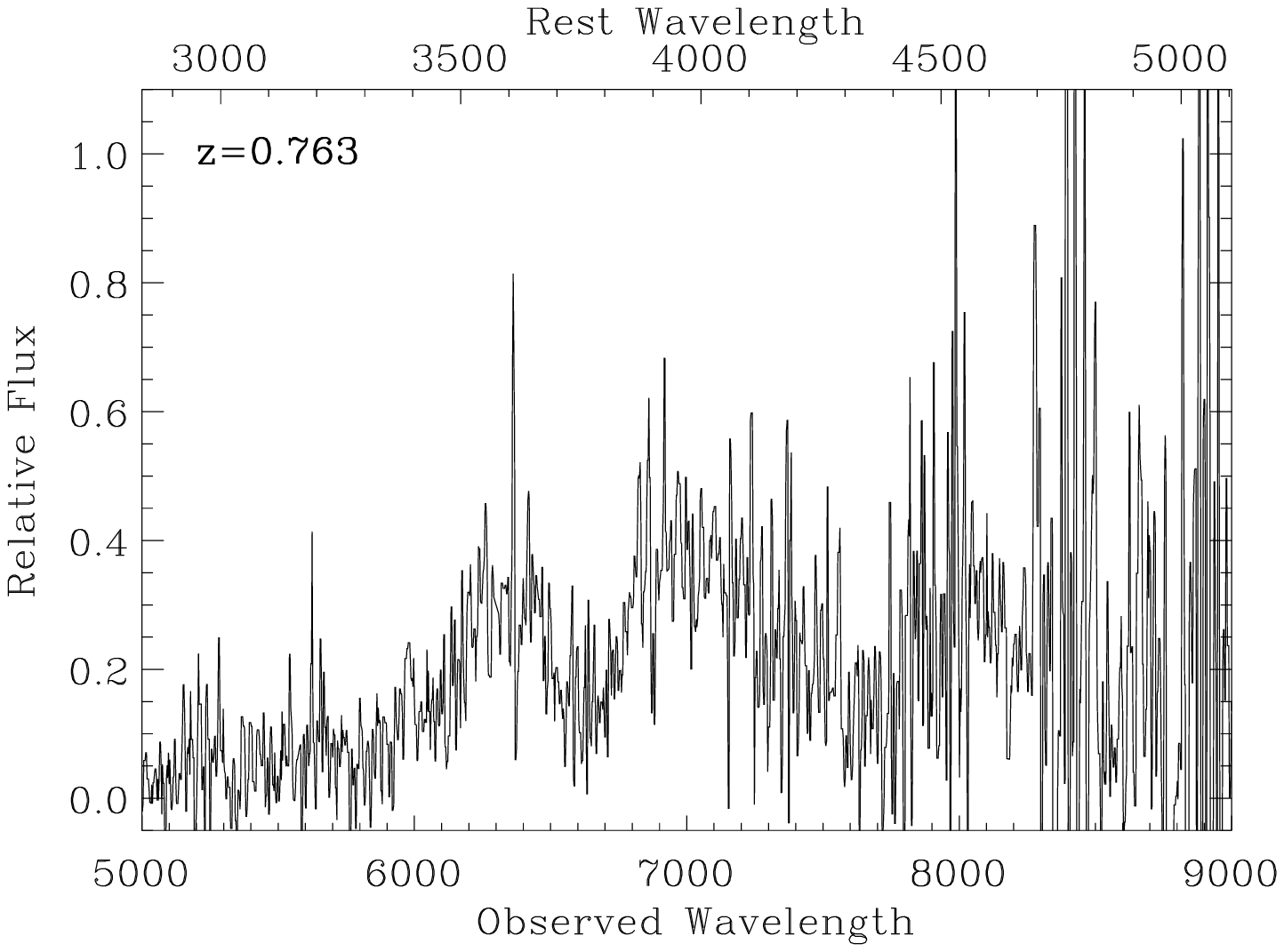}
\plotone{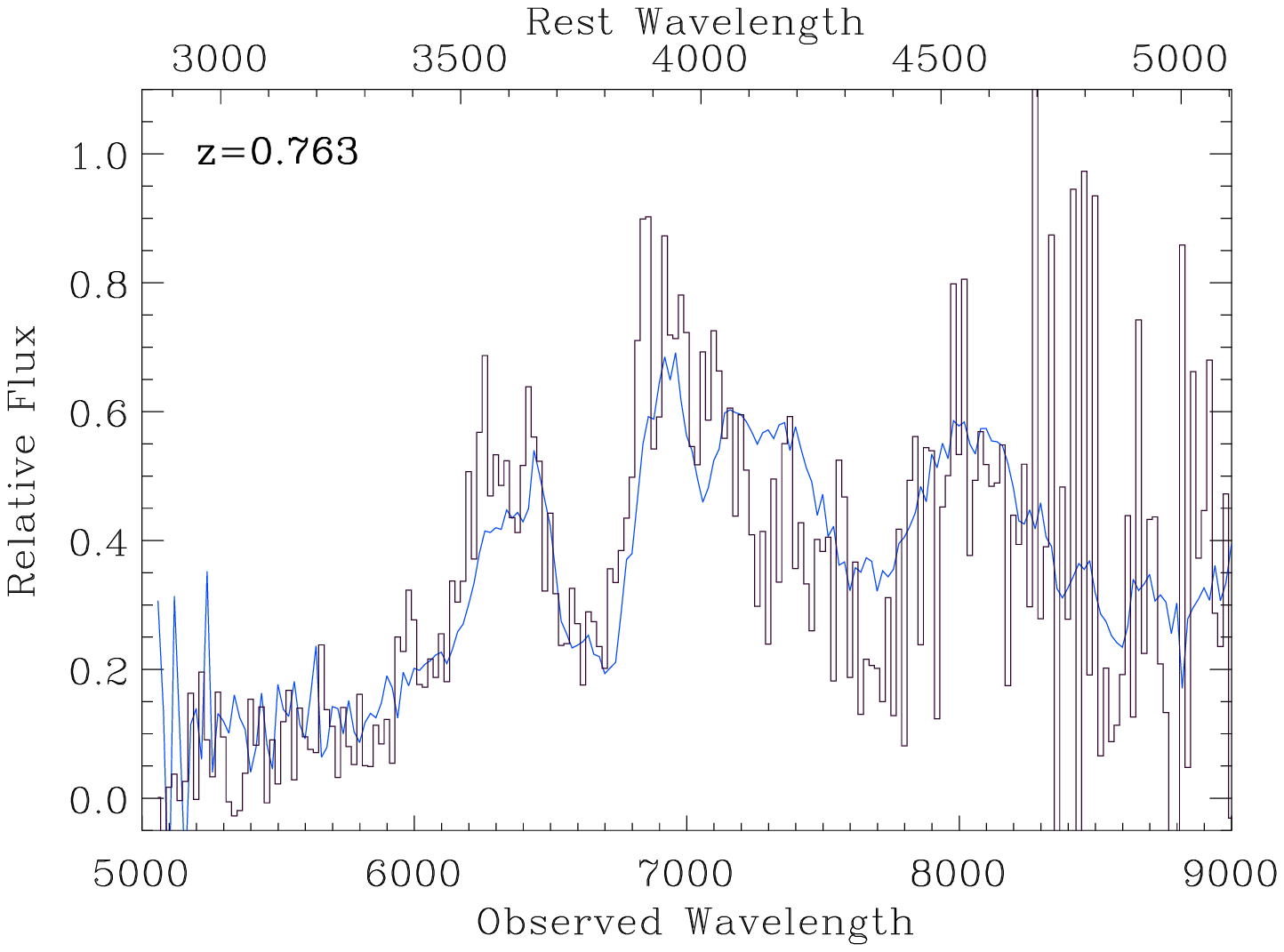}
\caption{(Upper) Lightly smoothed spectrum of neighbouring (host) galaxy
of SN 1997G (note that atmospheric absorption has not been corrected
in this spectrum). (Lower) Lightly smoothed spectrum of the supernova SN
1997G (lower) rebinned spectrum of SN 1997G (histogram) after interpolating
over poorly-subtracted sky lines and subtracting an Sa galaxy
template, compared with SN 1999bk at +4 days (smooth curve).}
\label{SN1997G_spec}
\end{figure}

\begin{figure} 
\plotone{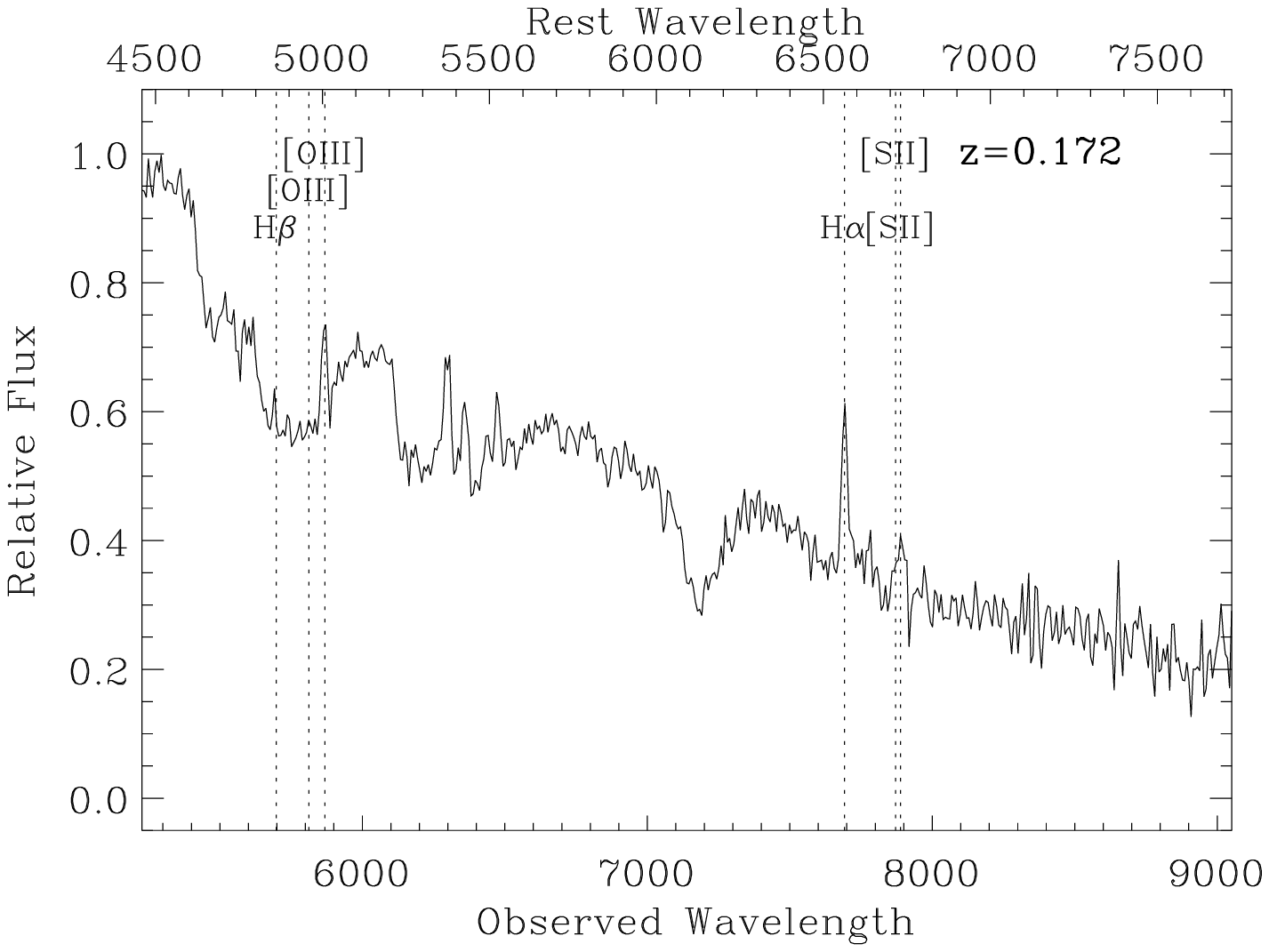}
\plotone{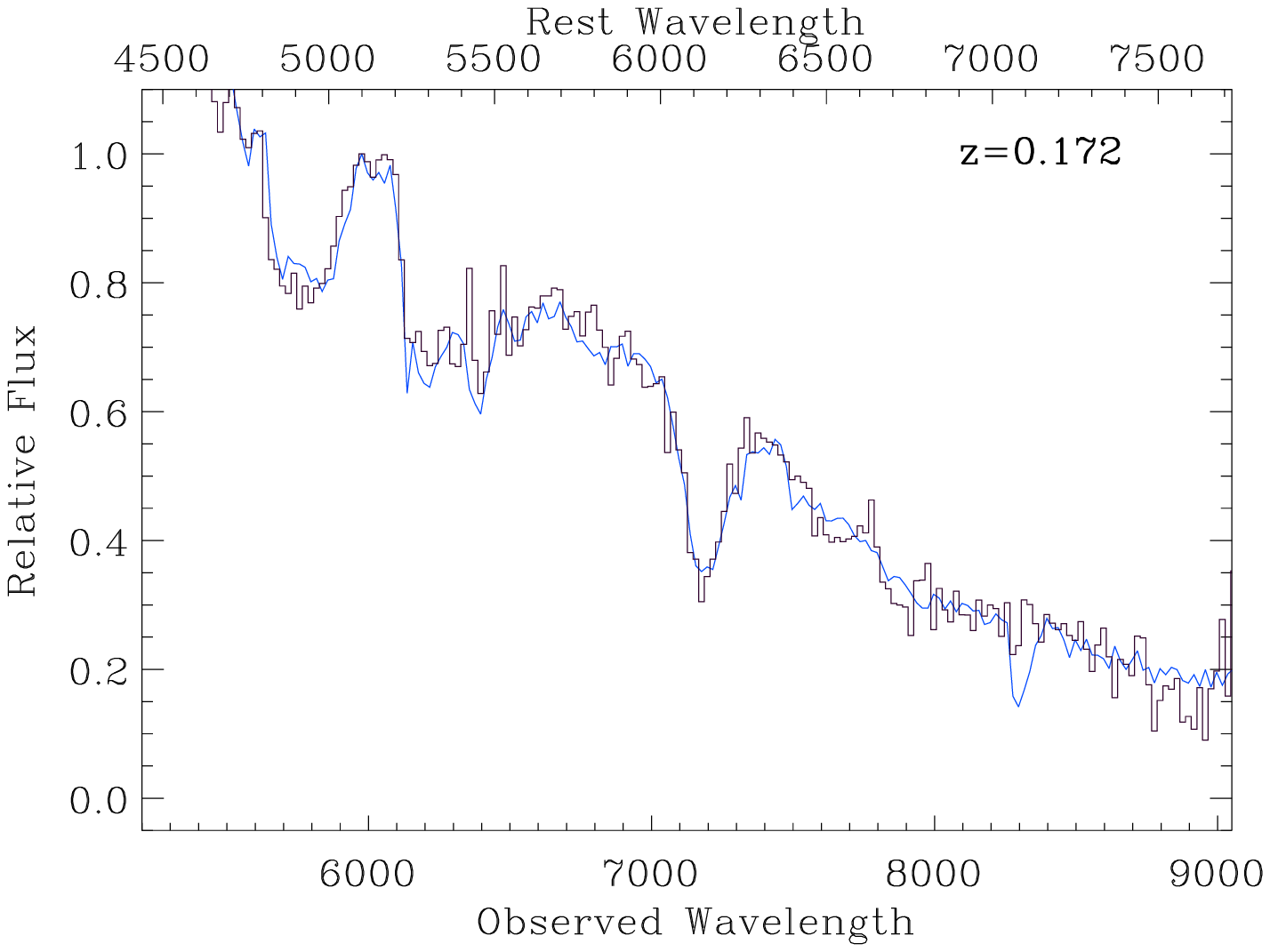}
\caption{(Upper) Unsmoothed spectrum of SN 1997I obtained at the ESO 3.6m
showing galaxy lines used to determine the redshift. (Lower) Rebinned
spectrum of SN 1997I (histogram) after interpolating over narrow
galaxy lines and subtraction of an Sb template galaxy spectrum,
compared with SN 1999bp at +1 day (smooth curve).}
\label{970spec}
\end{figure}

\begin{figure}
\plotone{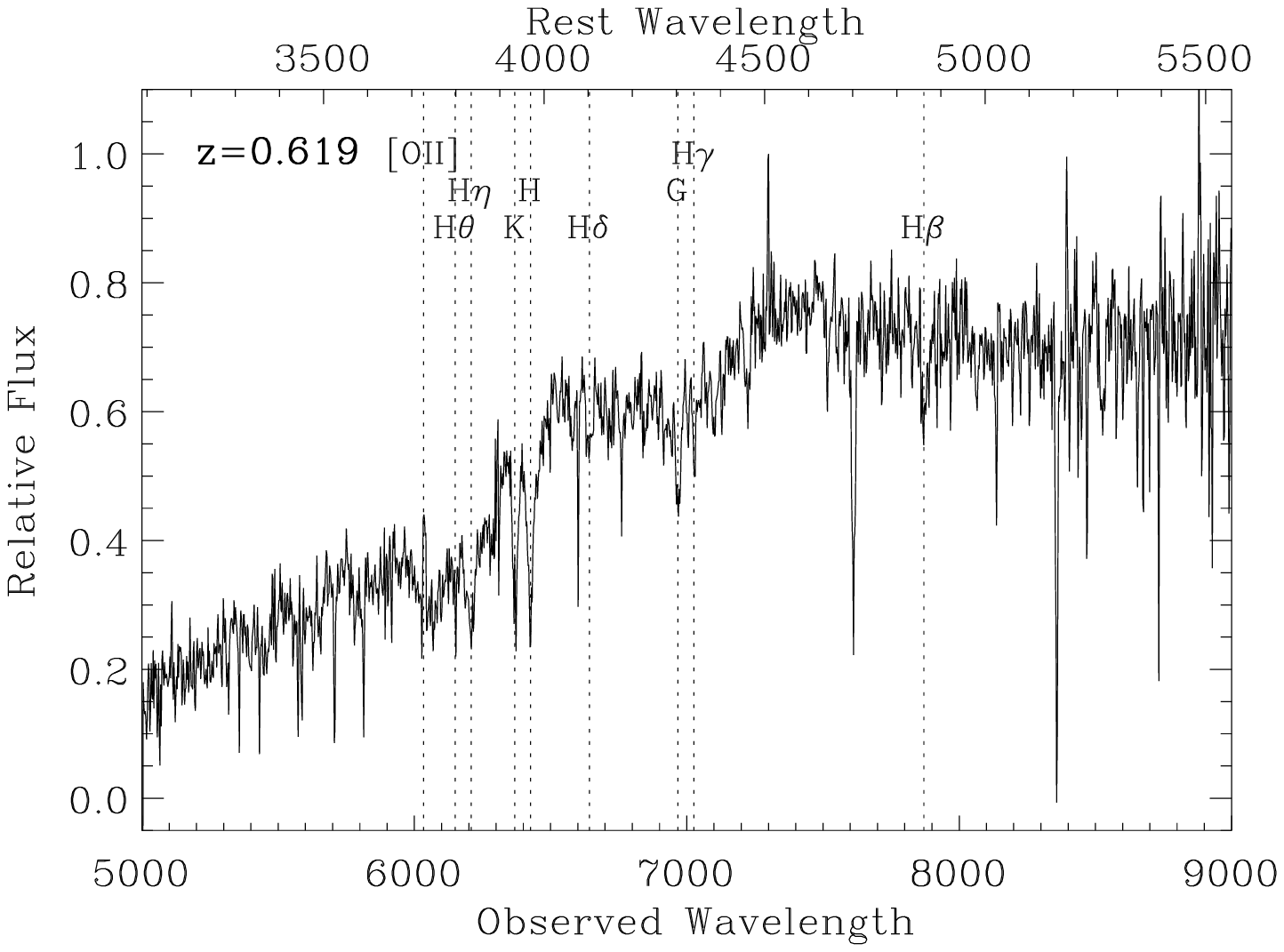}
\plotone{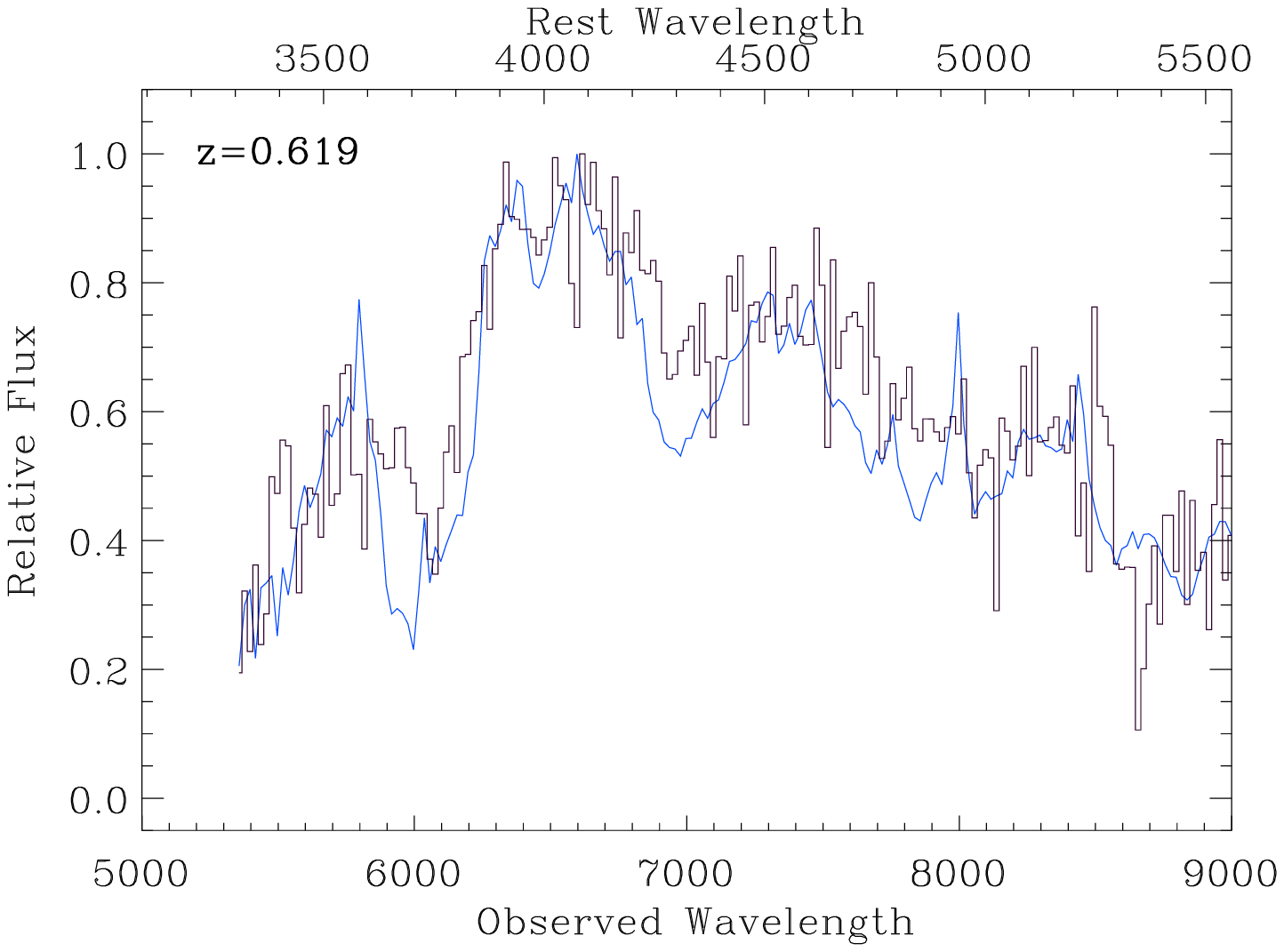}
\caption{(Upper) Unsmoothed spectrum of SN 1997J showing the narrow lines
used to determine the redshift. (Lower) rebinned spectrum of SN 1997J
(histogram) after subtraction of template S0 host galaxy, compared
with SN 1999bn at day +2 (smooth curve).}
\label{SN1997J_spec}
\end{figure}

\begin{figure} 
\plotone{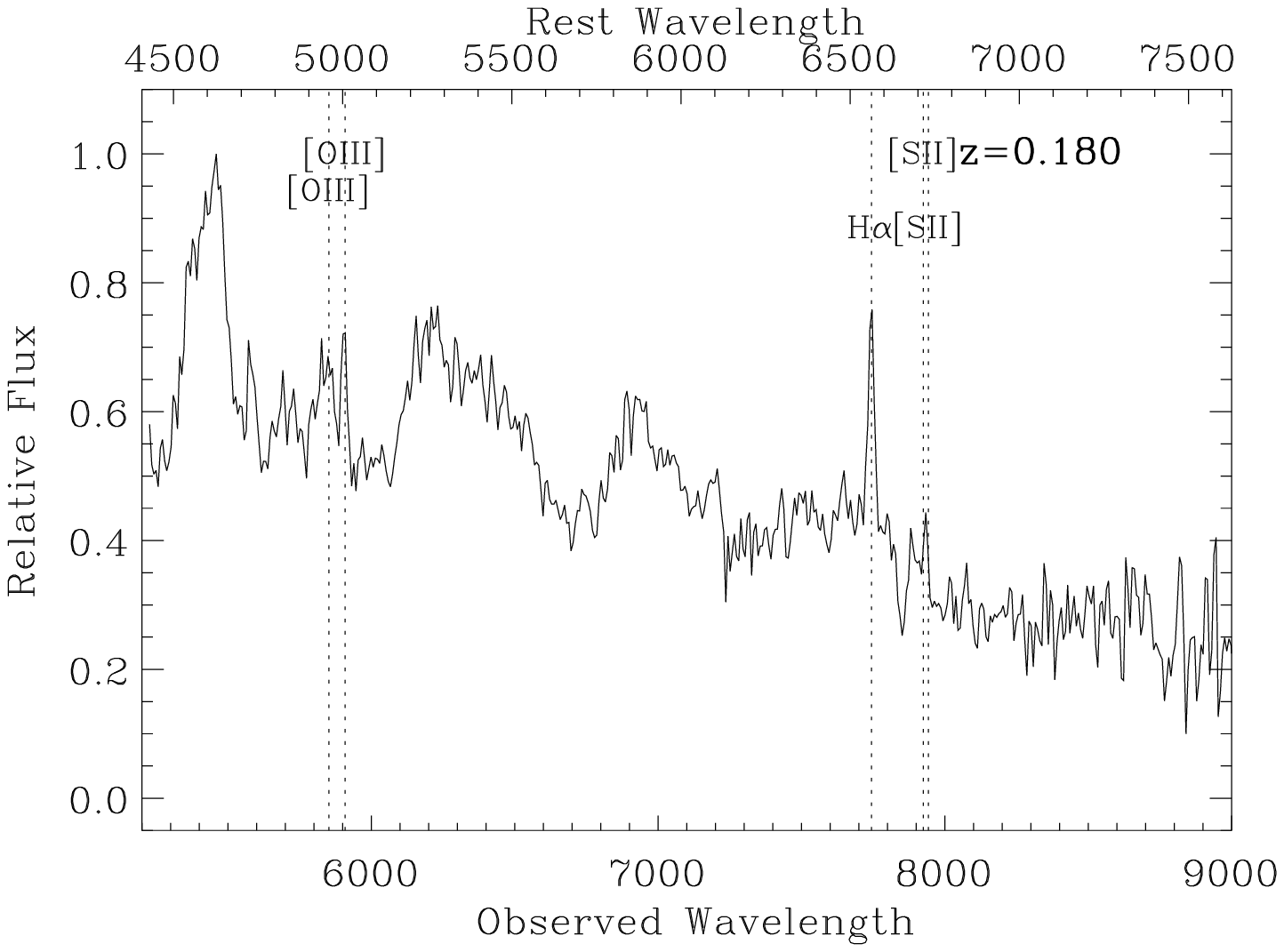}
\plotone{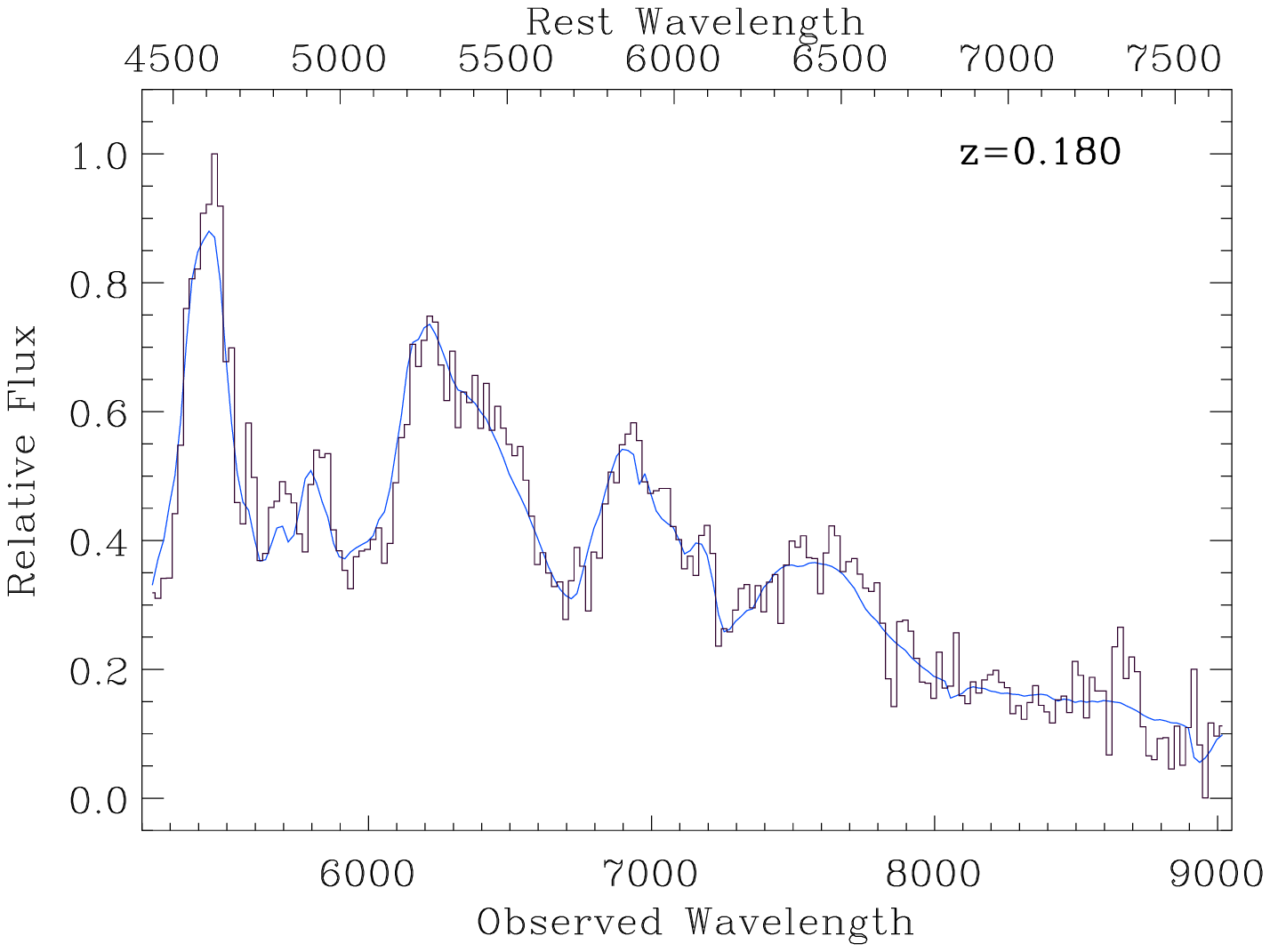}
\caption{(Upper) Unsmoothed spectrum of SN 1997N showing the narrow lines
used to determine the redshift. (Lower) Rebinned spectrum of SN~1997N
(histogram) after interpolating over strong emission lines of [OIII] and
H$\alpha$ and subtracting template SB2 host galaxy light, compared
with SN~1991T at +15 days (smooth curve).}
\label{9710_spec}
\end{figure}

\begin{figure}
\plotone{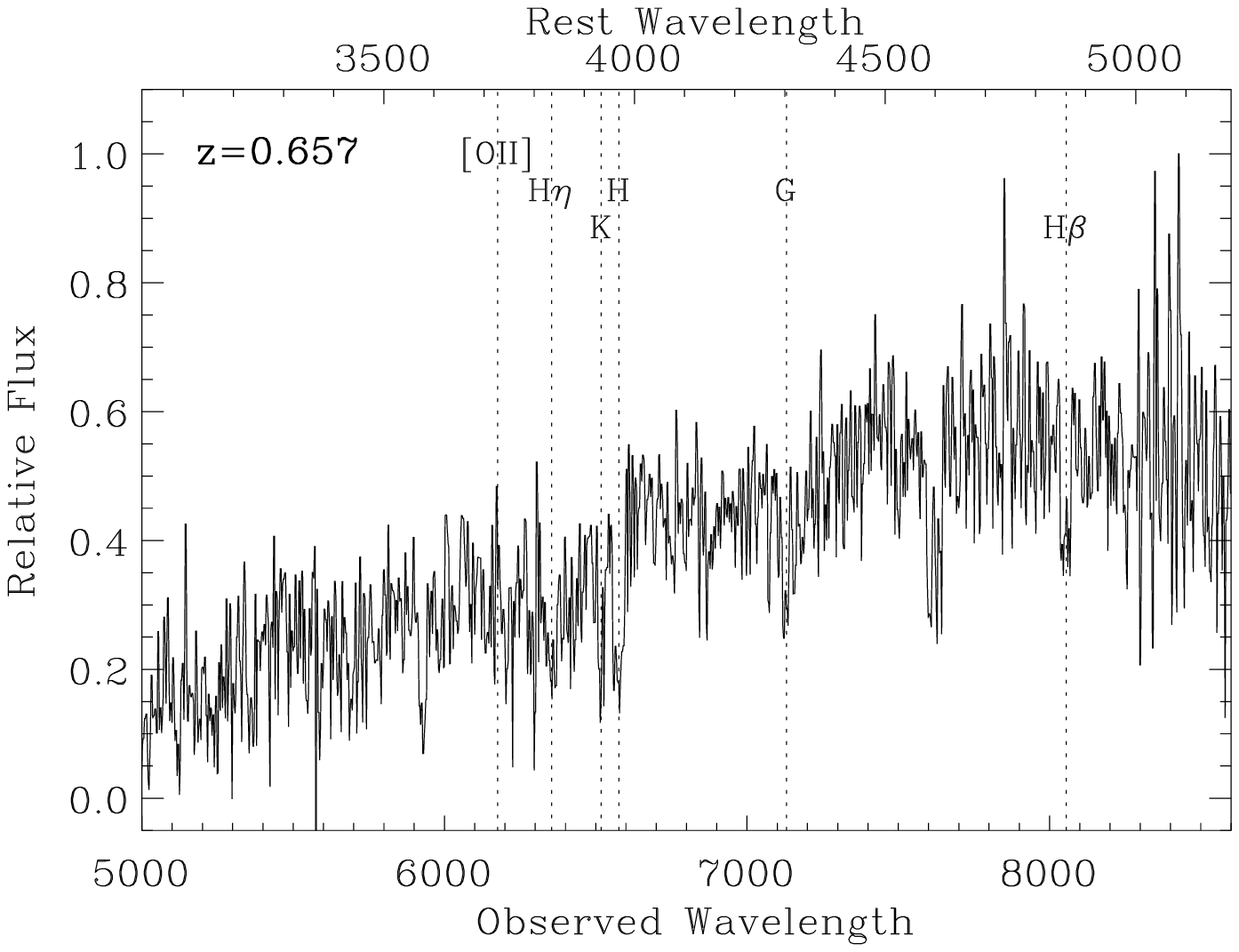}
\plotone{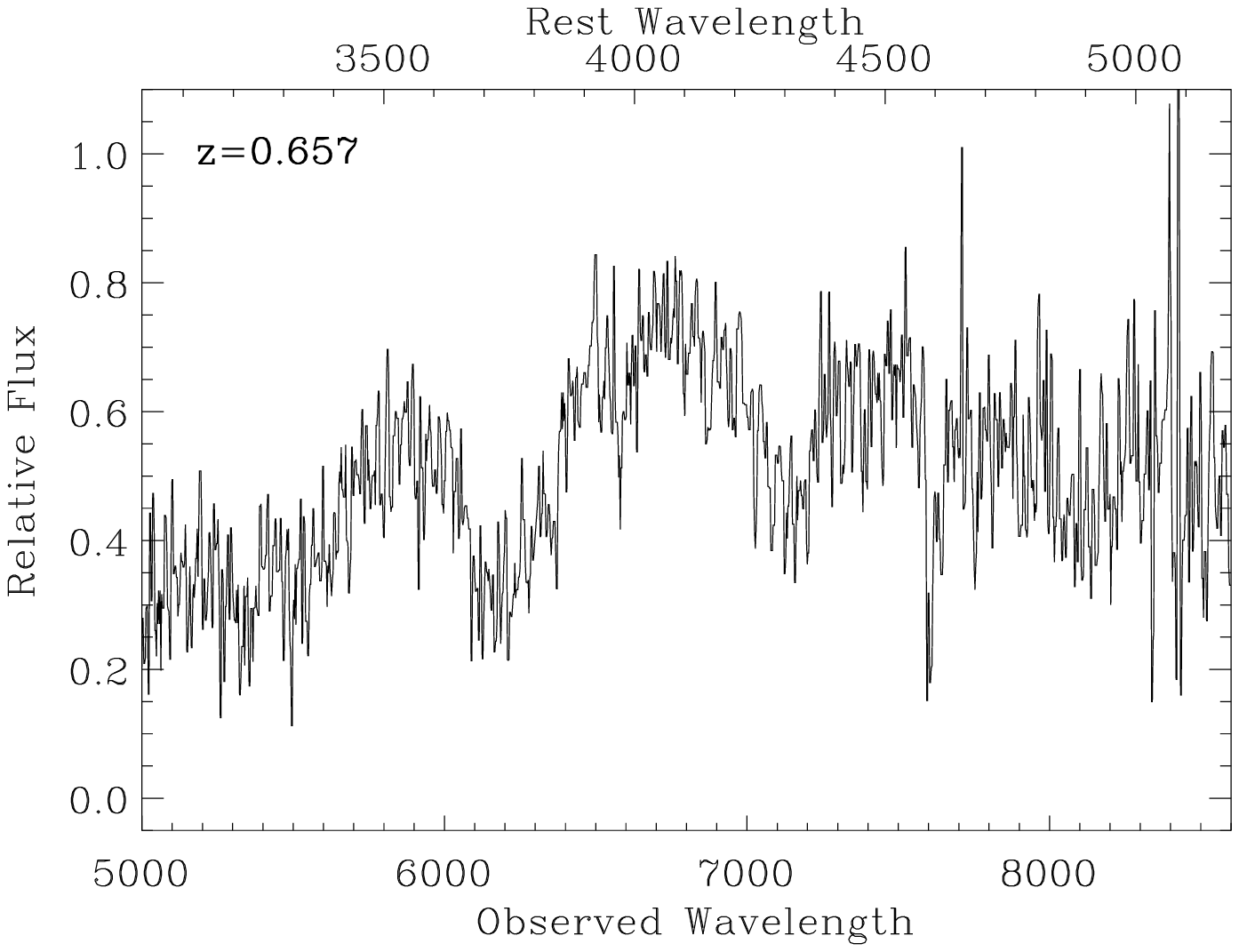}
\plotone{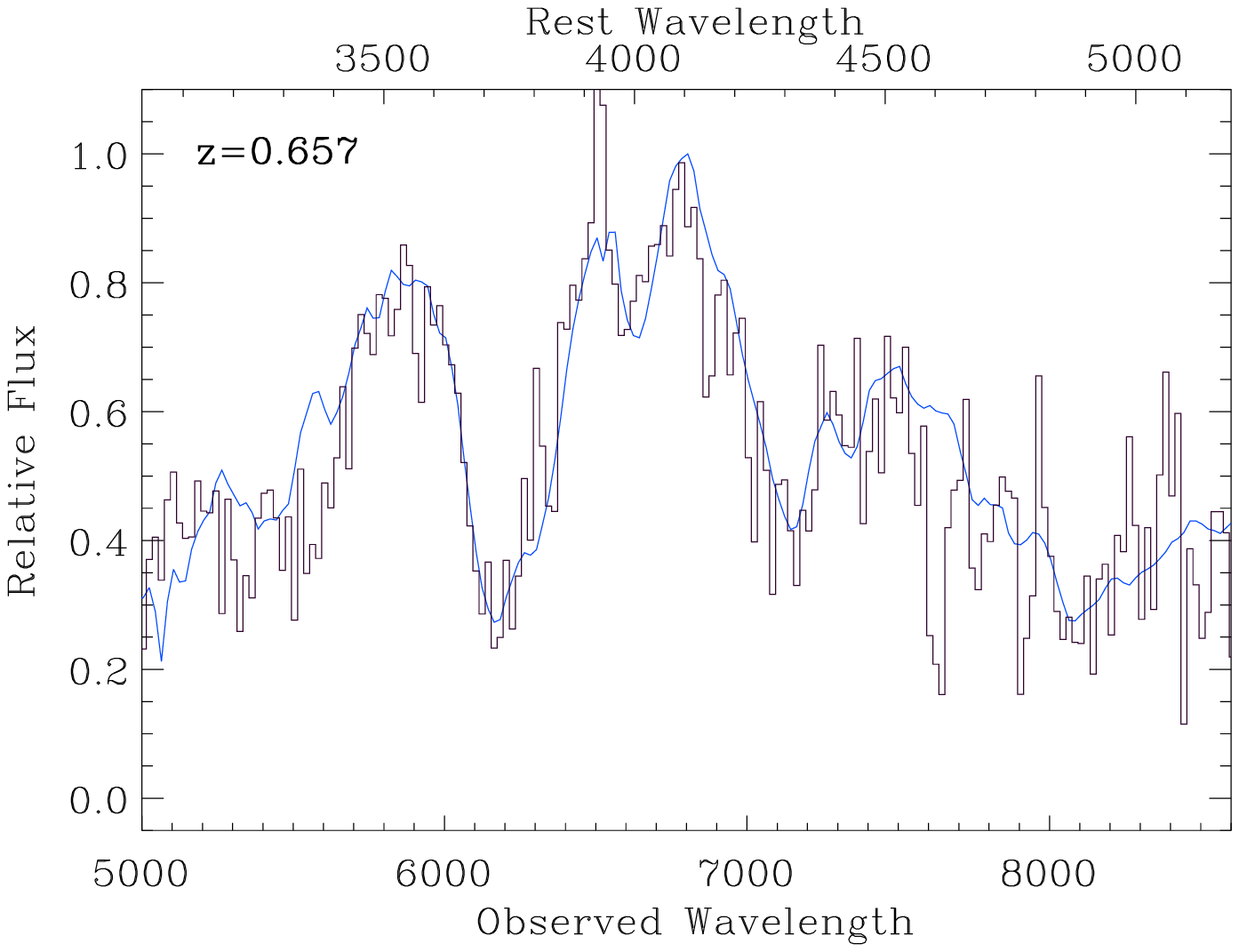}
\caption{(Upper) Lightly smoothed spectrum of the host (or possible
neighbour) galaxy of SN 1997R (atmospheric absorption has not been
corrected in this spectrum). (Middle) Lightly smoothed spectrum of
SN~1997R. (Lower) Rebinned spectrum of SN~1997R (histogram) after
interpolating over narrow host galaxy lines, rebinning to 20\AA\ and
subtracting template Sa galaxy, compared with with SN~1989B at $-5$ days
(smooth curve).}
\label{SN1997R_spec}
\end{figure}

\begin{figure} 
\plotone{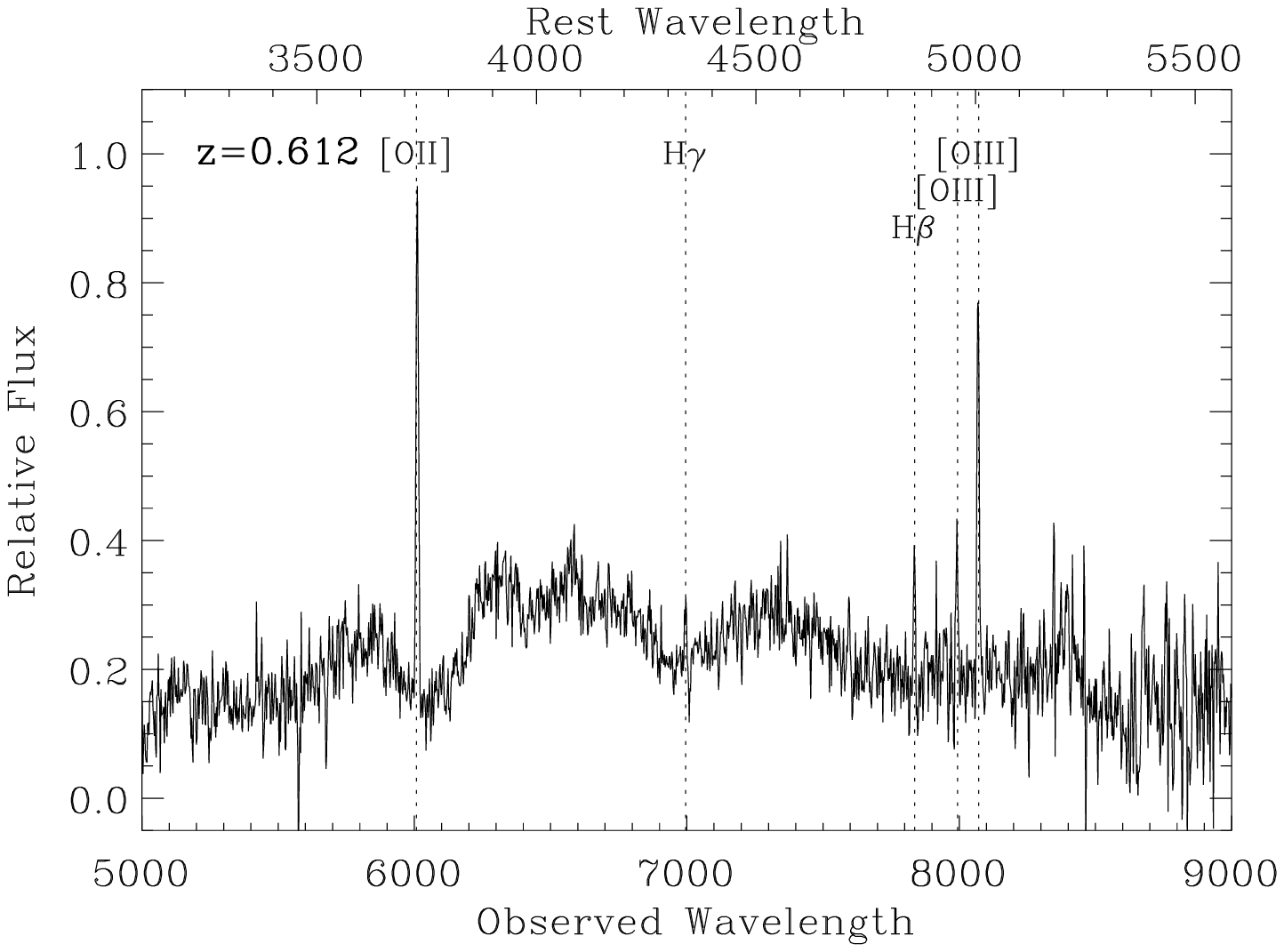}
\plotone{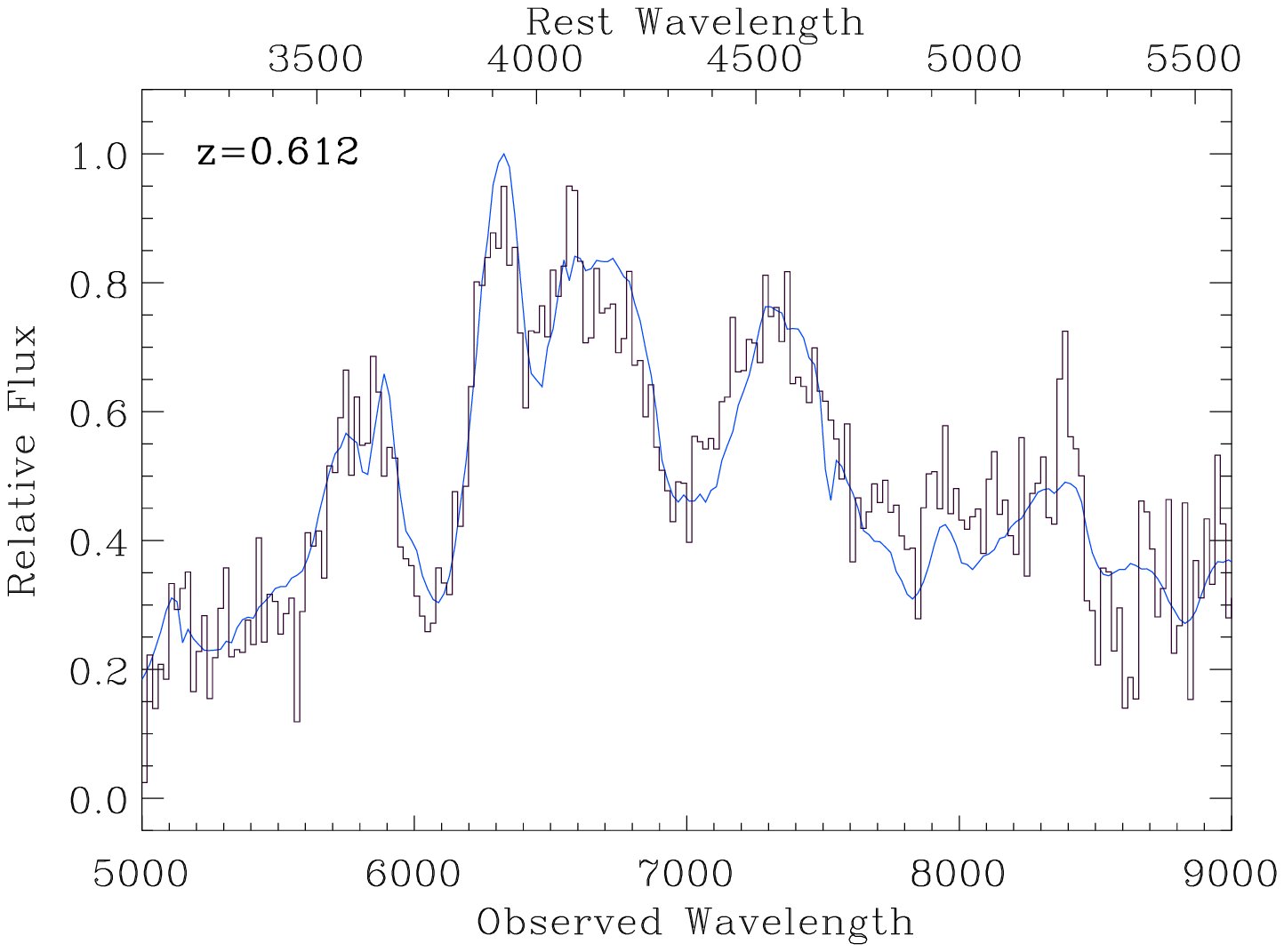}
\caption{(Upper) Unsmoothed spectrum of SN 1997S showing the narrow
lines used to determine the redshift. (Lower) rebinned spectrum of SN
1997S (histogram) after interpolating over strong emission lines of
[OII] and [OIII], $H\beta$ and $H\gamma$ and subtraction of a SB1 host
galaxy template, compared with SN~1992A at +5 days (smooth curve).}
\label{9739_spec}
\end{figure}

\begin{figure}
\plotone{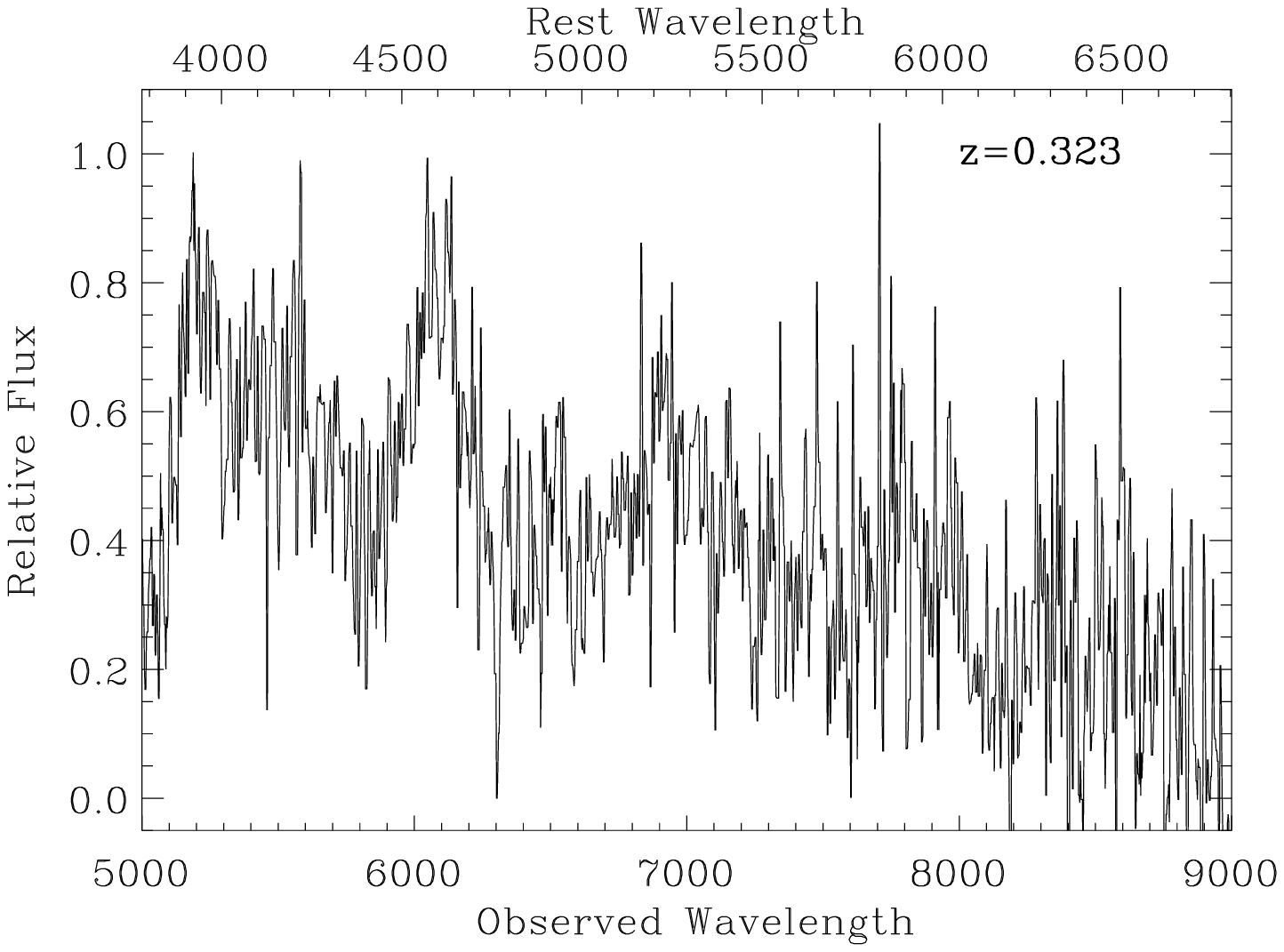}
\plotone{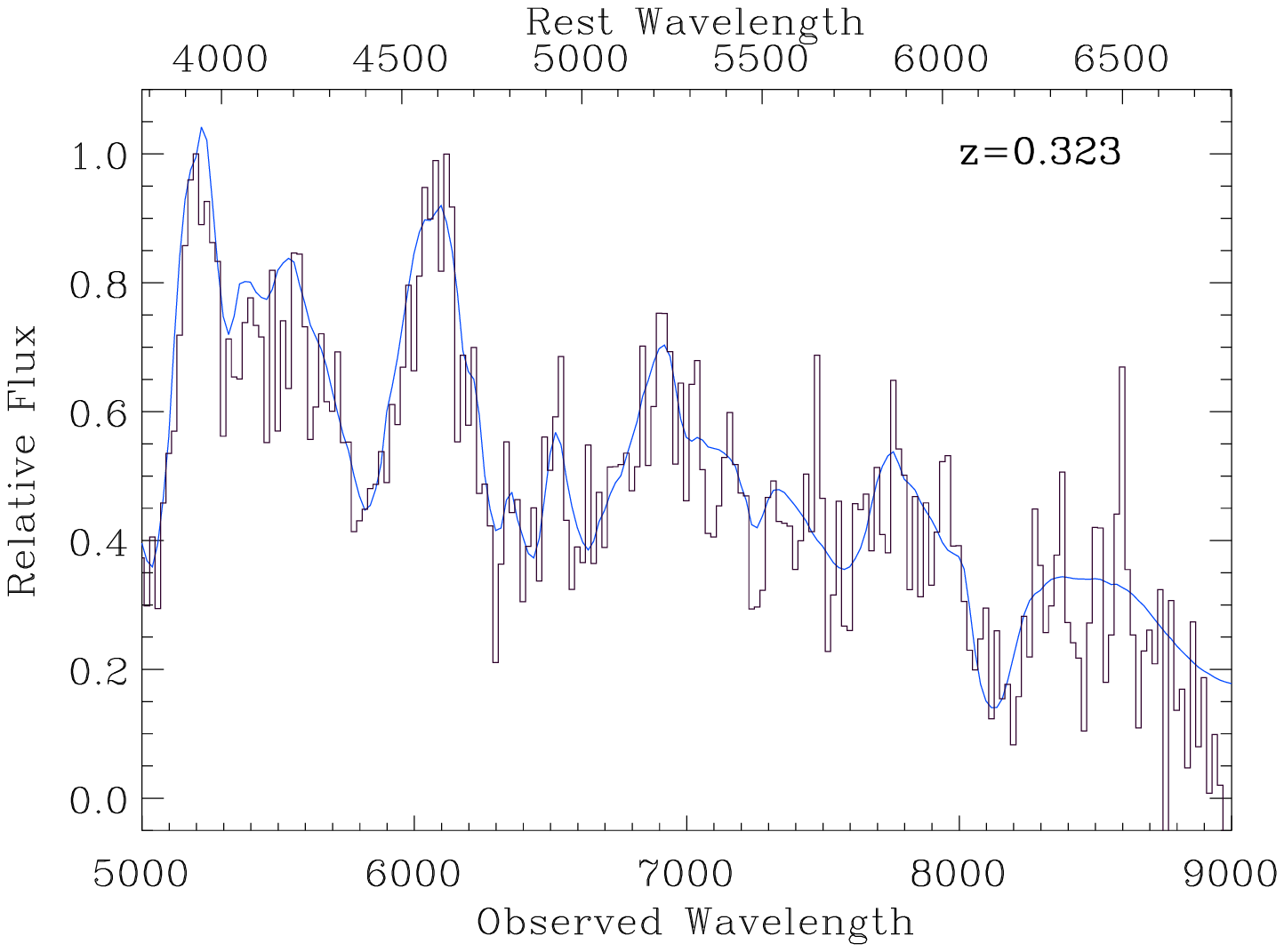}
\caption{(Upper) lightly smoothed spectrum of SN 1997ac. (Lower) rebinned
spectrum of SN 1997ac (histogram) compared with SN 1998bu at + 11 days
(smooth curve).}
\label{SN1997ac_spec}
\end{figure}

\begin{figure}
\plotone{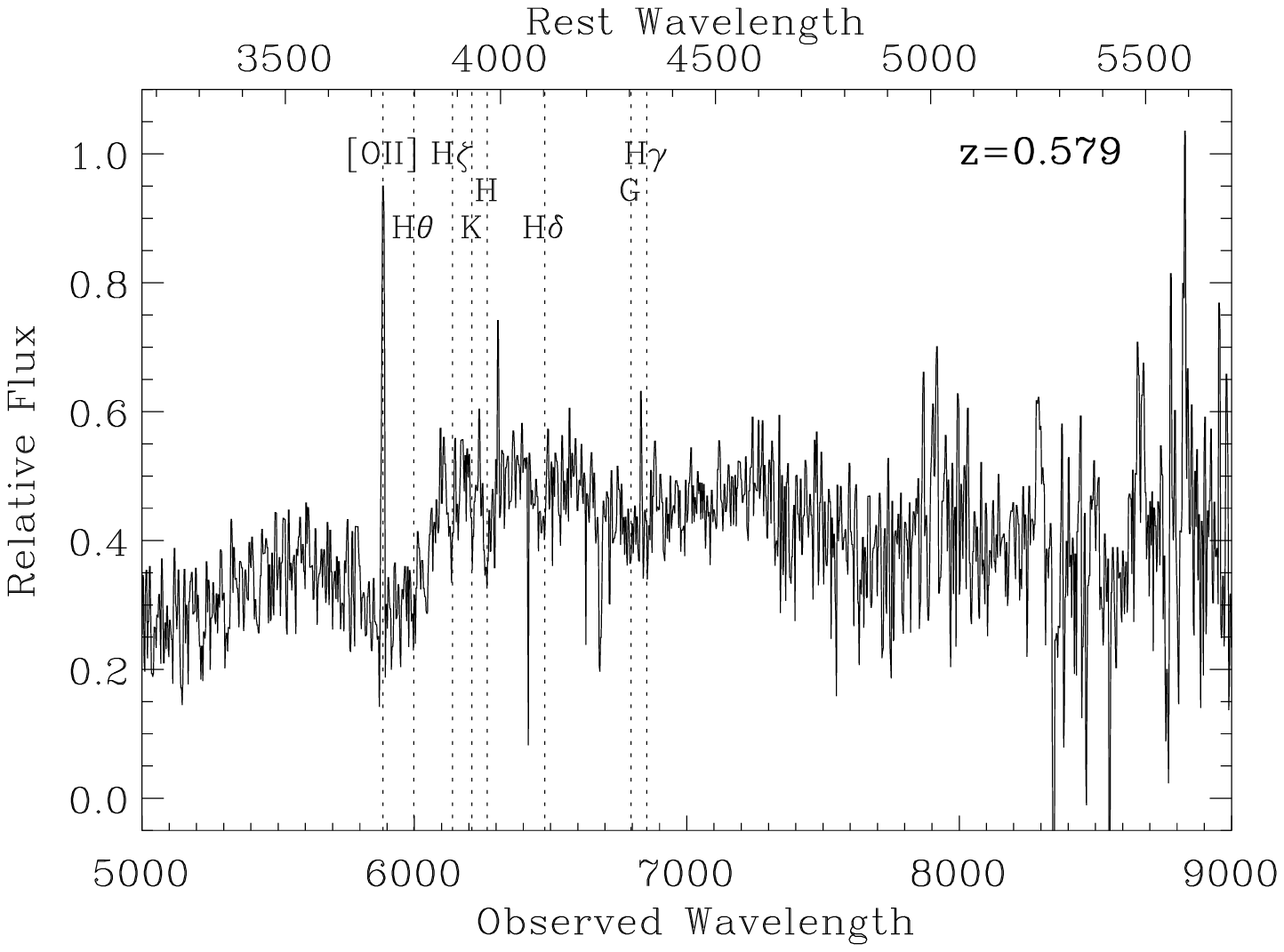}
\plotone{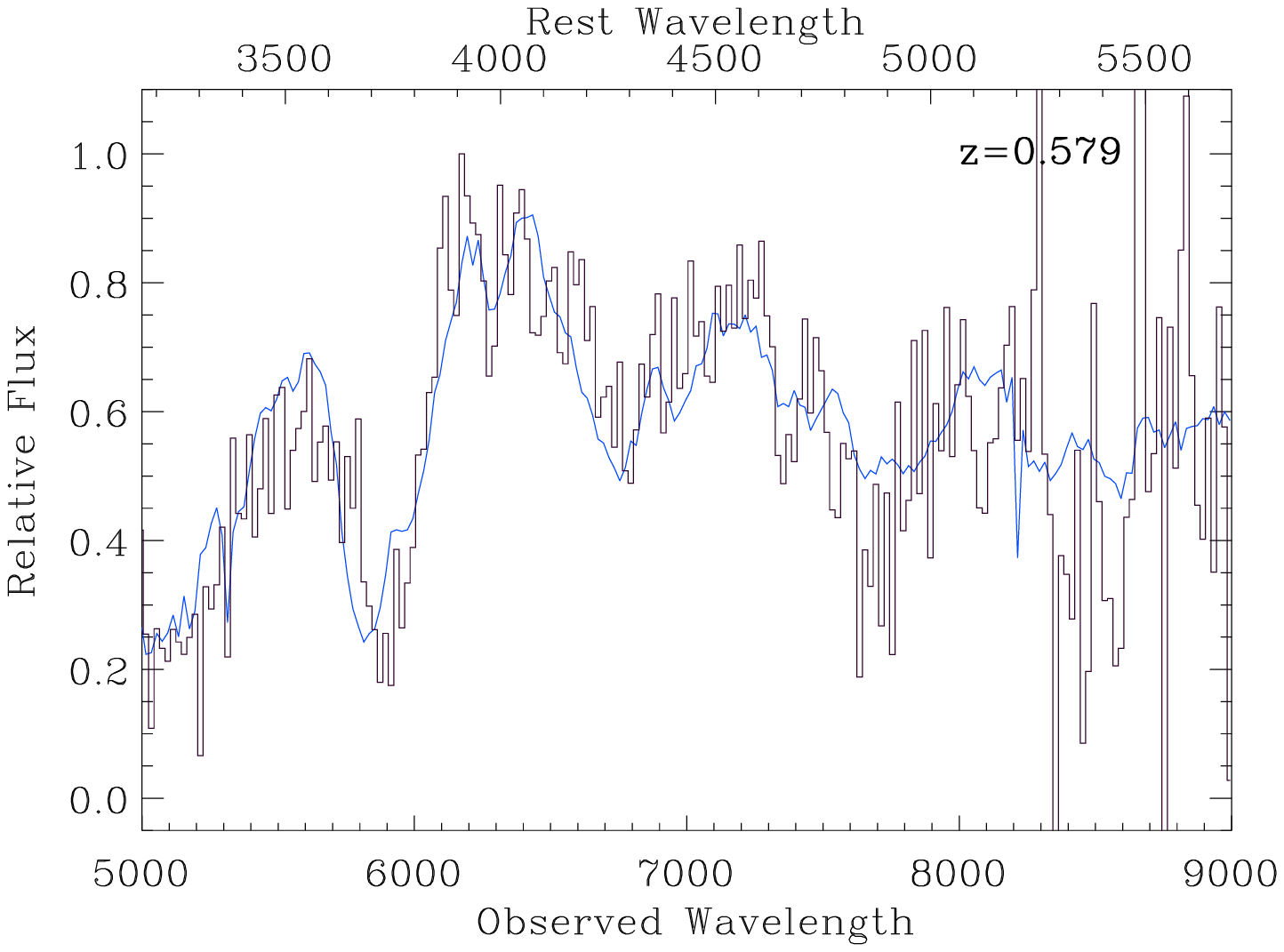}
\caption{(Upper) Unsmoothed spectrum of the host of SN~1997af showing the
narrow lines used to determine the redshift. (Lower) rebinned spectrum of
SN~1997af (histogram) after interpolating across narrow lines from the
host galaxy and subtracting an SB6 galaxy template, compared with the
Type Ia SN~1999bp at $-2$ days (smooth curve).}
\label{SN1997af_spec}
\end{figure}

\begin{figure}
\plotone{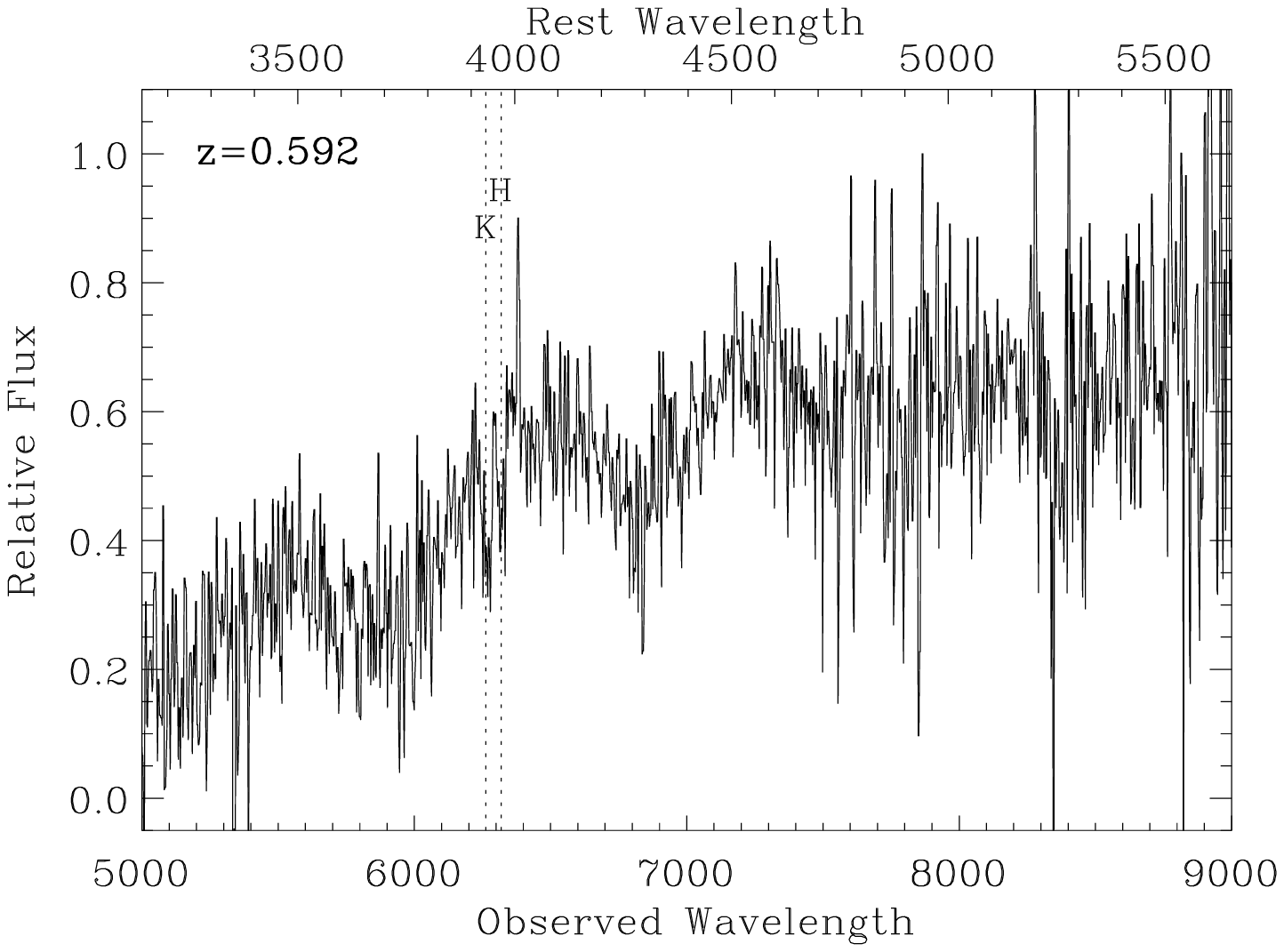}
\plotone{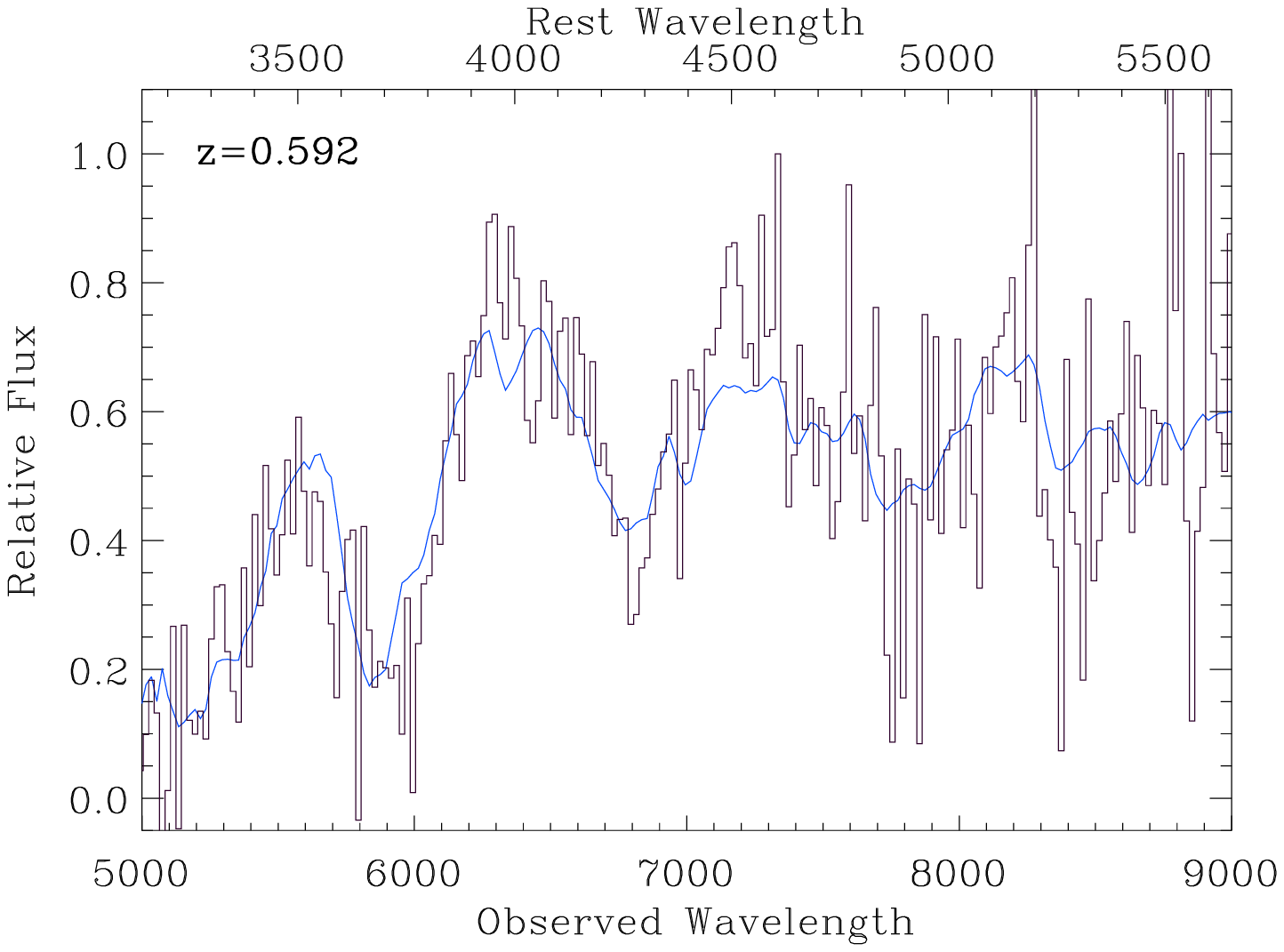}
\caption{(Upper) Lightly smoothed spectrum of SN~1997ag showing narrow
features used to determine the redshift. (Lower) rebinned spectrum
SN~1997ag (histogram) after interpolating over narrow galaxy lines,
compared with the Type Ia SN~1990N at $+7$ days before maximum (smooth
curve).}
\label{9777_spec}
\end{figure}

\begin{figure} 
\plotone{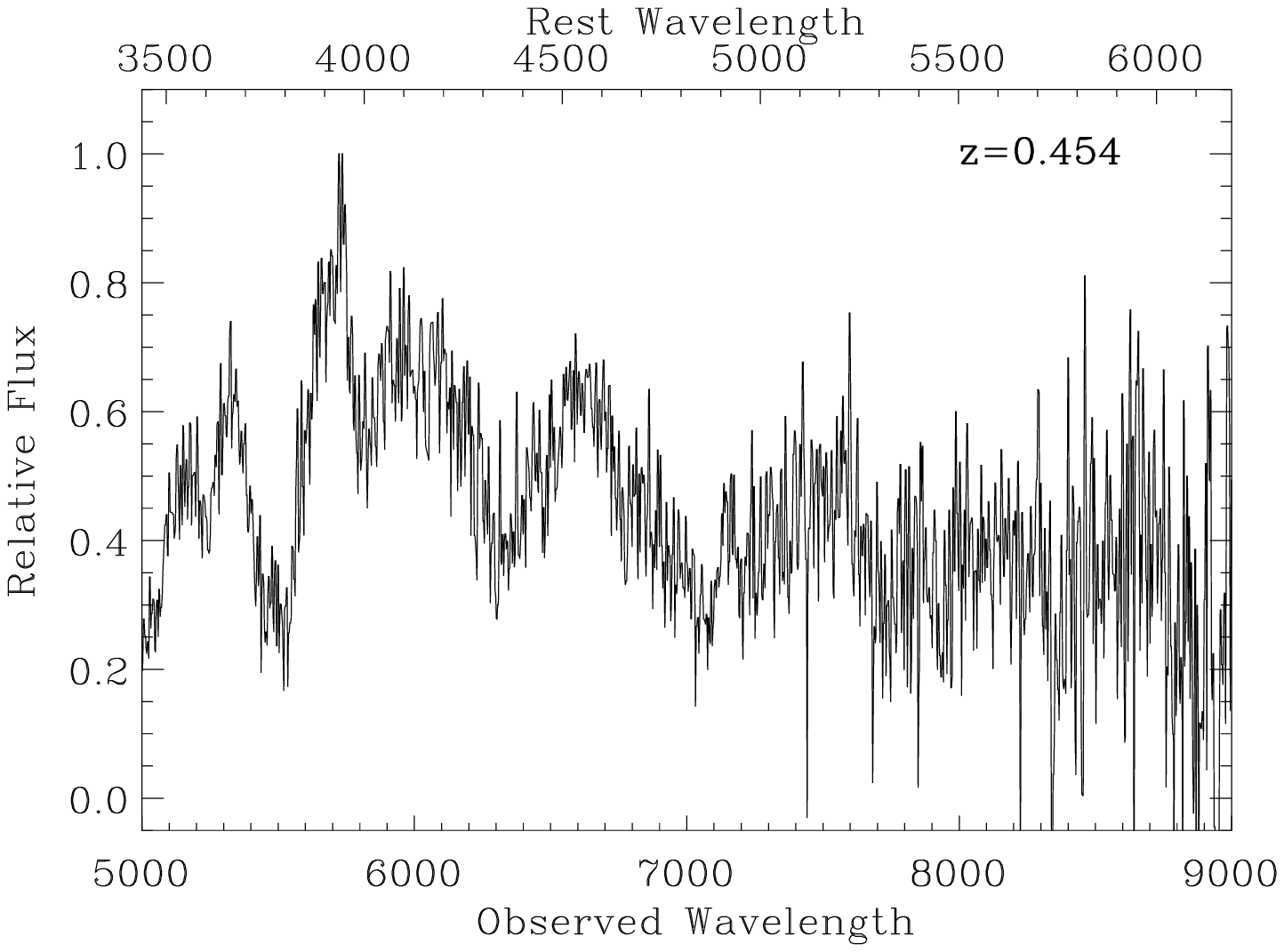}
\plotone{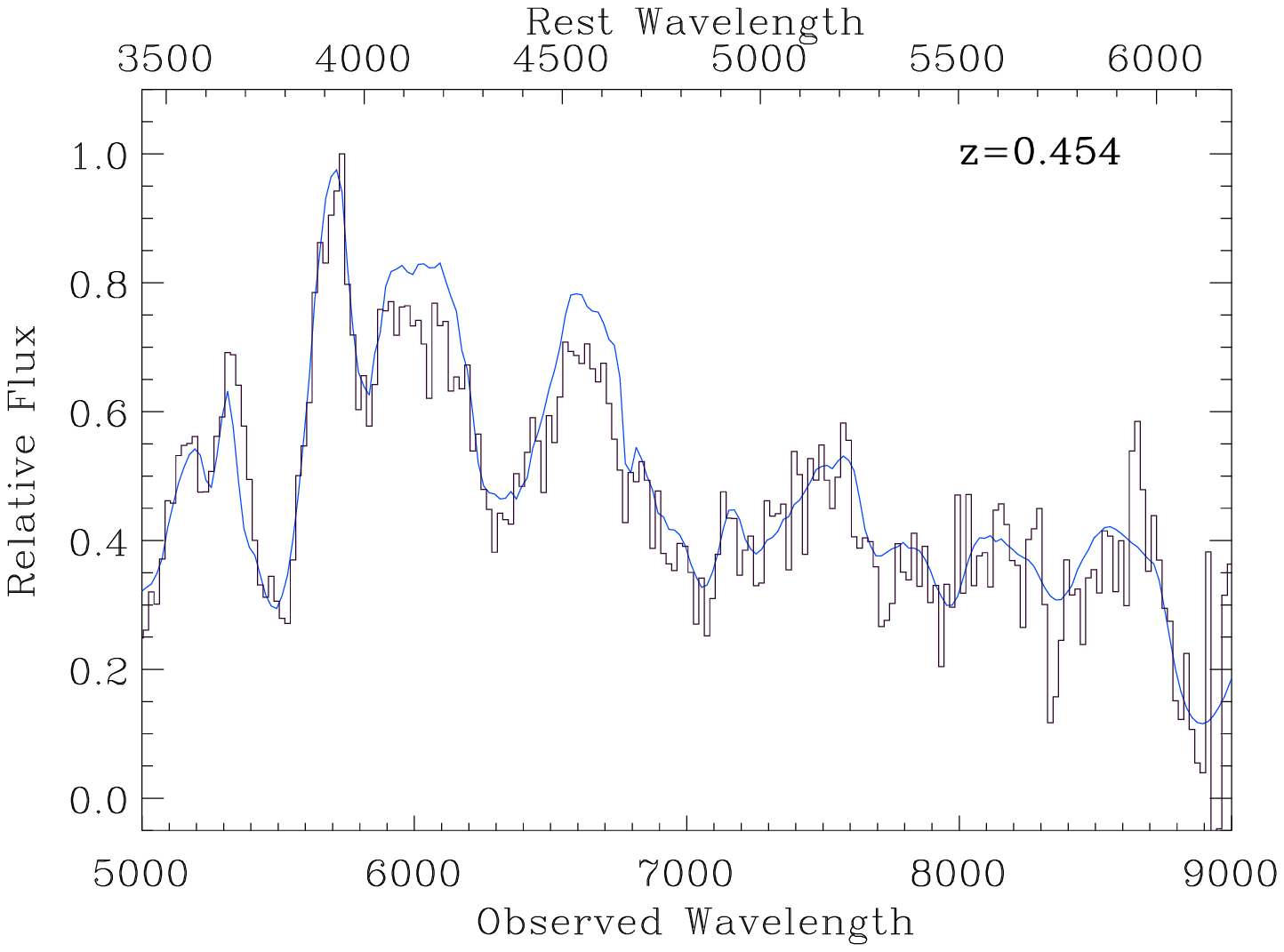}
\caption{(Upper) Lightly smoothed spectrum of SN~1997ai. (Lower) rebinned
spectrum of SN~1997ai (histogram) compared to the Type Ia SN 1992A at
$+5$ days (smooth curve).}
\label{SN1997ai_spec}
\end{figure}

\begin{figure}
\plotone{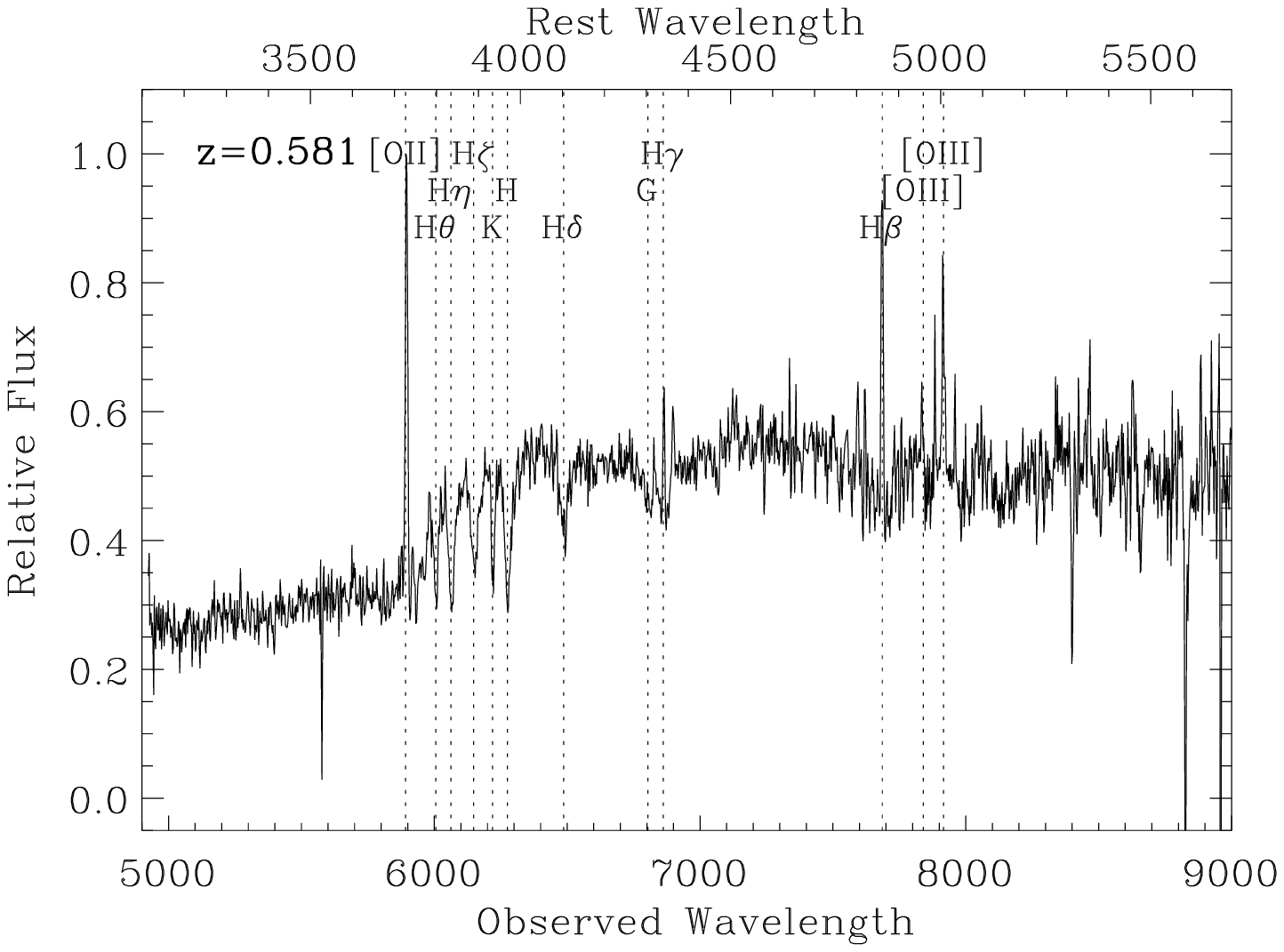}
\plotone{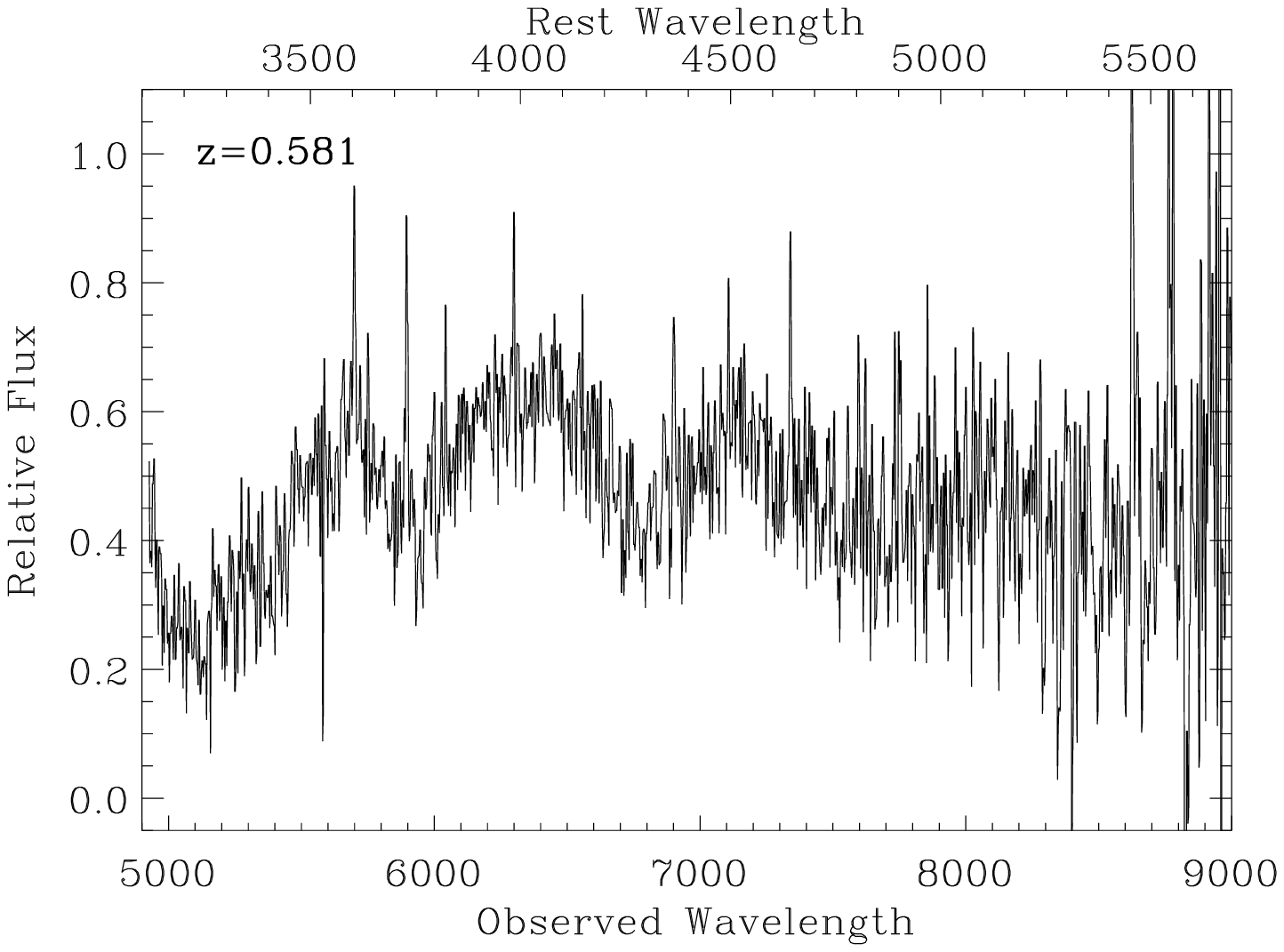}
\plotone{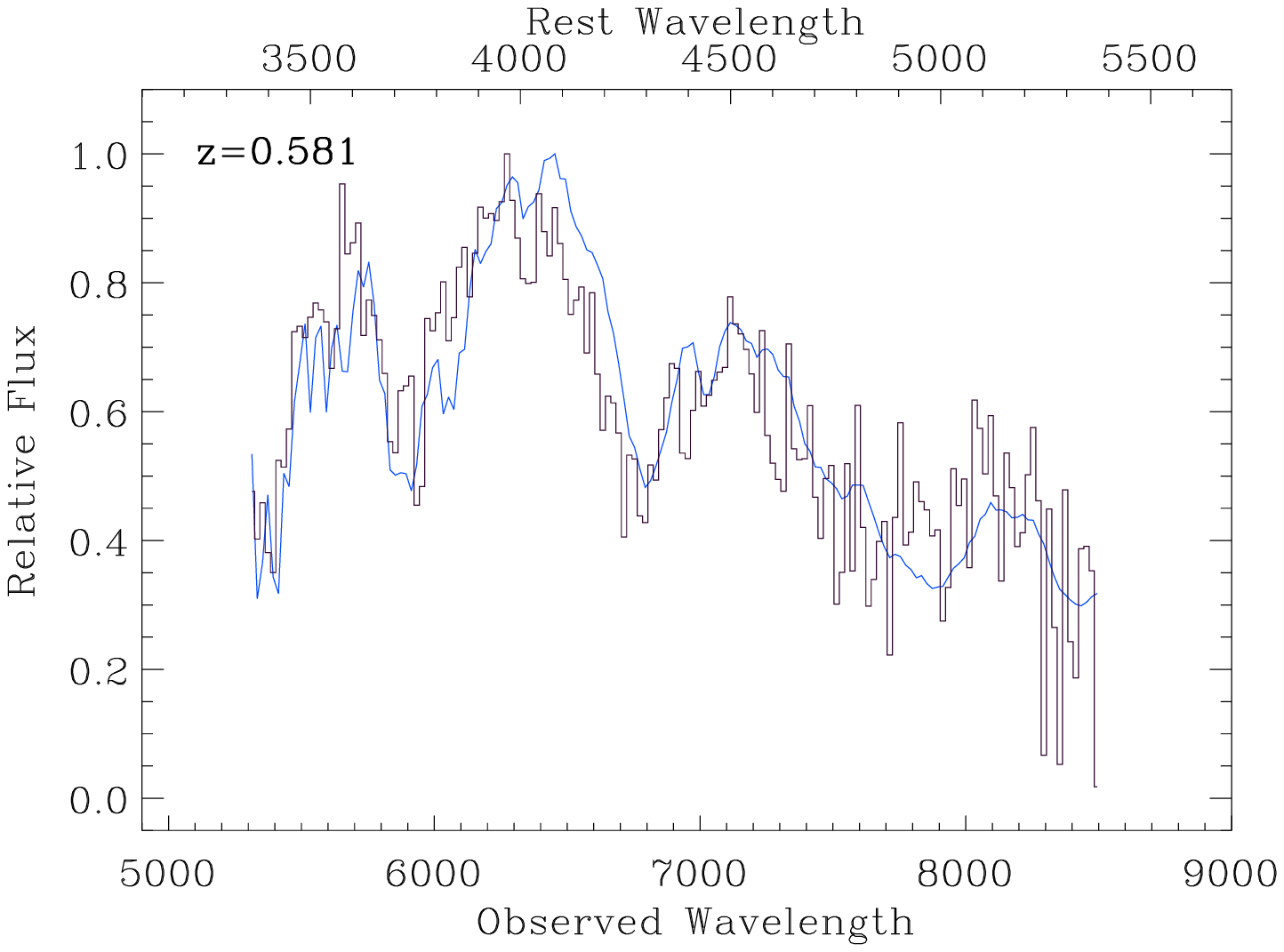}
\caption{(Upper) Unsmoothed spectrum of the host of SN 1997aj (a
separate extraction was possible although there are still narrow
galaxy lines in the SN extraction) showing the narrow lines used to
determine the redshift. (Middle) lightly smoothed spectrum of the
supernova SN 1997aj. (Lower) Rebinned spectrum of SN~1997aj
(histogram) after interpolating over narrow host galaxy lines of [OII]
and Ca H\&K and subtracting Sb galaxy light, compared with the Type Ia
SN 1999aa at $-3$ days (smooth curve).}
\label{SN1997aj_spec}
\end{figure}

Figures~\ref{974spec} to~\ref{9784_spec} show the spectra, all in
$F_{\lambda}$ units with arbitrary normalisation. The top panel shows
the unsmoothed spectrum which typically contains both supernova and
galaxy light.  In the cases where the supernova and galaxy light were
resolved and separate extractions were possible, the top panel shows
the unsmoothed data for the host galaxy.  The lower panels show the
results of the template matching described in the previous section. In
almost all cases clear supernova Ia features are visible in the
high-redshift supernova spectrum.

\begin{figure}
\plotone{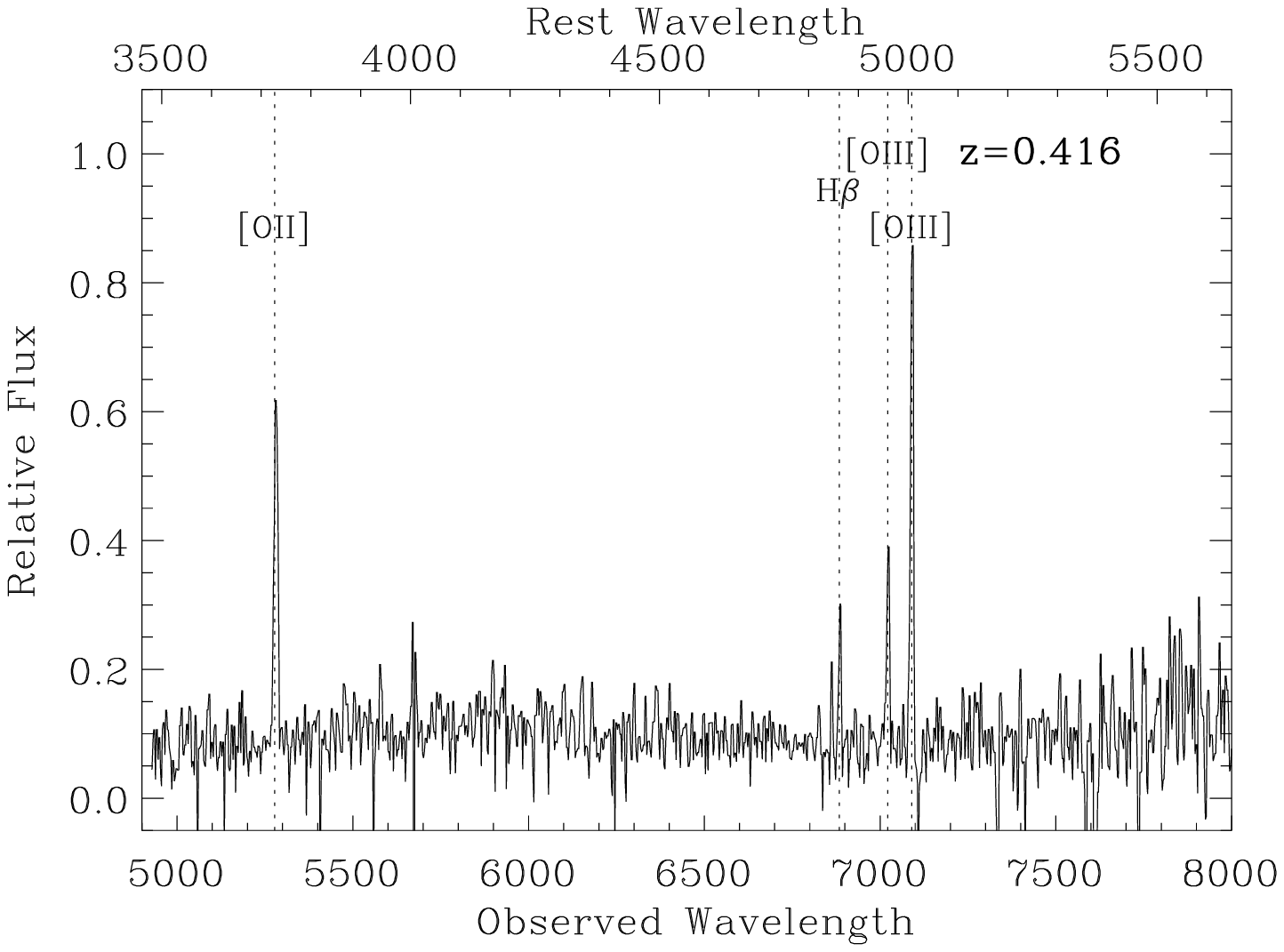}
\plotone{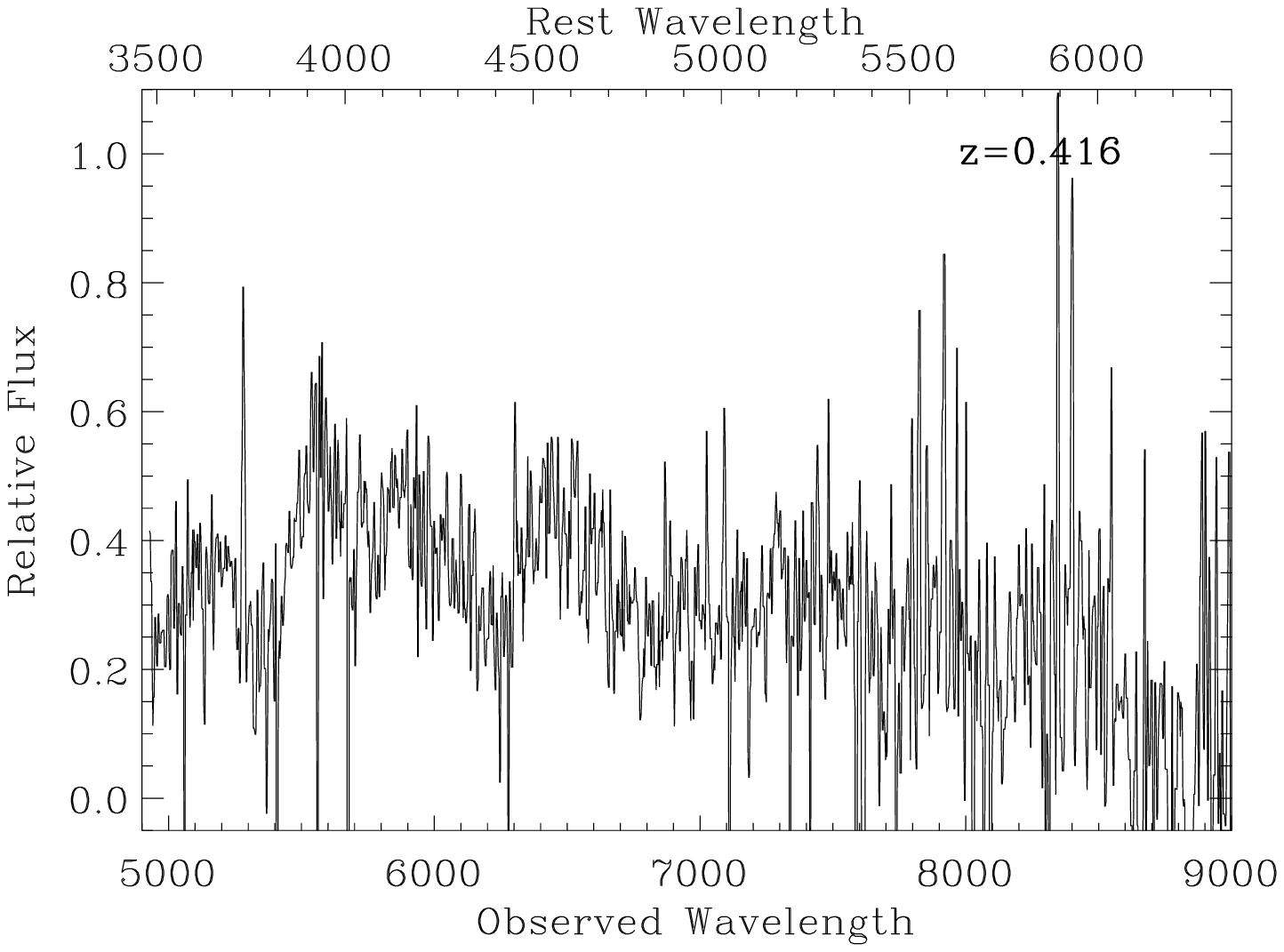}
\caption{(Upper) Unsmoothed spectrum of the host of SN 1997am showing the
narrow lines used to determine the redshift. (Lower) lightly smoothed
spectrum of SN 1997am.}
\label{9785_spec}
\end{figure}

\setcounter{figure}{14}

\begin{figure}
\plotone{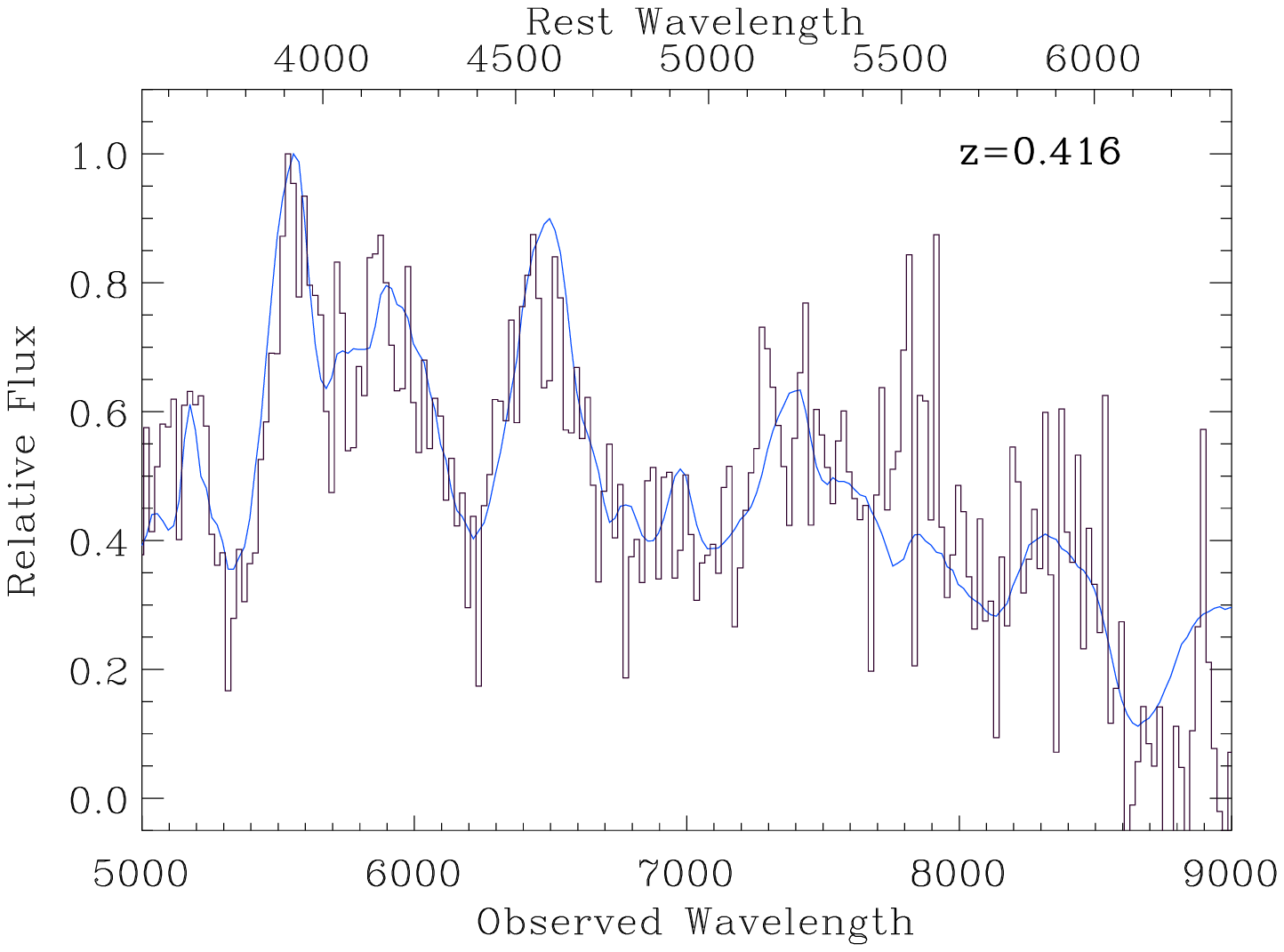}
\plotone{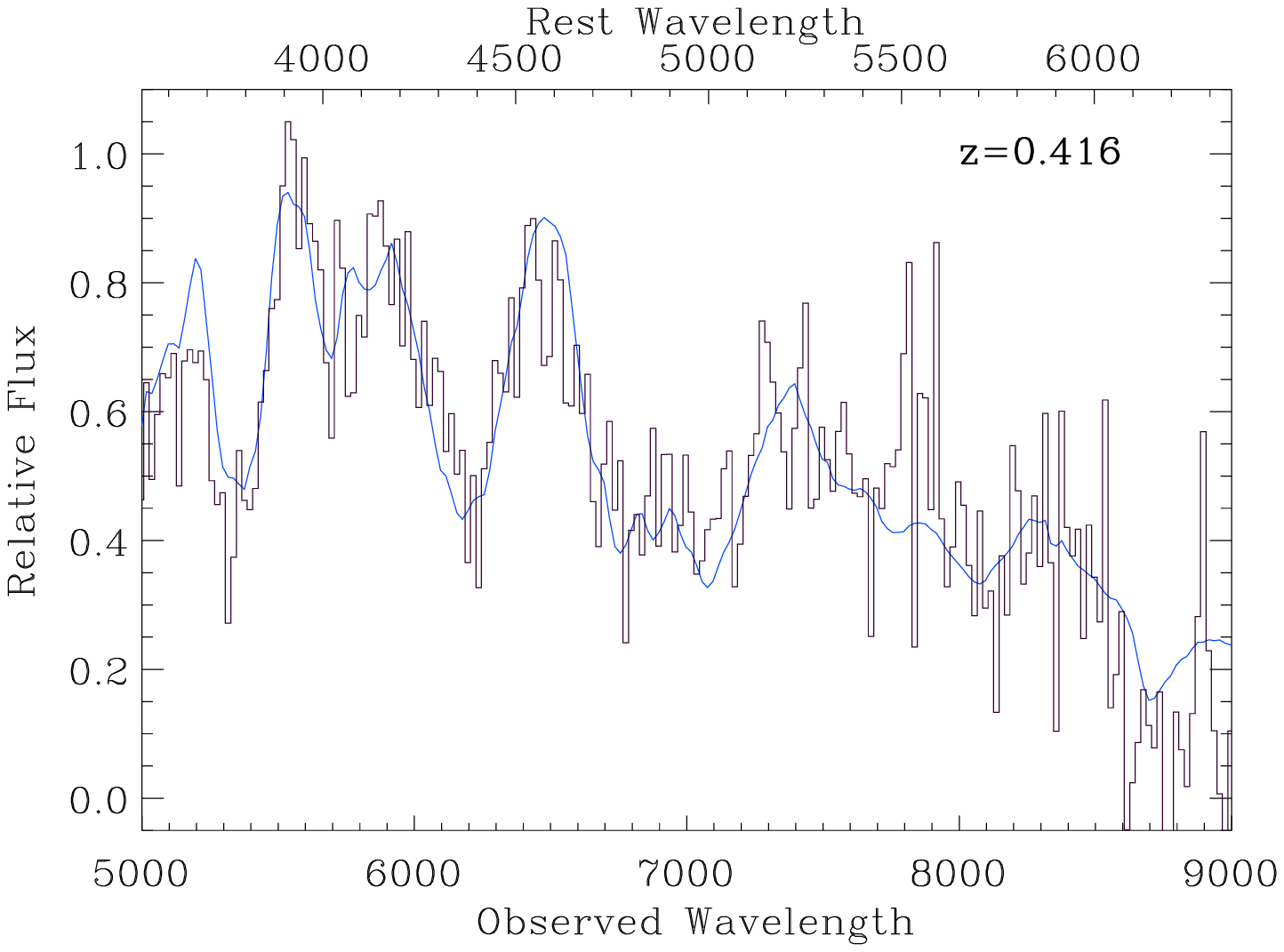}
\caption{cont. (Upper) Rebinned spectrum of SN~1997am (histogram) after
interpolating across cosmic rays and narrow lines from the host galaxy
and subtraction of a SB1 galaxy template, compared with the Type Ia
SN~1992A at +9 days (smooth curve). (Lower) previous plot but with the Type
Ia SN~1991T at +10 days for comparison.}
\end{figure}

\begin{figure} 
\plotone{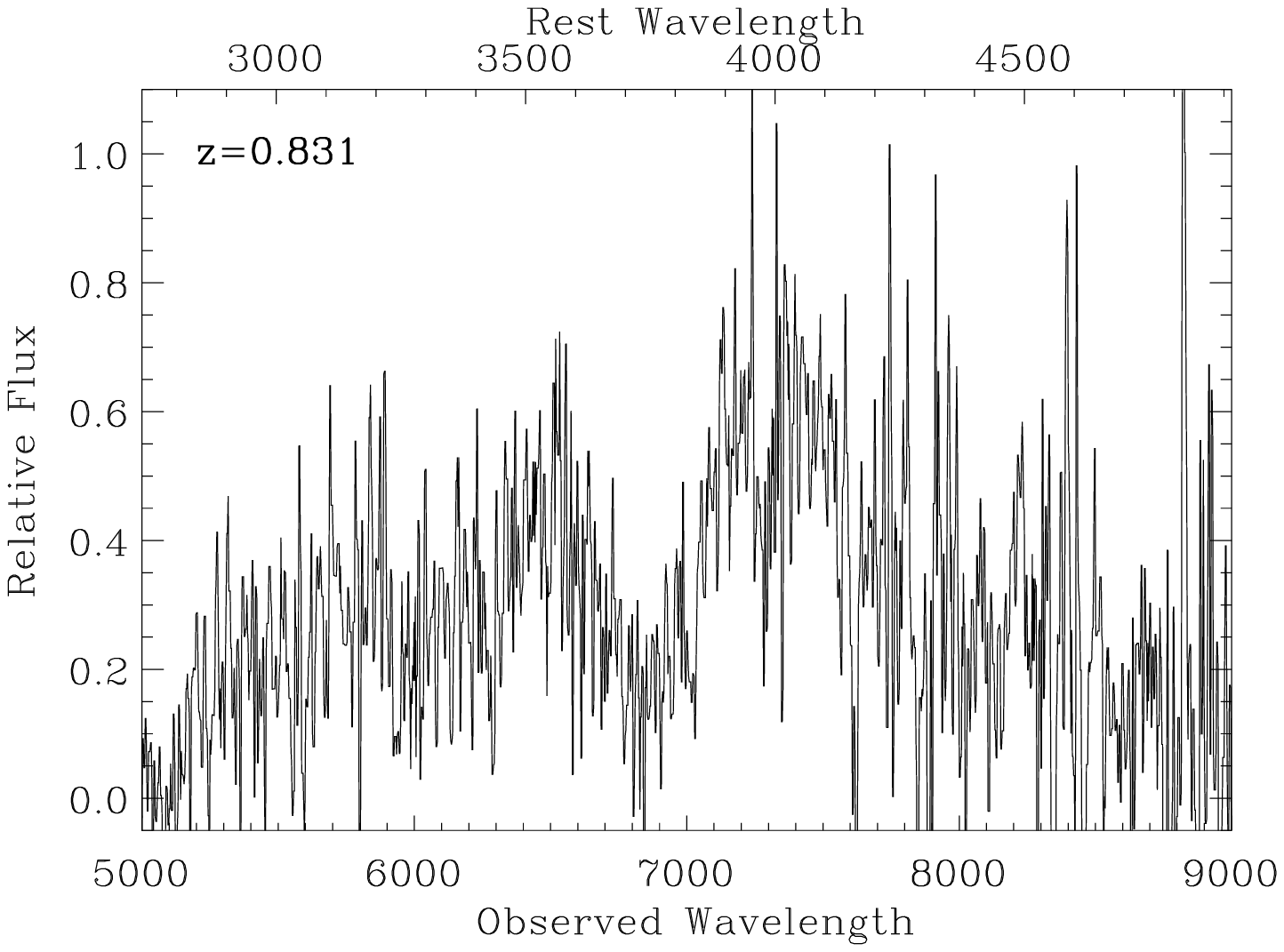}
\plotone{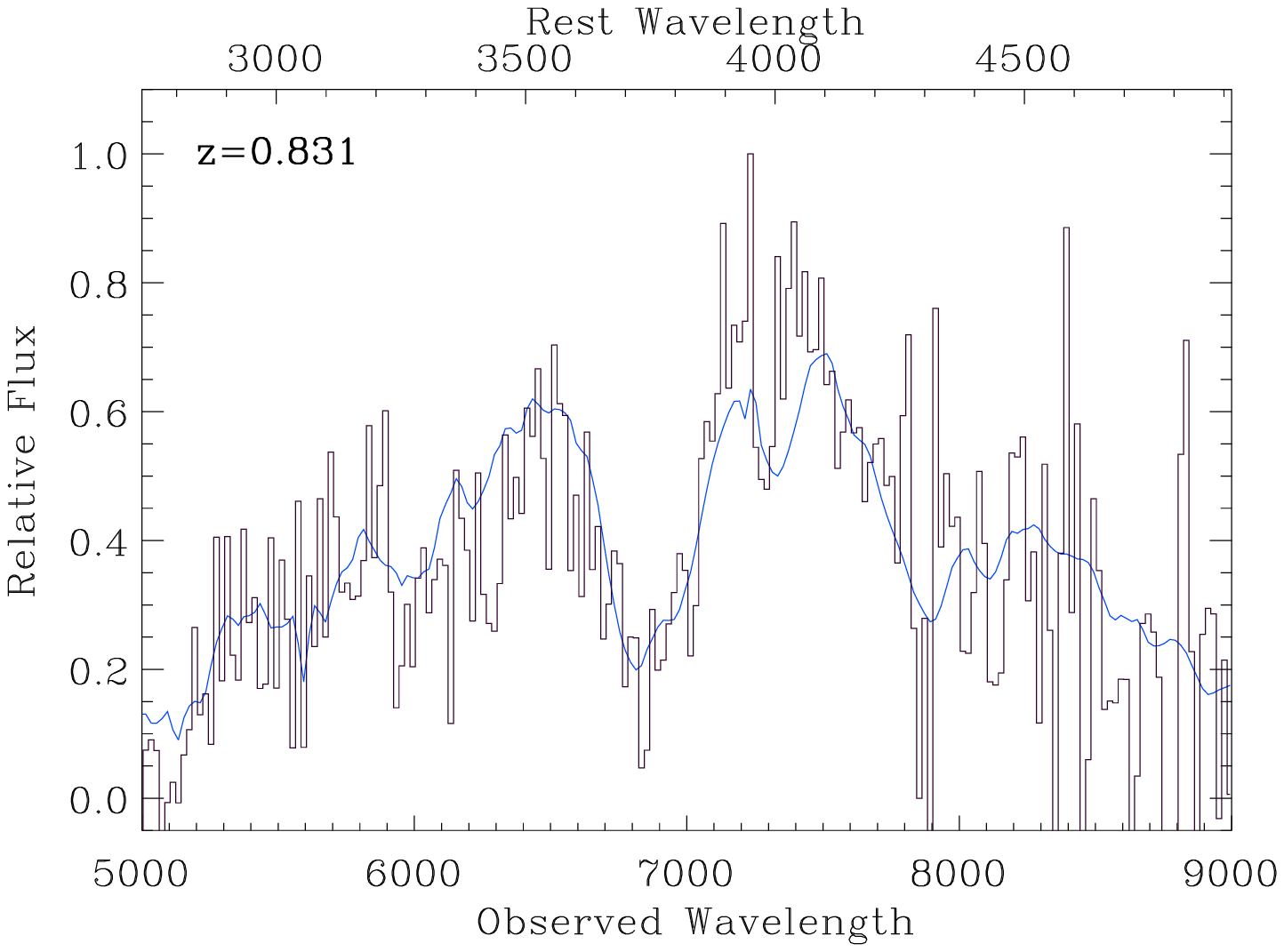}
\caption{(Upper) Lightly smoothed spectrum of SN 1997ap. (Lower) Rebinned
spectrum of SN 1997ap (histogram) after interpolating across the sky
absorption region at 7600\AA\ compared with the Type Ia SN~1989B at $-5$
days (smooth curve).}
\label{9784_spec}
\end{figure}

\subsection{Notes on Individual Objects}


\paragraph{SN~1997F} (Figure~\ref{974spec}).
This spectrum is a blend of galaxy and SN light (the host galaxy and
SN were not separable in the extraction).  The host galaxy has
$z=0.580\pm 0.001$, based on narrow lines from [OII], H$\gamma$,
G-band and [OIII].  After subtraction of a galaxy template, broad
\ion{Ca}{2} is visible.  In a SN Ia at this epoch ($-7$~days), other
features are weak, so it is difficult to distinguish them given the
S/N of this spectrum.  Though the spectrum is dominated by galaxy
light, two lines of evidence give us confidence that the residual
spectrum is a Type~Ia SN.  Firstly, the epoch determined by the
fitting program ($-6.5$~days) agrees well with the epoch determined
from the light curve ($-7.5$~days).  Secondly, a Type Ia is a much
better fit to the data than any other SN Type.

\paragraph{SN~1997G} (Figure~\ref{SN1997G_spec}).
The supernova is slightly merged with a brighter galaxy which has
$z=0.763\pm 0.001$. SN~Ia features can be seen at the same
redshift. The fact that the two objects were merged together made sky
subtraction difficult, and residuals remain at observed wavelengths
$\lambda_{obs}=5577$\AA\ and 6300\AA. These have been interpolated
across for the Figures.  While we consider a Type Ia identification to
be the most probable for this SN, a Type Ib/c identification cannot be
ruled out due to the lack of definitive \ion{Si}{2} and the poor S/N
redward of $\sim$8200\AA, corresponding to $\sim$4650\AA\ in the rest
frame.  Furthermore, there are few UV spectra of SNe Ib/c, so while
the bumps at 2900\AA\ and 3150\AA\ (rest frame) observed at
$\sim$5110\AA\ and $\sim$5550\AA\ match those in a Ia, the behaviour
of SNe Ib/c in this region is not well studied.


\paragraph{SN~1997I} (Figure~\ref{970spec}). 
The redshift of $z=0.172\pm 0.001$ was derived from galaxy lines
(H$\alpha$, H$\beta$, SII, [OIII]). The spectrum is an excellent
match to SN~Ia features, including the 6150\AA\ \ion{Si}{2} feature
seen at $\lambda_{obs} \sim 7200$\AA\ and SII.  The identification
of this SN as a Type Ia is unambiguous, as the \ion{Si}{2} and SII
lines can easily be seen.

\paragraph{SN~1997J} (Figure~\ref{SN1997J_spec}). 
Narrow absorption lines in the spectrum (\ion{Ca}{2} H\&K + 4000\AA\
break from the host galaxy) give $z=0.619\pm 0.001$. SN~Ia features at
this redshift are clearly visible, despite being affected by the
galaxy absorption lines. After subtracting a template S0 galaxy
spectrum the features become much more clear.  Here also, the Type Ia
identification is most probable, but not definitive.
 

\paragraph{SN~1997N} (Figure~\ref{9710_spec}). 
The redshift of $z= 0.180 \pm 0.001$ was derived from 
narrow galaxy lines (H$\alpha$ [OIII] and possibly SII). SN~Ia
features from $\sim$ 2 weeks after maximum are clearly visible in the
spectrum including the \ion{Si}{2} 6150\AA\ feature seen at
$\lambda_{obs}\sim 7250$\AA.  The spectra of the low-redshift supernovae
SN 1991T and SN 1994D both match this spectrum equally well (see
section~\ref{peculiar}).


\paragraph{SN~1997R} (Figure~\ref{SN1997R_spec}). 
The host (or a neighbouring) galaxy 2.6\arcsec\ away has $z=0.657\pm
0.001$ (identified from the Ca~H\&K lines and the 4000\AA\ break). The
supernova shows clear broad features matching a Type Ia at this
redshift including the 4000\AA\ \ion{Si}{2} feature seen at
$\lambda_{obs}\sim 6630$\AA, just redward of the \ion{Ca}{2} feature.
Narrow lines ($\rm H\eta$, H$\rm\delta$ and possibly $\rm H\gamma$)
are also visible in the spectrum, presumably from the host galaxy.

\paragraph{SN~1997S}  (Figure~\ref{9739_spec}). 
Narrow lines due to the host galaxy ([OIII], [OII], $\rm H\beta$, $\rm
H\gamma$) give $z=0.612\pm 0.001$. Very clear SN~Ia features matching
this redshift are visible including the \ion{Ca}{2},
\ion{Si}{2}(4000\AA) and \ion{Fe}{2} features.

\paragraph{SN~1997ac} (Figure~\ref{SN1997ac_spec}). 
The spectrum shows very clear SN~Ia features
at a redshift of $z=0.323 \pm 0.005$ at $\sim 10$ days after maximum
light. Since there are no clear features due to the host galaxy, the
redshift is based solely on the supernova features. The Si 6150\AA\
feature is clearly visible at an observed wavelength of $\sim$8100\AA,
providing an unambiguous identification of the object as a SN~Ia.

\paragraph{SN~1997af} (Figure~\ref{SN1997af_spec}). 
The supernova and host galaxy spectra could not be extracted
separately.  Narrow lines due to the host galaxy of [OII], Ca H\&K,
$\rm H\gamma$, G-band, H$\eta$, H$\theta$, are clear in the spectrum,
giving a redshift of $z=0.579\pm0.001$. The spectrum also shows broad
features consistent with the spectrum of a SN~Ia at the same redshift,
including the small \ion{Si}{2} feature (4000\AA\ rest wavelength) at
an observed wavelength of $\sim6350$\AA.


\paragraph{SN~1997ag} (Figure~\ref{9777_spec}). 
The supernova and host light are blended and
it was not possible to obtain separate extractions.  The Ca~H\&K lines
from the host galaxy are visible at $\lambda_{obs}=6266$\AA\ and
6317\AA\ giving a redshift of $z=0.592\pm 0.001$.  SN~Ia features are
clearly visible in the spectrum, especially broad \ion{Ca}{2} and
\ion{Fe}{2}.


\paragraph{SN~1997ai} (Figure~\ref{SN1997ai_spec}). 
The spectrum shows very clear SN~Ia  features
at a redshift of $z=0.454\pm 0.006$. Since there are no clear features
due to the host galaxy, the redshift is based solely on the supernova
features.  The lack of host galaxy contamination is confirmed by the
fitting program, which used no galaxy light in its fit of the
spectrum.  The \ion{Si}{2} (6150\AA) feature is clearly visible at an
observed wavelength of $\sim$8900\AA, providing an unambiguous
identification of the object as SN~Ia.


\paragraph{SN~1997aj} (Figure~\ref{SN1997aj_spec}). 
The supernova and host were separated by 2.5\arcsec\ on the slit and
their spectra could be extracted separately. The host galaxy spectrum
shows emission lines of [OII], [OIII] and H$\beta$ as well as
absorption lines of $\rm H\gamma$, $\rm H\delta$, $\rm H\theta$, $\rm
H\eta$, G-band and Ca H\& K, at a redshift of $z=0.581\pm 0.001$.  The
supernova spectrum is consistent with that of a Type Ia at $z=0.581$
showing the a strong, broad \ion{Fe}{2} 5000\AA\ feature seen at
$\lambda_{obs} \sim 7900$\AA, but a relatively narrow \ion{Ca}{2}
3800\AA\ feature seen at $\lambda_{obs}\sim 6000$\AA.


\paragraph{SN~1997am} (Figure~\ref{9785_spec}). 
Separate extractions of the host galaxy and supernova
were possible since they were separated by 2.3\arcsec\ on the
slit. The host galaxy spectrum shows emission lines of [OII], H$\beta$
and [OIII] at a redshift of $z=0.416\pm 0.001$. The supernova spectrum
is a good match to the normal Type Ia SN~1992A spectrum at day +9 and
clearly shows the presence of the \ion{Si}{2} (6150\AA) feature seen
at $\lambda_{obs}\sim 8710$\AA.  SN 1991T at day +10 is also a good
fit to the spectrum (see Section \ref{peculiar}).


\paragraph{SN~1997ap} (Figure~\ref{9784_spec}). 
The redshift of $z=0.831 \pm 0.007$ is based on supernova features
alone. This spectrum was discussed in detail in \citet{nature98}.  The
slightly improved redshift estimate presented here superceeds that of
earlier papers \citep{nature98,42SNe_98,knop03}. We note here that it
was identified as a SN~Ia by the presence of both the blue \ion{Ca}{2}
and \ion{Si}{2} features.


\section{Tests for Evolution}
Although these spectra were taken primarily for the purposes of redshift
measurement and SN classification, they also allow some basic tests of
supernova evolution with redshift.  We show that SNe Ia do not look
dramatically different at high redshift, that they have similar
elemental velocities, and that SNe~Ia fall into the same sub-classes
at high redshift.

\subsection{Spectral morphology}
\begin{figure*} 
\epsscale{1.9} 
\plotone{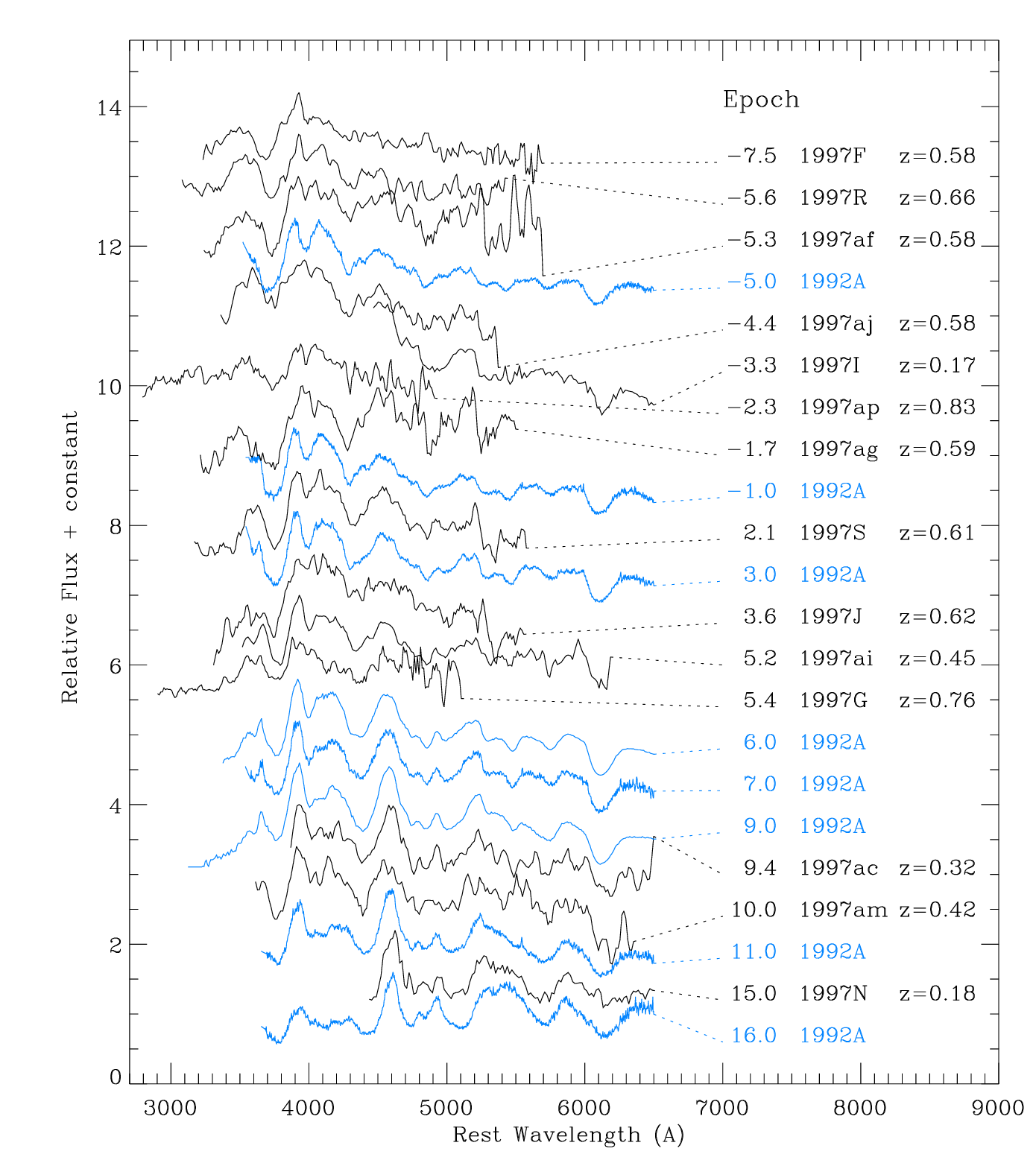}
\caption{The time sequence of high redshift supernova spectra in order
of rest-frame date relative to maximum light, as determined from the
light curve ($\tau_{lc}$). Spectra of the nearby Type Ia, SN
1992A are interspersed for comparison.}
\label{snseq}
\end{figure*}

High-redshift SNe Ia look similar to their low-redshift counterparts, but the
implications for constraints on evolution are complicated.
Figure~\ref{snseq} shows a selection of the high-redshift spectra
(those not significantly contaminated by host galaxy light) plotted in
order of rest frame days past maximum. These are interspersed with the
time sequence for the nearby Branch-normal Type Ia supernova SN
1992A. Note that because the high-redshift spectra are not
spectrophotometric measurements (see sections 3 and 5.2), differences
in overall slope of the spectra should not be considered
significant. Despite the lower S/N of the high-redshift spectra, and
the non-uniform wavelength coverage (since the spectra have been
de-redshifted by different amounts) it is clear that the overall
trends in the spectral evolution with light curve phase are the same
at low and high redshift. Furthermore, Figures \ref{974spec} to
\ref{9784_spec} show that high-redshift SNe resemble well observed
local SNe on an object-by-object basis.

\subsection{The incidence of spectroscopically peculiar SNe}
\label{peculiar}

\citet{li01a,li01b} make the case that, while slow-declining,
spectroscopically peculiar SNe like SN 1991T and SN 1999aa represent
20\% of SNe discovered in volume-limited local surveys, they are
strangely absent from high-redshift SN samples.  This is paradoxical,
since these SNe are over-luminous, and so should be seen in {\it
greater} numbers in flux-limited surveys.  The authors present one
possible solution --- that these SNe become spectroscopically normal
with time, and they appear normal by the time spectra are taken.

Perhaps supporting this idea, for two of our cases, both taken at
relatively late times after maximum light, the spectrum of SN 1991T
fits as well as the spectrum of a more normal supernova. SN 1997am has
a light curve stretch (a measure of the rise and decline time of the
supernova light curve, defined in \cite{42SNe_98}) of $s=1.03\pm~0.06$
\citep{knop03}, which is consistent with the stretch of SN 1991T
($s=1.08$). Since the spectrum was taken approximately 10 rest-frame
days after maximum light, by which time SN 1991T itself appears to be
fairly spectroscopically normal, there is no definitive spectroscopic
evidence that SN~1997am is a peculiar Type Ia, but it cannot be ruled
out. In the other case, SN~1997N, the spectrum was taken approximately
16 rest-frame days after maximum. It is possible that this SN looked
spectroscopically similar to SN 1991T at early times although again
there is no definitive spectroscopic evidence for this. The stretch of
SN~1997N is $s=1.03 \pm 0.02$, somewhat lower than that of
SN~1991T. However we note that the nearby SN 1997br also resembled SN
1991T spectroscopically, yet had a fairly normal light curve, with
$\Delta m_{15}(B)=1.00\pm 0.15$, $s=1.04$ \citep{li99}. Garavini et
al. (in preparation) have found definitive evidence for a SN
1991T-like supernova at high redshift. In that case the spectrum was
taken 7 days before maximum when the spectral peculiarities are more
evident.

The explanation for the supposed dearth of SN 1991T-like SNe at high
redshift proposed by Li et al. (that they are in the data set, but
appear spectroscopically normal when spectra are taken) appears to be
part of the answer to the paradox, although the situation may be more
complicated.  These SNe may not be as consistently over-luminous as
first thought \citep{li99,saha01,gibson01}, thus \citet{li01a} may
have over-predicted the expected numbers in a flux-limited survey.

At the other extreme, spectroscopically peculiar under-luminous SNe
like SN 1991bg \citep{filippenko92, leibundgut93, turatto96} and SN
1999by \citep{toth00, howell01a, vink01, hoeflich02} have not been
seen at high redshift.  While it is unlikely that these SNe will be
found in flux-limited searches, it is also possible that these SNe are
from such an old stellar population that they do not exist at $z >
0.5$ \citep{howell01b}.

\subsection{Calcium velocities}
While the depth of absorption lines gives some information about the
quantity of an ion in a SN, the velocity of the line gives information
about the ion's distribution within the photosphere.  Furthermore, the
velocity of the supernova ejecta is related to the overall kinetic
energy of the event. The kinetic energy has a direct influence on the
opacities and hence on the total brightness of the supernova and its
light-curve shape.

\begin{figure*} 
\epsscale{1.4}
\plotone{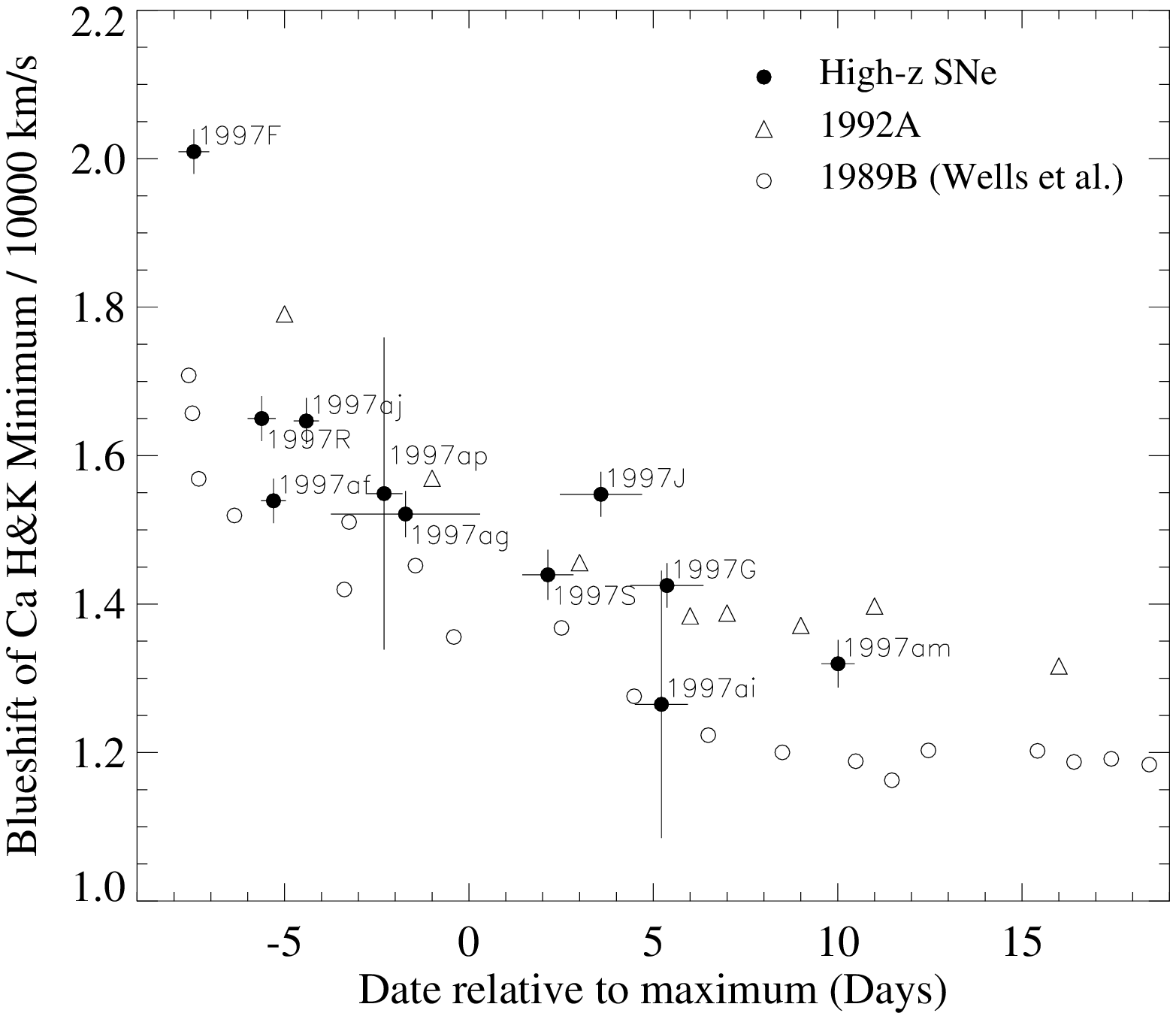}
\caption{Comparison of the velocities determined for the minima of the
\ion{Ca}{2} feature for two well observed nearby SNe~Ia (1992A
and 1989B) and the high-redshift SNe presented in this paper.} 
\label{ca2vel}
\end{figure*}

In Figure~\ref{ca2vel} we present a plot of the velocities of the
\ion{Ca}{2} minima for the SCP high-redshift supernovae compared
with two well-observed nearby supernovae. We note that while the
absorption feature at 3700\AA\ is dominated by \ion{Ca}{2},
\ion{Si}{2} and various ionization states of the iron-peak elements
\citep[see][]{hatano99} also contribute. For the
three lowest redshift supernovae in the SCP set, the \ion{Ca}{2}
feature is outside the observed wavelength range. Therefore this
analysis is restricted to the SCP supernovae with $z>0.38$.

The analysis was performed on the spectra after galaxy subtraction had
been done if this was necessary (as described in the previous
section). The \ion{Ca}{2} velocities were estimated by fitting a
Gaussian function to the region 3610\AA\ to 3870\AA\ in the rest
frame, followed by a second iteration using the region from -65\AA\ to
+100\AA\ either side of the previous estimate of the minimum. The
velocity shift was calculated from the wavelength of the minimum by
assuming a rest frame wavelength for this feature of 3945.28\AA.  The
same analysis was done on a spectral series for the normal Type~Ia
SN1992A \citep{kir92a}. Finally the published \ion{Ca}{2} velocities
for the normal Type~Ia supernova 1989B \citep{wells94} are also
shown.

The general trend of decreasing \ion{Ca}{2} velocities as a function
of time and the range in velocities at a given epoch seen in the data
is completely consistent with the range seen in nearby observations.
This result is confirmed by Garavini et al. (in preparation) in a more
recent independent set of high-redshift SN spectra.

Our high-redshift data is also broadly consistent with the recent
study of nearby SNe by \cite{Benetti05}, although the comparison is
complicated by the fact that Benetti et al. measure \ion{Si}{2}
velocities whereas we consider \ion{Ca}{2} velocities. Taking into
account the apparent offset between the two (as can be seen from
SN~1992A and SN~1989B for example), the velocities seen in our
high-redshift sample (including the large value for SN~1997F) are
broadly consistent with the range of velocities seen in nearby
SNe. Because of the larger error bars at high redshift, and only
having a single velocity measurement per SN, it is not possible to
determine into which of the \cite{Benetti05} classes each of our
objects would fall.

\section{Conclusions}

At the present time, the statistical and systematic errors in the
measurement of the cosmological parameters from Type Ia supernovae are
of similar size. As the quality and quantity of high-redshift
supernova observations grows it will become even more important to
constrain the systematic uncertainties. Non-Ia contamination and
evolution are two of the larger potential systematics we currently
face. Here we have presented some of the methods we employ to
reduce/study these uncertainties.

In this paper we have presented spectra for 14 high-redshift
supernovae and demonstrated that the spectra are consistent with Type
Ia.  In three cases at intermediate redshift (SN~1997ac, SN~1997ai and
SN~1997am at $z\sim 0.3-0.5$) in addition to two at lower redshift
(SN~1997I and SN~1997N) we have observed the \ion{Si}{2} 6150\AA\
feature and hence have unambiguously identified the SN as Type Ia. For
higher redshifts this \ion{Si}{2} feature becomes increasingly
difficult to observe as fringing and poor CCD response in the red, and
bright OH lines in the sky background makes the spectra very noisy
redward of $\sim 9500$\AA. Here we have used spectral matching
combined with the identification of specific spectral features to
identify the objects as SN~Ia, the most important being the
identification of \ion{Si}{2} features near 4000\AA.

We have carried out first-order quantitative tests to compare the
high-redshift spectra with their low-redshift counterparts.

We show that the spectral phase determined from spectral matching to
low-redshift Type Ia spectra is in very good agreement with the phase
determined from the high-redshift light curve.  Similarly the spectral
time series shows the same overall trends at low and high
redshift. Finally, quantitative measurements of the Calcium ejection
velocity in the high-redshift spectra are also consistent with those
measured from the spectra of low-redshift Branch-normal Type Ia
SNe. Therefore we have found no evidence for evolution in the
population of Type Ia SNe up to redshifts of $z\sim 0.8$.  While we
cannot prove that all SNe Ia at high redshift are identical to their
low-redshift counterparts, we can say that it is possible to choose a
set of high-redshift SNe Ia that are equivalent to low-redshift
counterparts to within the accuracy of these quantitative tests.

With larger samples, such as those now being collected by the
Supernova Legacy Survey (SNLS, \cite{pritchet05}) and the ESSENCE
project (\cite{Matheson} and references therein) it will be possible
to make more detailed comparisons in smaller redshift bins in the
range $0.2<z<1$ and in subsets based on light curve stretch, galaxy
host type and other factors. Such studies will provide further
confidence in the use of SNe~Ia as distance indicators, and, in cases
where quantitative measurements of spectral features are found to
correlate with luminosity or light curve stretch
(e.g. \citealt{nugseq95,Fol04}), these can be used to reduce scatter
in the Hubble diagram and thus improve the precision to which
cosmological parameters may be measured.

\section*{Acknowledgements} 

The authors acknowledge the help of the night assistants and support 
staff at the telescopes from which data for this paper were obtained.

We thank the anonymous referee for a very thorough reading of the
paper and helpful suggestions.

This paper makes use of light curve photometry collected at the Cerro
Tololo Inter-American Observatory, which is operated by Association of
Universities for Research in Astronomy, Inc. under a cooperative
agreement with the National Science Foundation. Based in part on
observations obtained at the WIYN Observatory, which is a joint
facility of the University of Wisconsin-Madison, Indiana University,
Yale University, and the National Optical Astronomy Observatory. We
also make use of observations made with the NASA/ESA Hubble Space
Telescope, obtained at the Space Telescope Science Institute, which is
operated by the Association of Universities for Research in Astronomy,
Inc., under NASA contract NAS 5-26555. These observations are
associated with programs DD-7590 and GO-7336.

This work was supported in part by the Royal Swedish Academy of
Sciences and by the Director, Office of Science, Office of High Energy
and Nuclear Physics, of the US Department of Energy under Contract
No. DE-AC03-76SF00098.  Support for this work was provided by NASA
through grant HST-GO-7336 from the Space Telescope Science Institute,
which is operated by the Association of Universities for Research in
Astronomy, Inc., under NASA contract NAS 5-26555.


\begin{thebibliography}{51}
\expandafter\ifx\csname natexlab\endcsname\relax\def\natexlab#1{#1}\fi
\expandafter\ifx\csname url\endcsname\relax
  \def\url#1{\texttt{#1}}\fi
\expandafter\ifx\csname urlprefix\endcsname\relax\def\urlprefix{URL }\fi
\providecommand{\eprint}[2][]{\url{#2}}

\bibitem[{{Baldwin} \& {Stone}(1984)}]{BS84}
{Baldwin}, J.~A., \& {Stone}, R.~P.~S. 1984, \mnras, 206, 241

\bibitem[{{Barris} et~al.(2004){Barris}, {Tonry}, {Blondin}, {Challis},
  {Chornock}, {Clocchiatti}, {Filippenko}, {Garnavich}, {Holland}, {Jha},
  {Kirshner}, {Krisciunas}, {Leibundgut}, {Li}, {Matheson}, {Miknaitis},
  {Phillips}, {Riess}, {Schmidt}, {Smith}, {Sollerman}, {Spyromilio}, {Stubbs},
  {Suntzeff}, {Aussel}, {Chambers}, {Connelley}, D., {Henry}, {Kaiser}, {Liu},
  {Martin}, \& {Wainscoat}}]{barris04}
{Barris}, B.~J., {Tonry}, J., {Blondin}, S., {Challis}, P., {Chornock}, R.,
  {Clocchiatti}, A., {Filippenko}, A.~V., {Garnavich}, P., {Holland}, S.~T.,
  {Jha}, S., {Kirshner}, R.~P., {Krisciunas}, K., {Leibundgut}, B., {Li}, W.,
  {Matheson}, T., {Miknaitis}, G., {Phillips}, M.~M., {Riess}, A.~G.,
  {Schmidt}, B., {Smith}, R.~C., {Sollerman}, J., {Spyromilio}, J., {Stubbs},
  C.~W., {Suntzeff}, N.~B., {Aussel}, H., {Chambers}, K.~C., {Connelley},
  M.~S., D., D., {Henry}, J., {Kaiser}, N., {Liu}, M., {Martin}, E., \&
  {Wainscoat}, R.~J. 2004, \apj, 602, 571

\bibitem[{{Benetti} et~al.(2005){Benetti}, {Cappellaro}, {Mazzali}, {Turatto},
  {Altavilla}, {Bufano}, {Elias-Rosa}, {Kotak}, {Pignata}, {Salvo}, \&
  {Stanishev}}]{Benetti05}
{Benetti}, S., {Cappellaro}, E., {Mazzali}, P.~A., {Turatto}, M., {Altavilla},
  G., {Bufano}, F., {Elias-Rosa}, N., {Kotak}, R., {Pignata}, G., {Salvo}, M.,
  \& {Stanishev}, V. 2005, \apj, 623, 1101

\bibitem[{{Buzzoni} et~al.(2004){Buzzoni}, {Delabre}, {Dekker}, {Dodorico},
  {Enard}, {Focardi}, {Gustafsson}, {Nees}, {Paureau}, \& {Reiss}}]{Buzzoni84}
{Buzzoni}, B., {Delabre}, B., {Dekker}, H., {Dodorico}, S., {Enard}, D.,
  {Focardi}, P., {Gustafsson}, B., {Nees}, W., {Paureau}, J., \& {Reiss}, R.
  2004, ESO Messenger

\bibitem[{{Cappellaro} et~al.(1995){Cappellaro}, {Turatto}, \& {Fernley}}]{IUE}
{Cappellaro}, E., {Turatto}, M., \& {Fernley}, J. 1995, {IUE-ULDA Access Guide
  No. 6: Supernovae} (ESA SCIENTIFIC PUBLICATION ESA-SP 1189)

\bibitem[{Cardelli et~al.(1989)Cardelli, Clayton, \& Mathis}]{card89}
Cardelli, J.~A., Clayton, G.~C., \& Mathis, J.~S. 1989, ApJ, 345, 245

\bibitem[{{Clocchiatti} et~al.(1996){Clocchiatti}, {Wheeler}, {Brotherton},
  {Cochran}, {Wills}, {Barker}, \& {Turatto}}]{1996ApJ...462..462C}
{Clocchiatti}, A., {Wheeler}, J.~C., {Brotherton}, M.~S., {Cochran}, A.~L.,
  {Wills}, D., {Barker}, E.~S., \& {Turatto}, M. 1996, \apj, 462, 462

\bibitem[{{Coil} et~al.(2000){Coil}, {Matheson}, {Filippenko}, {Leonard},
  {Tonry}, {Riess}, {Challis}, {Clocchiatti}, {Garnavich}, {Hogan}, {Jha},
  {Kirshner}, {Leibundgut}, {Phillips}, {Schmidt}, {Schommer}, {Smith},
  {Soderberg}, {Spyromilio}, {Stubbs}, {Suntzeff}, \& {Woudt}}]{coil00}
{Coil}, A.~L., {Matheson}, T., {Filippenko}, A.~V., {Leonard}, D.~C., {Tonry},
  J., {Riess}, A.~G., {Challis}, P., {Clocchiatti}, A., {Garnavich}, P.~M.,
  {Hogan}, C.~J., {Jha}, S., {Kirshner}, R.~P., {Leibundgut}, B., {Phillips},
  M.~M., {Schmidt}, B.~P., {Schommer}, R.~A., {Smith}, R.~C., {Soderberg},
  A.~M., {Spyromilio}, J., {Stubbs}, C., {Suntzeff}, N.~B., \& {Woudt}, P.
  2000, \apjl, 544, L111

\bibitem[{{Filippenko}(1997)}]{1997ARAA..35..309F}
{Filippenko}, A.~V. 1997, \araa, 35, 309

\bibitem[{{Filippenko} et~al.(1992){Filippenko}, {Richmond}, {Branch},
  {Gaskell}, {Herbst}, {Ford}, {Treffers}, {Matheson}, {Ho}, {Dey}, {Sargent},
  {Small}, \& {van Breugel}}]{filippenko92}
{Filippenko}, A.~V., {Richmond}, M.~W., {Branch}, D., {Gaskell}, M., {Herbst},
  W., {Ford}, C.~H., {Treffers}, R.~R., {Matheson}, T., {Ho}, L.~C., {Dey}, A.,
  {Sargent}, W.~L.~W., {Small}, T.~A., \& {van Breugel}, W.~J.~M. 1992, \aj,
  104, 1543

\bibitem[{Folatelli(2004)}]{Fol04}
Folatelli, G. 2004, Ph.D. thesis, Stockholm University

\bibitem[{Garnavich et~al.(1998)}]{garn_w_98}
Garnavich, P., et~al. 1998, ApJ, 509, 74

\bibitem[{{Gibson} \& {Stetson}(2001)}]{gibson01}
{Gibson}, B.~K., \& {Stetson}, P.~B. 2001, \apjl, 547, L103

\bibitem[{{H{\" o}flich} et~al.(2002){H{\" o}flich}, {Gerardy}, {Fesen}, \&
  {Sakai}}]{hoeflich02}
{H{\" o}flich}, P., {Gerardy}, C.~L., {Fesen}, R.~A., \& {Sakai}, S. 2002,
  \apj, 568, 791

\bibitem[{{Hamuy}(1994)}]{Hamuy94}
{Hamuy}, M. 1994, \pasp, 106, 566

\bibitem[{{Hatano} et~al.(1999){Hatano}, {Branch}, {Fisher}, {Millard}, \&
  {Baron}}]{hatano99}
{Hatano}, K., {Branch}, D., {Fisher}, A., {Millard}, J., \& {Baron}, E. 1999,
  ApJS, 121, 233

\bibitem[{{Howell}(2001)}]{howell01b}
{Howell}, D.~A. 2001, \apjl, 554, L193

\bibitem[{{Howell} et~al.(2001){Howell}, {H{\" o}flich}, {Wang}, \&
  {Wheeler}}]{howell01a}
{Howell}, D.~A., {H{\" o}flich}, P., {Wang}, L., \& {Wheeler}, J.~C. 2001,
  \apj, 556, 302

\bibitem[{{Howell} \& {Wang}(2002)}]{howell02}
{Howell}, D.~A., \& {Wang}, L. 2002, American Astronomical Society Meeting,
  201, 0

\bibitem[{{Kinney} et~al.(1993){Kinney}, {Bohlin}, {Calzetti}, {Panagia}, \&
  {Wyse}}]{kinney93}
{Kinney}, A.~L., {Bohlin}, R.~C., {Calzetti}, D., {Panagia}, N., \& {Wyse},
  R.~F.~G. 1993, \apjs, 86, 5

\bibitem[{Kirshner et~al.(1993)}]{kir92a}
Kirshner, R., et~al. 1993, ApJ, 415, 589

\bibitem[{{Knop} et~al.(2003){Knop}, {Aldering}, {Amanullah}, {Astier},
  {Blanc}, {Burns}, {Conley}, {Deustua}, {Doi}, {Fabbro}, {Folatelli},
  {Fruchter}, {Garavini}, {Gibbons}, {Goldhaber}, {Goobar}, {Groom}, {Hardin},
  {Hook}, {Howell}, {Irwin}, {Kim}, {Knop}, {Lidman}, {McMahon}, {Mendez},
  {Nobili}, {Nugent}, {Pain}, {Panagia}, {Pennypacker}, {Perlmutter}, {Quimby},
  {Raux}, {Regnault}, {Ruiz-Lapuente}, {Schaefer}, {Schahmaneche}, {Spadafora},
  {Walton}, {Wang}, {Wood-Vasey}, \& {Yasuda}}]{knop03}
{Knop}, R., {Aldering}, G., {Amanullah}, R., {Astier}, P., {Blanc}, G.,
  {Burns}, M.~S., {Conley}, A., {Deustua}, S.~E., {Doi}, M., {Fabbro}, S.,
  {Folatelli}, G., {Fruchter}, A.~S., {Garavini}, G., {Gibbons}, R.,
  {Goldhaber}, G., {Goobar}, A., {Groom}, D.~E., {Hardin}, D., {Hook}, I.,
  {Howell}, D.~A., {Irwin}, M., {Kim}, A.~G., {Knop}, R.~A., {Lidman}, C.,
  {McMahon}, R., {Mendez}, J., {Nobili}, S., {Nugent}, P.~E., {Pain}, R.,
  {Panagia}, N., {Pennypacker}, C.~R., {Perlmutter}, S., {Quimby}, R., {Raux},
  J., {Regnault}, N., {Ruiz-Lapuente}, P., {Schaefer}, B., {Schahmaneche}, K.,
  {Spadafora}, A.~L., {Walton}, N.~A., {Wang}, L., {Wood-Vasey}, W.~M., \&
  {Yasuda}, N. 2003, \apj, 598, 102

\bibitem[{{Leibundgut} et~al.(1993){Leibundgut}, {Kirshner}, {Phillips},
  {Wells}, {Suntzeff}, {Hamuy}, {Schommer}, {Walker}, {Gonzalez}, {Ugarte},
  {Williams}, {Williger}, {Gomez}, {Marzke}, {Schmidt}, {Whitney}, {Coldwell},
  {Peters}, {Chaffee}, {Foltz}, {Rehner}, {Siciliano}, {Barnes}, {Cheng},
  {Hintzen}, {Kim}, {Maza}, {Parker}, {Porter}, {Schmidtke}, \&
  {Sonneborn}}]{leibundgut93}
{Leibundgut}, B., {Kirshner}, R.~P., {Phillips}, M.~M., {Wells}, L.~A.,
  {Suntzeff}, N.~B., {Hamuy}, M., {Schommer}, R.~A., {Walker}, A.~R.,
  {Gonzalez}, L., {Ugarte}, P., {Williams}, R.~E., {Williger}, G., {Gomez}, M.,
  {Marzke}, R., {Schmidt}, B.~P., {Whitney}, B., {Coldwell}, N., {Peters}, J.,
  {Chaffee}, F.~H., {Foltz}, C.~B., {Rehner}, D., {Siciliano}, L., {Barnes},
  T.~G., {Cheng}, K.-P., {Hintzen}, P.~M.~N., {Kim}, Y.-C., {Maza}, J.,
  {Parker}, J.~W., {Porter}, A.~C., {Schmidtke}, P.~C., \& {Sonneborn}, G.
  1993, \aj, 105, 301

\bibitem[{{Li} et~al.(2001{\natexlab{a}}){Li}, {Filippenko}, \&
  {Riess}}]{li01a}
{Li}, W., {Filippenko}, A.~V., \& {Riess}, A.~G. 2001{\natexlab{a}}, \apj, 546,
  719

\bibitem[{{Li} et~al.(2001{\natexlab{b}}){Li}, {Filippenko}, {Treffers},
  {Riess}, {Hu}, \& {Qiu}}]{li01b}
{Li}, W., {Filippenko}, A.~V., {Treffers}, R.~R., {Riess}, A.~G., {Hu}, J., \&
  {Qiu}, Y. 2001{\natexlab{b}}, \apj, 546, 734

\bibitem[{{Li} et~al.(1999){Li}, {Qiu}, {Qiao}, {Zhu}, {Hu}, {Richmond},
  {Filippenko}, {Treffers}, {Peng}, \& {Leonard}}]{li99}
{Li}, W.~D., {Qiu}, Y.~L., {Qiao}, Q.~Y., {Zhu}, X.~H., {Hu}, J.~Y.,
  {Richmond}, M.~W., {Filippenko}, A.~V., {Treffers}, R.~R., {Peng}, C.~Y., \&
  {Leonard}, D.~C. 1999, \aj, 117, 2709

\bibitem[{{Lidman} et~al.(2005)}]{lidman05}
{Lidman}, C., et~al. 2005, A\&A, 430, 843

\bibitem[{Matheson et~al.(2005)}]{Matheson}
Matheson, T., et~al. 2005, AJ, 129, 2352

\bibitem[{{Meikle} et~al.(1996){Meikle}, {Cumming}, {Geballe}, {Lewis},
  {Walton}, {Balcells}, {Cimatti}, {Croom}, {Dhillon}, {Economou}, {Jenkins},
  {Knapen}, {Meadows}, {Morris}, {Perez-Fournon}, {Shanks}, {Smith}, {Tanvir},
  {Veilleux}, {Vilchez}, {Wall}, \& {Lucey}}]{1996MNRAS.281..263M}
{Meikle}, W.~P.~S., {Cumming}, R.~J., {Geballe}, T.~R., {Lewis}, J.~R.,
  {Walton}, N.~A., {Balcells}, M., {Cimatti}, A., {Croom}, S.~M., {Dhillon},
  V.~S., {Economou}, F., {Jenkins}, C.~R., {Knapen}, J.~H., {Meadows}, V.~S.,
  {Morris}, P.~W., {Perez-Fournon}, I., {Shanks}, T., {Smith}, L.~J., {Tanvir},
  N.~R., {Veilleux}, S., {Vilchez}, J., {Wall}, J.~V., \& {Lucey}, J.~R. 1996,
  \mnras, 281, 263

\bibitem[{Nugent et~al.(1995)Nugent, Phillips, Baron, Branch, \&
  Hauschildt}]{nugseq95}
Nugent, P., Phillips, M., Baron, E., Branch, D., \& Hauschildt, P. 1995, ApJ,
  455, L147

\bibitem[{{O'Donnell}(1994)}]{ODonn94}
{O'Donnell}, J.~E. 1994, ApJ, 422, 158

\bibitem[{{Oke} et~al.(1995){Oke}, {Cohen}, {Carr}, {Cromer}, {Dingizian},
  {Harris}, {Labrecque}, {Lucinio}, {Schaal}, {Epps}, \& {Miller}}]{B095}
{Oke}, J., {Cohen}, J., {Carr}, M., {Cromer}, J., {Dingizian}, A., {Harris},
  F., {Labrecque}, S., {Lucinio}, R., {Schaal}, W., {Epps}, H., \& {Miller}, J.
  1995, \pasp, 375

\bibitem[{{Oke} \& {Gunn}(1983)}]{oke_abmag_83}
{Oke}, J.~B., \& {Gunn}, J.~E. 1983, ApJ, 266, 713

\bibitem[{{Perlmutter} et~al.(1995){Perlmutter}, {Pennypacker}, {Goldhaber},
  {Goobar}, {Muller}, {Newberg}, {Desai}, {Kim}, {Kim}, {Small}, {Boyle},
  {Crawford}, {McMahon}, {Bunclark}, {Carter}, {Irwin}, {Terlevich}, {Ellis},
  {Glazebrook}, {Couch}, {Mould}, {Small}, \& {Abraham}}]{sn92bi}
{Perlmutter}, S., {Pennypacker}, C.~R., {Goldhaber}, G., {Goobar}, A.,
  {Muller}, R.~A., {Newberg}, H.~J.~M., {Desai}, J., {Kim}, A.~G., {Kim},
  M.~Y., {Small}, I.~A., {Boyle}, B.~J., {Crawford}, C.~S., {McMahon}, R.~G.,
  {Bunclark}, P.~S., {Carter}, D., {Irwin}, M.~J., {Terlevich}, R.~J., {Ellis},
  R.~S., {Glazebrook}, K., {Couch}, W.~J., {Mould}, J.~R., {Small}, T.~A., \&
  {Abraham}, R.~G. 1995, \apjl, 440, L41

\bibitem[{Perlmutter et~al.(1997)}]{perl96}
Perlmutter, S., et~al. 1997, in Thermonuclear Supernova, edited by
  P.~Ruiz-Lapuente, R.~Canal, \& J.Isern (Dordrecht: Kluwer), 749

\bibitem[{Perlmutter et~al.(1998)}]{nature98}
--- 1998, Nature, 391

\bibitem[{Perlmutter et~al.(1999)}]{42SNe_98}
--- 1999, ApJ, 517, 565

\bibitem[{{Pritchet}(2005)}]{pritchet05}
{Pritchet}, C. 2005, in Observing Dark Energy, edited by S.~Wolff, \& T.~Lauer
  (Provo: BYU Press)

\bibitem[{{Richardson} et~al.(2002){Richardson}, {Thomas}, {Casebeer},
  {Branch}, \& {Baron}}]{Richardson02}
{Richardson}, D., {Thomas}, R., {Casebeer}, D., {Branch}, D., \& {Baron}, E.
  2002, American Astronomical Society, 201st AAS Meeting, \#56.09; Bulletin of
  the American Astronomical Society, 34, 1205

\bibitem[{Riess et~al.(1998)}]{riess_acc_98}
Riess, A., et~al. 1998, AJ, 116, 1009

\bibitem[{{Riess} et~al.(2004){Riess}, {Strolger}, {Tonry}, {Casertano},
  {Ferguson}, {Mobasher}, {Chalis}, {Filippenko}, {Li}, {Chornock}, {Kirshner},
  {Leibundgut}, {Dickinson}, {Livio}, {Giavalisco}, {Steidel}, {Benites}, \&
  {Tsvetanov}}]{reiss04}
{Riess}, A.~G., {Strolger}, L.-G., {Tonry}, J., {Casertano}, S., {Ferguson},
  H.~C., {Mobasher}, B., {Chalis}, P., {Filippenko}, S., A.~V.and~{Jha}, {Li},
  W., {Chornock}, R., {Kirshner}, R.~P., {Leibundgut}, B., {Dickinson}, M.,
  {Livio}, M., {Giavalisco}, M., {Steidel}, C.~C., {Benites}, T., \&
  {Tsvetanov}, Z. 2004, apj, 607, 665

\bibitem[{{Saha} et~al.(2001){Saha}, {Sandage}, {Thim}, {Labhardt}, {Tammann},
  {Christensen}, {Panagia}, \& {Macchetto}}]{saha01}
{Saha}, A., {Sandage}, A., {Thim}, F., {Labhardt}, L., {Tammann}, G.~A.,
  {Christensen}, J., {Panagia}, N., \& {Macchetto}, F.~D. 2001, \apj, 551, 973

\bibitem[{{Schmidt} et~al.(1998){Schmidt}, {Suntzeff}, {Phillips}, {Schommer},
  {Clocchiatti}, {Kirshner}, {Garnavich}, {Challis}, {Leibundgut},
  {Spyromilio}, {Riess}, {Filippenko}, {Hamuy}, {Smith}, {Hogan}, {Stubbs},
  {Diercks}, {Reiss}, {Gilliland}, {Tonry}, {Maza}, {Dressler}, {Walsh}, \&
  {Ciardullo}}]{schmidt_98}
{Schmidt}, B.~P., {Suntzeff}, N.~B., {Phillips}, M.~M., {Schommer}, R.~A.,
  {Clocchiatti}, A., {Kirshner}, R.~P., {Garnavich}, P., {Challis}, P.,
  {Leibundgut}, B., {Spyromilio}, J., {Riess}, A.~G., {Filippenko}, A.~V.,
  {Hamuy}, M., {Smith}, R.~C., {Hogan}, C., {Stubbs}, C., {Diercks}, A.,
  {Reiss}, D., {Gilliland}, R., {Tonry}, J., {Maza}, J., {Dressler}, A.,
  {Walsh}, J., \& {Ciardullo}, R. 1998, ApJ, 507, 46

\bibitem[{{Stone} \& {Baldwin}(1983)}]{SB83}
{Stone}, R.~P.~S., \& {Baldwin}, J.~A. 1983, \mnras, 204, 347

\bibitem[{{Sullivan} et~al.(2003){Sullivan}, {Ellis}, {Aldering}, {Amanullah},
  {Astier}, {Blanc}, {Burns}, {Conley}, {Deustua}, {Doi}, {Fabbro},
  {Folatelli}, {Fruchter}, {Garavini}, {Gibbons}, {Goldhaber}, {Goobar},
  {Groom}, {Hardin}, {Hook}, {Howell}, {Irwin}, {Kim}, {Knop}, {Lidman},
  {McMahon}, {Mendez}, {Nobili}, {Nugent}, {Pain}, {Panagia}, {Pennypacker},
  {Perlmutter}, {Quimby}, {Raux}, {Regnault}, {Ruiz-Lapuente}, {Schaefer},
  {Schahmaneche}, {Spadafora}, {Walton}, {Wang}, {Wood-Vasey}, \&
  {Yasuda}}]{sullivan03}
{Sullivan}, M., {Ellis}, R.~S., {Aldering}, G., {Amanullah}, R., {Astier}, P.,
  {Blanc}, G., {Burns}, M.~S., {Conley}, A., {Deustua}, S.~E., {Doi}, M.,
  {Fabbro}, S., {Folatelli}, G., {Fruchter}, A.~S., {Garavini}, G., {Gibbons},
  R., {Goldhaber}, G., {Goobar}, A., {Groom}, D.~E., {Hardin}, D., {Hook}, I.,
  {Howell}, D.~A., {Irwin}, M., {Kim}, A.~G., {Knop}, R.~A., {Lidman}, C.,
  {McMahon}, R., {Mendez}, J., {Nobili}, S., {Nugent}, P.~E., {Pain}, R.,
  {Panagia}, N., {Pennypacker}, C.~R., {Perlmutter}, S., {Quimby}, R., {Raux},
  J., {Regnault}, N., {Ruiz-Lapuente}, P., {Schaefer}, B., {Schahmaneche}, K.,
  {Spadafora}, A.~L., {Walton}, N.~A., {Wang}, L., {Wood-Vasey}, W.~M., \&
  {Yasuda}, N. 2003, \mnras, 340, 1057

\bibitem[{{Tonry} et~al.(2003){Tonry}, {Schmidt}, {Barris}, {Candia},
  {Challis}, {Clocchiatti}, {Coil}, {Filippenko}, {Garnavich}, {Hogan},
  {Holland}, {Jha}, {Kirshner}, {Krisciunas}, {Leibundgut}, {Li}, {Matheson},
  {Phillips}, {Riess}, {Schommer}, {Smith}, {Sollerman}, {Spyromilio},
  {Stubbs}, \& {Suntzeff}}]{tonry03}
{Tonry}, J.~L., {Schmidt}, B.~P., {Barris}, B., {Candia}, P., {Challis}, P.,
  {Clocchiatti}, A., {Coil}, A.~L., {Filippenko}, A.~V., {Garnavich}, P.,
  {Hogan}, C., {Holland}, S.~T., {Jha}, S., {Kirshner}, R.~P., {Krisciunas},
  K., {Leibundgut}, B., {Li}, W., {Matheson}, T., {Phillips}, M.~M., {Riess},
  A.~G., {Schommer}, R., {Smith}, R.~C., {Sollerman}, J., {Spyromilio}, J.,
  {Stubbs}, C.~W., \& {Suntzeff}, N.~B. 2003, apj, 594, 1

\bibitem[{{Toth} \& {Szab{\' o}}(2000)}]{toth00}
{Toth}, I., \& {Szab{\' o}}, R. 2000, \aap, 361, 63

\bibitem[{{Turatto} et~al.(1996){Turatto}, {Benetti}, {Cappellaro}, {Danziger},
  {della Valle}, {Gouiffes}, {Mazzali}, \& {Patat}}]{turatto96}
{Turatto}, M., {Benetti}, S., {Cappellaro}, E., {Danziger}, I.~J., {della
  Valle}, M., {Gouiffes}, C., {Mazzali}, P.~A., \& {Patat}, F. 1996, \mnras,
  283, 1

\bibitem[{{Vink{\' o}} et~al.(2001){Vink{\' o}}, {Kiss}, {Cs{\' a}k}, {F{\H
  u}r{\' e}sz}, {Szab{\' o}}, {Thomson}, \& {Mochnacki}}]{vink01}
{Vink{\' o}}, J., {Kiss}, L.~L., {Cs{\' a}k}, B., {F{\H u}r{\' e}sz}, G.,
  {Szab{\' o}}, R., {Thomson}, J.~R., \& {Mochnacki}, S.~W. 2001, \aj, 121,
  3127

\bibitem[{{Wells} et~al.(1994){Wells}, {Phillips}, {Suntzeff}, {Heathcote},
  {Hamuy}, {Navarrete}, {Fernandez}, {Weller}, {Schommer}, {Kirshner},
  {Leibundgut}, {Willner}, {Peletier}, {Schlegel}, {Wheeler}, {Harkness},
  {Bell}, {Matthews}, {Filippenko}, {Shields}, {Richmond}, {Jewitt}, {Luu},
  {Tran}, {Appleton}, {Robson}, {Tyson}, {Guhathakurta}, {Eder}, {Bond},
  {Potter}, {Veilleux}, {Porter}, {Humphreys}, {Janes}, {Williams}, {Costa},
  {Ruiz}, {Lee}, {Lutz}, {Rich}, {Winkler}, \& {Tyson}}]{1994AJ....108.2233W}
{Wells}, L.~A., {Phillips}, M.~M., {Suntzeff}, B., {Heathcote}, S.~R., {Hamuy},
  M., {Navarrete}, M., {Fernandez}, M., {Weller}, W.~G., {Schommer}, R.~A.,
  {Kirshner}, R.~P., {Leibundgut}, B., {Willner}, S.~P., {Peletier}, S.~P.,
  {Schlegel}, E.~M., {Wheeler}, J.~C., {Harkness}, R.~P., {Bell}, D.~J.,
  {Matthews}, J.~M., {Filippenko}, A.~V., {Shields}, J.~C., {Richmond}, M.~W.,
  {Jewitt}, D., {Luu}, J., {Tran}, H.~D., {Appleton}, P.~N., {Robson}, E.~I.,
  {Tyson}, J.~A., {Guhathakurta}, P., {Eder}, J.~A., {Bond}, H.~E., {Potter},
  M., {Veilleux}, S., {Porter}, A.~C., {Humphreys}, R.~M., {Janes}, K.~A.,
  {Williams}, T.~B., {Costa}, E., {Ruiz}, M.~T., {Lee}, J.~T., {Lutz}, J.~H.,
  {Rich}, R.~M., {Winkler}, P.~F., \& {Tyson}, N.~D. 1994, \aj, 108, 2233

\bibitem[{Wells et~al.(1994)}]{wells94}
Wells, L.~A., et~al. 1994, AJ, 108, 2233

\end{thebibliography}

\clearpage

\end{document}